\documentclass[fleqn,usenatbib,onecolumn]{mnras}

\usepackage{newtxtext,newtxmath}

\usepackage[T1]{fontenc}

\DeclareRobustCommand{\VAN}[3]{#2}
\let\VANthebibliography\thebibliography
\def\thebibliography{\DeclareRobustCommand{\VAN}[3]{##3}\VANthebibliography}

\usepackage{graphicx}	
\usepackage{amsmath}	
\usepackage{multirow}

\def\araa{ARA\&A}
\def\apj{ApJ}
\def\apjl{ApJ}
\def\apjs{ApJS}
\def\ao{Appl.~Opt.}
\def\aap{A\&A}
\def\aaps{A\&AS}
\def\mnras{MNRAS}
\def\pasj{PASJ}
\def\nat{Nature}
\def\physrep{Phys.~Rep.}

\newcommand{\be}{\begin{equation}}
\newcommand{\ee}{\end{equation}}
\newcommand{\bary}{\begin{eqnarray}}
\newcommand{\eary}{\end{eqnarray}}

\title[Reverse shock, closure relations and LAT GRBs]{Closure Relations of Synchrotron Self-Compton from Reverse shock and \textit{Fermi}-LAT GRBs}

\author[Fraija et al.]{
Nissim Fraija,$^{1}$\thanks{E-mail: nifraija@astro.unam.mx}
B. Betancourt Kamenetskaia,$^{2}$
Antonio Galv\'an G\'amez$^{3}$
\newauthor
Maria G. Dainotti$^{4,5,6,7,8}$
and Hermes Le\'on Vargas$^{3}$
\\
$^{1}$Instituto de Astronom\' ia, Universidad Nacional Aut\'onoma de M\'exico, Circuito Exterior, C.U., A. Postal 70-264, 04510 M\'exico City, M\'exico\\
$^{2}$Cosmology, Gravity, and Astroparticle Physics Group, Center for Theoretical Physics of the Universe,
Institute for Basic Science (IBS), Daejeon, 34126, Korea\\
$^{3}$Instituto de F\' isica, Universidad Nacional Aut\'onoma de M\'exico, Circuito Exterior, C.U., A. Postal 70-264, 04510 M\'exico City, M\'exico\\
$^{4}$Division of Science, National Astronomical Observatory of Japan, 2-21-1 Osawa, Mitaka, Tokyo 181-8588, Japan \\
$^{5}$The Graduate University for Advanced Studies (SOKENDAI), Shonankokusaimura, Hayama, Miura District, Kanagawa 240-0115, Japan \\
$^{6}$Space Science Institute, 4765 Walnut St, Suite B, Boulder, CO 80301, USA 4 Nevada \\
$^{7}$Center for Astrophysics, University of Nevada, 4505 Maryland Parkway, Las Vegas, NV 89154, USA \\
$^{8}$Bay Environmental Institute, P.O. Box 25 Moffett Field, CA 94035, California \\
}

\date{Accepted XXX. Received YYY; in original form ZZZ}

\pubyear{\the\year{}}

\begin{document}
\label{firstpage}
\pagerange{\pageref{firstpage}--\pageref{lastpage}}
\maketitle

\begin{abstract}
Synchrotron radiation from the reverse- and forward-shock regions typically describes the evolution of temporal and spectral features given by the closure relations (CRs) during the late and long-/short-lasting emission in the afterglow phase of Gamma-ray bursts (GRBs). Although synchrotron photons are restricted to keV and a few MeV energies, the synchrotron self-Compton (SSC) mechanism can disperse them above hundreds of MeV energies. We present the CRs of the SSC process radiated from the reverse-shock region for the case of a thick and thin shell, considering that the reverse shock lies in the adiabatic regime and evolves in an environment with a homogeneous and stratified medium.   We analyze these CRs with the spectral and temporal characteristics of the bursts described in the second \textit{Fermi}-LAT GRB catalog (2FLGC) and found that i) the thin shell case is preferred over a thick shell and a constant-density medium over a stellar-wind environment,  ii) bursts with an atypical and hard spectral index could be successfully described by this scenario in different cooling conditions, iii) the early optical flash and GeV emission exhibited in GRB 160625B and 180720B were generated from the same accelerated region and electron population, concluding that LAT emission originated during the early afterglow, and iv) the maximum synchrotron energy radiated from the reverse-shock scenario could explain only a few photons with the exception of a pair of bursts, so that scattered photons by the SSC process must be required.

\end{abstract}




\section{Introduction}

The most powerful cosmic explosions, known as gamma-ray bursts (GRBs), can potentially be attributed to the mergers of binary compact objects such as two neutron stars (NS) or a black hole (BH) -NS \citep{1992ApJ...395L..83N, 1992ApJ...392L...9D, 1992Natur.357..472U, 1994MNRAS.270..480T, 2011MNRAS.413.2031M}, and the core collapse of colossal stars \citep{1993ApJ...405..273W, 1998Natur.395..670G, 1999Natur.401..453B, 2001ApJ...550..410M, 2003Natur.423..847H, 2006ARA&A..44..507W, 2012grbu.book..169H}.  The core collapse of massive stars has been attributed to long GRBs (lGRBs), whereas the merging of binary compact objects has been linked to short GRBs (sGRB). Regardless of their progenitor, observations have definitively demonstrated that highly relativistic and collimated outflows cause the prompt and afterglow events interpreted in the fireball scenario \citep{2002ApJ...571..779P, 2004ApJ...609L...1T, 2015PhR...561....1K}.    Because relativistic outflows transmit a significant amount of energy to the external circumstellar medium when they come into contact with it, forward and reverse shocks are generated during the afterglow phase \citep{1995ApJ...455L.143S, 1998ApJ...497L..17S, 1999ApJ...513..669K, 2000ApJ...545..807K, 2003ApJ...597..455K, 2016ApJ...818..190F}. Nonthermal electrons are accelerated in the front shocks, which produces synchrotron and synchrotron self-Compton (SSC) radiation in both regions.

The mechanisms can be identified by the evolution of the spectral ($\beta$) and temporal ($\alpha$) power-law (PL) indexes, along with the characteristics of the closure relations profile ($\propto t^{-\alpha}\nu^{-\beta}$). While synchrotron photons are limited to keV and a few MeV energies, the SSC process can disperse them to energies exceeding hundreds of MeV \citep{2001ApJ...559..110Z, 2009ApJ...706L.138A, 2011MNRAS.412..522B, 2019ApJ...883..162F}. The profile of the CRs depends on the circumstellar environment, energy injection, adiabatic/radiative regime, and variation of microphysical parameters, among others. 

High-energy gamma-ray emissions, lasting hundreds to thousands of seconds and exceeding 100 MeV, are often attributed to the synchrotron forward-shock model. This model predicts that such emissions should adhere to CRs specific to forward-shock dynamics \citep{2009MNRAS.400L..75K,2010MNRAS.409..226K}. The LAT instrument onboard the \textit{Fermi} Gamma-ray Space Telescope \citep{2009ApJ...697.1071A} has been instrumental in detecting and analyzing GRBs with high-energy emissions spanning from 20 MeV to beyond 300 GeV. Research into CRs within the gamma-ray spectrum has been extensive. One pertinent example is that of~\cite{2019ApJ...883..134T}, who examined 59 GRBs detected by \textit{Fermi}-LAT, selecting those with precise measurements of temporal and spectral indices. The synchrotron forward-shock model adequately accounts for numerous cases; however, a significant subset of bursts does not conform to this model.  GRBs that did not align with any canonical relations demonstrated a gradual decay characterized by a temporal decay index of less than 1. \cite{Ajello_2019} also identified the occurrences of sustained plateau emission at elevated energies. Another relevant study is that of \cite{2021ApJS..255...13D}, who analyzed three bursts GRB 090510A, 090902B, and 160509A and determined they followed a slow-cooling regime rather than a fast-cooling regime. It held regardless of whether the surrounding environment was a constant-density interstellar medium or a stellar wind medium~\citep{1998ApJ...497L..17S}. Expanding on this work, \cite{2023Galax..11...25D} examined a broader sample of 86 GRBs, incorporating both broken power-law (BPL) and simple PL fitting—unlike \cite{2019ApJ...883..134T}, who used only simple PL fitting. They also used a frequentist approach instead of the Bayesian one by \cite{2019ApJ...883..134T}. Despite these differences, the findings were consistent: most GRBs conformed to the CRs within the slow-cooling regime, irrespective of their environment being a constant-density or stellar wind medium. Among the 86 gamma-ray bursts analyzed, 74 exhibited consistency with at least one CR. Nonetheless, 12 GRBs did not conform to any CRs, identifying them as noteworthy cases. Further analysis suggested a preference for CRs without energy injection over those that included it. Specifically, 35 GRBs satisfied at least one CR without energy injection. Within a subset of 21 GRBs fitted with a BPL, eight (090926A, 091003, 110731A, 130504C, 160509A, 160816A, 171010A, and 171120A) exhibited temporal and spectral values aligning with the fast- and slow-cooling regime across all  PL values of stratification medium ($k$). Contrastingly, models with energy injection were less effective, where 15 GRBs failed to satisfy any CRs.  To explain the GeV flare observed in some bursts from 2FLGC, the SSC process has been required. For instance, \cite{2020ApJ...905..112F} and \cite{2024MNRAS.527.1674F} derived the SSC light curves from the reverse shock region for a constant density and a stellar-wind medium when the electron distribution was described with spectral $p<2$. The authors described an early GeV flare that observed some LAT-detected bursts, such as GRB 160509A, GRB 160625B, and 180720B, among others. \cite{2024NatAs...8..134A} required the SSC process from the reverse shock to model the GeV gamma-ray emission during the initial few hundred seconds of GRB 180720B.

This manuscript derives the CRs of the SSC mechanism generated by a relativistic electron population accelerated in the reverse-shock region during the adiabatic phase and characterized by a hard ($1<p<2$) and a soft spectral index ($2<p$). We examine the development of the reverse shock in both thick- and thin-shell scenarios inside a stratified environment created by the stellar wind launched by the progenitor and a uniform medium. We performed an analysis that compared these CRs with the spectral and temporal features of the bursts recorded in 2FLGC. The paper is organized as follows. In Section~2, we show the closure relations of SSC from reverse shock evolving in the thick- and thin-shell case for a stellar wind-type-like environment and ISM. In Section 3, we show the data analysis for evaluating the CRs and present our results of this section. In Section 4, we present and analyze the optical and GeV correlations, and finally, Section 5 provides a summary.

\section{SSC Closure relations from reverse shock }

When the relativistic jet encounters the circumburst environment, a reverse shock forms and propagates back into the outflow \citep{1997ApJ...476..232M, 1999ApJ...517L.109S, 2000ApJ...542..819K,2019ApJ...881...12B,2019ApJ...887..254B}. Electrons described by a PL distribution ($\propto \gamma_e^{-p}\,d\gamma_e$) with $\gamma_{\rm m, r}\leq \gamma_{\rm e}$ are accelerated during reverse shock and cooled mainly by synchrotron and SSC emission.\footnote{The subindex ``r" refers to the quantities derived in the reverse-shock region.}.     The electron minimum Lorentz factor ($\gamma_{\rm m, r}$) is defined depending on the range of the spectral index; $\gamma_{\rm m, r}=\left[\tilde{g}(p)\frac{m_p}{m_e} \varepsilon_{e, r}\Gamma \chi_{\rm e}^{-p+1} \gamma_{\rm max}\right]^{\frac{1}{p-1}}$ for $1<p<2$ \citep{2001ApJ...558L.109D} and $\gamma_{\rm m, r}=g(p)\frac{m_p}{m_e}\varepsilon_{e, r}\Gamma\chi_{\rm e}^{-1}$ for $2 < p$ \citep{2006MNRAS.369..197F}. The synchrotron and SSC light curves in the fast and slow cooling regimes depend on the dynamics of the reverse shock (given the critical Lorentz factor ($\Gamma_c$), the shock crossing time ($t_{\rm x}$), and the duration of the burst ($T_{90}$)), the hardness of the spectral index of the electron distribution ($1<p<2$ and $2 < p$), and the density profile of the circumburst environment ($n(r)=A_{\rm k}r^{-\rm k}$), with $k=0$ the constant density medium ($A_{\rm 0}=n$)   and $k=2$ the stellar wind. In the previous case, the density corresponds to $n(r) = \frac{\dot{M}_{\rm W}}{4\pi m_p\,v_{\rm W}} A_{\rm W} r^{-2} =3.0\times 10^{35}\,{\rm cm^{-1}}\,A_{\rm W}r^{-2}$.    Depending on the values of the critical Lorentz factor, the shock crossing time and the duration of the burst,  the reverse shock would evolve in the thick-shell regime with the condition $\Gamma_c \lesssim \Gamma$ or $t_{\rm x} \lesssim T_{90}$, or in the thin-shell regime with the condition $\Gamma < \Gamma_c$ or $T_{90} < t_{\rm x}$.

According to the results of the observational studies, the electron spectral index differs between bursts and follows a normal distribution. Instead of having a universal value, these values fall anywhere between 1.5 and 3.5 \citep{2002ApJ...581.1248P, 2008MNRAS.388..144R, 2015ApJ...811...83Z, 2006MNRAS.371.1441S, 2008ApJ...672..433S, 2009MNRAS.395..580C, 2019ApJ...883..134T}. \cite{2008MNRAS.388..144R} required a double-slope electron energy distribution to describe the afterglow observation of GRB 010222, GRB 020813 and GRB 041006. They found a hard value of spectral index of $p=1.47^{+0.004}_{-0.003}$, $1.40^{+0.007}_{-0.004}$ and $1.29 - 1.32$ for GRB 010222, GRB 020813 and GRB 041006, respectively. \cite{2015ApJ...811...83Z} modeled GRB 091127 finding evidence of a hard value of the electron spectral index of $p=1.5\pm 0.01$ at early times. \cite{2002ApJ...581.1248P} performed an analysis
using the first spectroscopy catalog of GRBs observed by BATSE\footnote{The Burst and Transient Source Experiment} aboard the \textit{Compton} Gamma Ray Observatory (CGRO) which consisted of 5021 spectra. Requiring several PL configurations of the electron spectral index, they showed via the histograms that, irrespective of the PL configuration, a substantial number of spectra had a spectral index ranging $1<p<2$. \cite{2006MNRAS.371.1441S} reported no universal value for the electron spectral index in a large sample of GRB afterglows. The authors considered 13, 14, and 28 afterglow observations collected from the HETE-2, BeppoSAX, and Swift satellites, respectively, finding that a large fraction of spectral data was described with values $1<p<2$.  \cite{2019ApJ...883..134T} conducted an extensive analysis of the closure relations in a sample of 59 selected LAT-detected bursts, considering both temporal and spectral indices. Regardless of the long or short clasification, and if they are described with a stratified or constant medium,  the authors found that the closure relations of seven bursts satisfied the spectral indexes with $1<p<2$.

\subsection{Light curves in the thick-shell scenario}

The reverse shock in the thick-shell scenario is ultra-relativistic and can decelerate the shell dramatically. The shock crossing time is defined by $t_{\rm x}=31.5\,{\rm s}\,\left(\frac{1+z}{1.5}\right)\Delta_{11.8}$. {The critical Lorentz factor is not a fixed value; rather, it is a threshold defined by the corresponding kinetic energy and density. This assesses the nature of the reverse shock and determines whether it is Newtonian or relativistic.}  In this relativistic scenario, the critical Lorentz factor becomes
{\small
\begin{eqnarray}\nonumber
\label{fcssc_t}
\Gamma_c\equiv \begin{cases} 
2.0\times 10^2\left(\frac{1+z}{1.5} \right)^{\frac38} E^{\frac18}_{53} n_{-1}^{-\frac18} T_{90}^{-\frac38}\hspace{0.75cm}{\rm for}\hspace{0.2cm}{\rm ISM}\,\,(\rm k=0) \cr
1.2\times 10^2\left(\frac{1+z}{1.5} \right)^{\frac14} E^{\frac14}_{53} A_{\rm W,-2}^{-\frac14} T_{90}^{-\frac14}\hspace{0.5cm}{\rm for}\hspace{0.16cm}{\rm wind}\,\,(\rm k=2)\,.
\end{cases}
\end{eqnarray}
}
In the following, we use the convention $Q_{\rm x}=Q/10^{\rm x}$ in c.g.s. units. 

\subsubsection{Constant-density medium ($k=0$)}

\paragraph{Light curves for $t < t_{\rm x}$.}

Before the reverse shock crosses the shell, the magnetic field, minimum electron Lorentz factor, and cooling Lorentz factor evolve for $1<p<2$ ($2 < p$) as $B'_{\rm r}\propto t^{-\frac14}$, $\gamma_{\rm m,r}\propto t^{\frac{p}{8(p-1)}}\,(t^{\frac14})$ and $\gamma_{\rm c,r}\propto t^{-\frac14}$, respectively.  Moreover, the synchrotron spectral breaks and the maximum synchrotron flux in terms of the observed time are $\nu^{\rm syn}_{\rm m, r}\propto t^{\frac{2-p}{4(p-1)}}\,(t^0)$ and $\nu^{\rm syn}_{\rm c, r}\propto t^{-1}$ and $F^{\rm syn}_{\rm max,r }\propto t^{\frac12}$, respectively.
At the deceleration radius ($r$), the same electron population can up-scatter synchrotron photons up to higher energies as $h\nu^{\rm ssc}_{\rm i, r}\simeq\gamma^2_{\rm i, r} h\nu^{\rm syn}_{\rm i, r}$ with ${\rm i=m}$ and ${\rm c}$,  reaching a maximum flux of $F^{\rm ssc}_{\rm max,r}\sim\, \frac{4\sigma_Tn r}{3g(p)}\,F^{\rm syn}_{\rm max,r}$ \citep{2024MNRAS.527.1674F}. The quantities $\gamma_{\rm m, r}$ and $\gamma_{\rm c, r}$ correspond to the minimum and the cooling electron Lorentz factors, respectively, and $\nu^{\rm syn}_{\rm m, r}$ and $\nu^{\rm syn}_{\rm c, r}$ to
the characteristic and cooling spectral breaks of synchrotron emission, which are calculated explicitly in appendix A.  Therefore, the SSC spectral breaks and the maximum SSC flux for $1<p<2$  and $2 < p$ are

{\small
\begin{eqnarray}\label{break_thick_aft_k0}
h \nu^{\rm ssc}_{\rm m, r} &\simeq& \begin{cases}
7.3\times10^{-7}\,{\rm GeV}\,\left(\frac{1+z}{1.5}\right)^{\frac{1-2p}{2(p-1)}} g^{\frac{4}{p-1}}(1.9) \chi_{\rm e,-1}^{-4} \varepsilon^{\frac{4}{p-1}}_{\rm e_r,-1} \varepsilon^{\frac{3-p}{2(p-1)}}_{\rm B_r,-2}\,\Gamma_{2.5}^{\frac{4}{p-1}}\, n^{\frac{7-2p}{4(p-1)}}_{-1}\Delta^{\frac{1}{4(p-1)}}_{11.8}\,E^{-\frac{1}{4(p-1)}}_{53} t^{\frac{1}{2(p-1)}}_1\hspace{1.4cm}{\rm for} \hspace{0.1cm} { 1<p<2 }\cr
5.4\times10^{-4}\,{\rm GeV}\,\left(\frac{1+z}{1.5}\right)^{-\frac32} g^4(2.2) \chi_{\rm e,-1}^{-4} \varepsilon^4_{\rm e_r,-1} \varepsilon^{\frac12}_{\rm B_r,-2} \Gamma^4_{2.5} n^{\frac{3}{4}}_{-1}\Delta^{\frac{1}{4}}_{11.8}\,E^{-\frac{1}{4}}_{53} t^{\frac{1}{2}}_1\hspace{4.65cm}{\rm for} \hspace{0.1cm} {2<p }
\end{cases}\cr
h\nu^{\rm ssc}_{\rm c, r}&\simeq& 1.6\times 10^{2}\,{\rm GeV}\,\left(\frac{1+z}{1.5}\right)^{\frac{1}{2}}  \left(\frac{1+Y_{\rm r}}{2} \right)^{-4}\varepsilon^{-\frac72}_{\rm B_r,-2} n^{-\frac{9}{4}}_{-1}\Delta^{\frac{5}{4}}_{11.8} E^{-\frac{5}{4}}_{53}\,t^{-\frac{3}{2}}_1,\cr
F^{\rm ssc}_{\rm max,r} &\simeq& 1.8\times 10^{-4}\,{\rm mJy}\, \,g^{-1}(2.2)\,\chi_{\rm e,-1} \varepsilon^{\frac12}_{\rm B_r,-2}\,n_{-1} \, d^{-2}_{\rm z,28.3}\,\Gamma^{-1}_{2.5}\,\Delta^{-\frac{3}{2}}_{11.8}\,E^{\frac{3}{2}}_{53} t_1\,,\,\,\,\,\,\,\,\,
\eary
}
respectively.

The Klein-Nishina (KN) impact on the SSC spectrum directly causes the attenuation of up-scattered photons generated by synchrotron radiation. However, the dominant nature of SSC photons and the cooling of certain injected electrons with variable Lorentz factors are associated effects.  The KN breaks  for $1<p<2$ and $2 < p$ become
%
%
{\small
\bary
h \nu^{\rm ssc}_{\rm KN, m, r} &\simeq&\begin{cases}
9.5\,{\rm GeV}\,\left(\frac{1+z}{1.5}\right)^{\frac{6-7p}{8(p-1)}} g^{\frac{1}{p-1}}(2.2)\,\chi_{e,-1}^{-1}\,\varepsilon^{\frac{1}{p-1}}_{\rm e_r,-1}\varepsilon^{\frac{2-p}{4(p-1)}}_{\rm B_r,-2}\,\Gamma_{2.5}^{\frac{1}{p-1}}\,n_{-1}^{\frac{5(2-p)}{16(p-1)}}\,E^{-\frac{2-p}{16(p-1)}}_{53}\Delta^{\frac{2-p}{16(p-1)}}_{11.8}\,t^{\frac{2-p}{8(p-1)}}_{1} \hspace{1.25cm}{\rm for} \hspace{0.1cm} { 1<p<2 }\cr
4.9\times 10\,{\rm GeV}\,\left(\frac{1+z}{1.5}\right)^{-1} g(2.2)\,\chi_{e,-1}^{-1}\,\varepsilon_{\rm e_r,-1}\,\Gamma_{2.5},\hspace{7.03cm}{\rm for} \hspace{0.1cm} {2<p }
\end{cases}
\cr
h \nu^{\rm ssc}_{\rm KN, c, r}&\simeq& 1.2\times 10^{3}\,{\rm GeV}\,\left(\frac{1+z}{1.5}\right)^{-\frac12} \left(\frac{1+Y_{\rm r}}{2} \right)^{-1}\,\varepsilon^{-1}_{\rm B_r,-2}\,n^{-\frac{3}{4}}_{-1} \,E^{-\frac{1}{4}}_{53}\,\Delta^{\frac{1}{4}}_{11.8}\,t^{-\frac{1}{2}}_{1},\,\,
\eary
}
respectively.

\paragraph{Light curves for $t_{\rm x}<t$.}
Once the reverse shock crosses the shell,  the magnetic field, the minimum and cooling electron Lorentz factors, the synchrotron spectral breaks and the maximum synchrotron flux in terms of the observed time for $1<p<2$ ($2 < p$) are $B'\propto t^{-\frac{13}{24}}$, $\gamma_{\rm m,r}\propto t^{-\frac{68-21p}{96(p-1)}}\,(t^{-\frac{13}{48}})$, $\gamma_{\rm c,r}\propto t^{\frac{25}{48}}$, $\nu^{\rm syn}_{\rm m, r}\propto t^{-\frac{21+26p}{48(p-1)}}\,(t^{-\frac{73}{48}})$, $\nu^{\rm syn}_{\rm c, r}\propto t^{\frac{1}{16}}$ and $F^{\rm syn}_{\rm max,r}\propto t^{-\frac{47}{48}}$, respectively.\footnote{Hereafter, we adopt the unprimed and primed variables for the observer and comoving frames, respectively.}
As the fluid expands adiabatically, the cutoff frequency evolves as $\nu^{\rm ssc}_{\rm cut, r}=\nu^{\rm ssc}_{\rm c, r}(t_{\rm x})\,\left(\frac{t}{t_{\rm x}} \right)^{-\frac{33}{16}}$.  The SSC spectral breaks and the maximum SSC flux for a constant-density medium for $1<p<2$ and $2 < p$ are given by
{\small
\begin{eqnarray}\label{break_thick_aft_k0}
h \nu^{\rm ssc}_{\rm m, r} &\simeq& \begin{cases}
9.6\times 10^{-8}\,{\rm GeV}  \, \left(\frac{1+z}{1.5} \right)^{\frac{137-43p}{48(p-1)}}\,\tilde{g}^{\frac{4}{p-1}}(1.9)\,\chi_{\rm e,-1}^{-4} \varepsilon^{\frac{4}{p-1}}_{\rm e_r,-1} \varepsilon^{\frac{3-p}{2(p-1)}}_{\rm B_r,-2}n^{\frac{7-2p}{4(p-1)}}_{-1}\,\Delta_{11.8}^{\frac{5(25+p)}{48(p-1)}}\,\Gamma_{2.5}^{\frac{4}{p-1}}\,E^{-\frac{1}{4(p-1)}}_{53}t^{-\frac{89+5p}{48(p-1)}}_{2}\hspace{0.7cm}{\rm for} \hspace{0.1cm} { 1<p<2 }\cr
8.7\times 10^{-5}\,{\rm GeV}  \, \left(\frac{1+z}{1.5} \right)^{\frac{17}{16}}\,g^4(2.2)\,\chi_{\rm e,-1}^{-4}\, \,\varepsilon^4_{\rm e_r,-1} \varepsilon^{\frac12}_{\rm B_r,-2}  n^{\frac{3}{4}}_{-1}\,\Gamma^4_{2.5}\Delta_{11.8}^{\frac{45}{16}}\,E^{-\frac{1}{4}}_{53}t^{-\frac{33}{16}}_{2}\hspace{4.1cm}{\rm for} \hspace{0.1cm} {2<p }
\end{cases}\cr
h \nu^{\rm ssc}_{\rm cut, r}&\simeq& 1.1\times10\,{\rm GeV}\, \left(\frac{1+z}{1.5} \right)^{\frac{17}{16}}\,\left(\frac{1+Y_{\rm r}}{2}\right)^{-4}\,\varepsilon^{-\frac72}_{\rm B_r,-2} n^{-\frac{9}{4}}_{-1}\,     \Delta^{\frac{29}{16}}_{11.8}\,E^{-\frac{5}{4}}_{53}\,t^{-\frac{33}{16}}_{2},\cr
F^{\rm ssc}_{\rm max, r}&\simeq& 2.1\times 10^{-4}\,{\rm mJy}\,\left(\frac{1+z}{1.5} \right)^{\frac{89}{48}}\,g^{-1}(2.2)\,\chi_{e,-1}\,   \varepsilon^{\frac12}_{\rm B_r,-2}\,  \Gamma^{-1}_{2.5}\,n_{-1}\,  \Delta_{11.8}^{\frac{17}{48}} \,d^{-2}_{\rm z,27.9} \,E^{\frac{3}{2}}_{53}\,t^{-\frac{41}{48}}_{2}, \,\,\,\,\,\,\,\,
\eary
}
respectively. The SSC light curve may exhibit a steeper temporal decay index during this interval due to the electrons' inability to reaccelerate within the reverse-shock region, which results in the ceasing of emission from this region. However, the abrupt disappearance is mitigated by the emission produced at significant angles in relation to the jet axis, a phenomenon known as angular-time delay. In such a situation, the SSC flux would evolve as $F_\nu\propto t^{-(\beta+2)}\nu^{-\beta}$ \citep{2000ApJ...543...66P, 2003ApJ...597..455K}.

For $\nu^{\rm ssc}_{\rm c, r}<\nu^{\rm ssc}_{\rm m, r}$ and $\nu^{\rm ssc}_{\rm m, r}<\nu^{\rm ssc}_{\rm c, r}$, the spectral breaks in the KN regime for $1<p<2$ and $2 < p$ are
{\small
\bary
h \nu^{\rm ssc}_{\rm KN, m, r} &\simeq&\begin{cases}
4.0\,{\rm GeV}\,\left(\frac{1+z}{1.5}\right)^{\frac{122-75p}{96(p-1)}} \tilde{g}^{\frac{1}{p-1}}(1.9)\,\chi_{e,-1}^{-1}\,\varepsilon^{\frac{1}{p-1}}_{\rm e_r,-1}\,\,\varepsilon^{\frac{2-p}{4(p-1)}}_{\rm B_r,-2}\,n^{\frac{5(2-p)}{16(p-1)}}_{-1}\,E^{\frac{p-2}{16(p-1)}}_{53}\Delta^{\frac{62+3p}{96(p-1)}}_{11.8}\,\Gamma_{2.5}^{\frac{1}{p-1}}\,t^{-\frac{26+21p}{96(p-1)}}_{2} \hspace{0.7cm}{\rm for} \hspace{0.1cm} { 1<p<2 }\cr
2.2\times 10\,{\rm GeV}\,\left(\frac{1+z}{1.5}\right)^{-\frac{7}{24}} g(2.2)\,\chi_{e,-1}^{-1}\,\varepsilon_{\rm e_r,-1}\,\Gamma_{2.5}\Delta^{\frac{17}{24}}_{11.8}\,t^{-\frac{17}{24}}_{2},\hspace{5.4cm}{\rm for} \hspace{0.1cm} {2<p }
\end{cases}
\cr
h \nu^{\rm ssc}_{\rm KN, c, r}&\simeq& 7.4\times 10^2\,{\rm GeV}\,\left(\frac{1+z}{1.5}\right)^{-\frac{13}{12}} \left(\frac{1+Y_{\rm r}}{2} \right)^{-1}\,\varepsilon^{-1}_{\rm B_r,-2}\,n^{-\frac{3}{4}}_{-1}\,E^{-\frac{1}{4}}_{53}\,\Delta^{-\frac{1}{3}}_{11.8}\,t^{\frac{1}{12}}_{2},\,\,
\eary
}
respectively.

\subsubsection{Stellar-wind medium ($k=2$)}

\paragraph{Light curves for $t < t_{\rm x}$.}
The reverse shock in the thick-shell scenario is ultra-relativistic and can decelerate the shell dramatically. Before the reverse shock crosses the shell,  the magnetic field, the minimum and cooling electron Lorentz factors evolve for $1<p<2$ ($2< p$) as $B'_{\rm r}\propto t^{-1}$, $\gamma_{\rm m,r}\propto t^{-\frac{2-p}{2(p-1)}}\,(t^0)$ and $\gamma_{\rm c,r}\propto t$, respectively.  Moreover, the synchrotron spectral breaks and the maximum synchrotron flux in term of the observed time are $\nu^{\rm syn}_{\rm m, r}\propto t^{-\frac{1}{p-1}}\,(t^{-1})$ and $\nu^{\rm syn}_{\rm c, r}\propto t$ and $F^{\rm syn}_{\rm max,r }\propto t^0$, respectively.
For the stellar-wind environment the maximum SSC flux evolve as $F^{\rm ssc}_{\rm max,r}\propto\, \sigma_T A_{\rm W} r^{-1} \,F^{\rm syn}_{\rm max,r}$. Therefore, the SSC spectral breaks and the maximum SSC flux for $1<p<2$ and $2 < p$ are

{\small
\begin{eqnarray}\label{break_thick_aft_k2}
h \nu^{\rm ssc}_{\rm m, r} &\simeq& \begin{cases}
1.7\times 10^{-4}\,{\rm GeV}\,\left(\frac{1+z}{1.5}\right)^{\frac{2(2-p)}{p-1}} g^{\frac{4}{p-1}}(1.9) \chi_{\rm e,-1}^{-4} \varepsilon^{\frac{4}{p-1}}_{\rm e_r,-1} \varepsilon^{\frac{3-p}{2(p-1)}}_{\rm B_r,-2}\,\Gamma_{2.5}^{\frac{4}{p-1}} \, A^{\frac{7-2p}{2(p-1)}}_{\rm W,-2} \Delta^{\frac{4-p}{2(p-1)}}_{11.8}\,E^{-\frac{4-p}{2(p-1)}}_{53} t^{-\frac{3-p}{p-2}}_1\hspace{0.7cm}{\rm for} \hspace{0.1cm} { 1<p<2 }\cr
5.2\times 10^{-2}\,{\rm GeV}\,g^4(2.2) \chi_{\rm e,-1}^{-4} \varepsilon^4_{\rm e_r,-1} \varepsilon^{\frac12}_{\rm B_r,-2}\, \Gamma^4_{2.5} A^{\frac{3}{2}}_{\rm W,-2} \Delta_{11.8}\,E^{-1}_{53} t^{-1}_1\hspace{4.6cm}{\rm for} \hspace{0.1cm} { 2<p }\cr
\end{cases}\cr
h \nu^{\rm ssc}_{\rm c, r}&\simeq& 3.9\times 10^{-6}\,{\rm eV}\,\left(\frac{1+z}{1.5}\right)^{-4}  \left(\frac{1+Y_{\rm r}}{2} \right)^{-4}\varepsilon^{-\frac72}_{\rm B_r,-2} A^{-\frac{9}{2}}_{\rm W,-2} \Delta^{-1}_{11.8}\,E_{53}\,t^3_1,\,\cr
F^{\rm ssc}_{\rm max,r} &\simeq& 1.4\times 10^{-1}\,{\rm mJy}\, \left(\frac{1+z}{1.5} \right)^2\,g^{-1}(2.2)\,\chi_{\rm e,-1} \varepsilon^{\frac12}_{\rm B_r,-2}\,A_{\rm W,-2}^2 d^{-2}_{\rm z,28.3}\, \,\Gamma^{-1}_{2.5}\,\Delta^{-\frac12}_{11.8}\,E^{\frac12}_{53} t^{-1}_1\,,
\eary
}
respectively.

For $\nu^{\rm ssc}_{\rm c, r}<\nu^{\rm ssc}_{\rm m, r}$ and $\nu^{\rm ssc}_{\rm m, r}<\nu^{\rm ssc}_{\rm c, r}$,  the KN breaks for $1<p<2$ and $2 < p$ become

{\small
\bary
h \nu^{\rm ssc}_{\rm KN, m, r}&\simeq&\begin{cases}
1.1\times 10\,{\rm GeV}\,\left(\frac{1+z}{1.5}\right)^{\frac{4-3p}{2(p-1)}} g^{\frac{1}{p-1}}(1.9) \chi_{\rm e,-1}^{-1} \varepsilon^{\frac{1}{p-1}}_{\rm e_r,-1} \,\varepsilon^{\frac{2-p}{4(p-1)}}_{\rm B_r,-2}\Gamma_{2.5}^{\frac{1}{p-1}} A^{\frac{5(2-p)}{8(p-1)}}_{\rm W,-2} \Delta^{\frac{3(2-p)}{8(p-1)}}_{11.8}\,E^{-\frac{3(2-p)}{8(p-1)}}_{53} t^{-\frac{2-p}{2(p-1)}}_1\,\hspace{0.7cm}{\rm for} \hspace{0.1cm} { 1<p<2 }\cr
4.9\times 10\,{\rm GeV}\,\left(\frac{1+z}{1.5}\right)^{-1}g(2.2)\chi_{e,-1}^{-1} \,\varepsilon_{\rm e_r,-1}\,\Gamma_{2.5}\,\hspace{6.9cm}{\rm for} \hspace{0.1cm} { 2<p }
\end{cases}\cr
h \nu^{\rm ssc}_{\rm KN, c, r}&\simeq& 4.6\,{\rm GeV}\,\left(\frac{1+z}{1.5}\right)^{-2} \left(\frac{1+Y_{\rm r}}{2} \right)^{-1}\,\varepsilon^{-1}_{\rm B_r,-2}\,A^{-\frac32}_{\rm W,-2}\,\Delta^{-\frac{1}{2}}_{11.8}\,E^{\frac{1}{2}}_{53}\,t_1\,,\,\,
\eary
}
respectively.

\paragraph{Light curves for $t_{\rm x} < t$.}
The scaling of the magnetic field, the breaks of the electron Lorentz factors, the spectral breaks of the synchrotron process and the maximum synchrotron flux evolve as ${B'}_{\rm r}\propto t^{-\frac34}$, $\gamma_{\rm m,r}\propto t^{-\frac{16-5p}{16(p-1)}}\,(t^{-\frac38})$ and $\gamma_{\rm c,r}\propto t^{\frac78}$, $\nu^{\rm syn}_{\rm m, r}\propto t^{-\frac{7+4p}{8(p-1)}}\,(t^{-\frac{15}{8}})$ and $\nu^{\rm syn}_{\rm c, r}\propto t^{\frac58}$ and $F^{\rm syn}_{\rm max,r}\propto t^{-\frac98}$, respectively.   The SSC spectral breaks and the maximum flux in the stellar-wind environment for $1<p<2$ and $2 < p$ can be written as

{\small
\begin{eqnarray}\label{break_thick_aft_k2}
h \nu^{\rm ssc}_{\rm m, r} &\simeq& \begin{cases}
1.4\times 10^{-6}\,{\rm GeV}\,\left(\frac{1+z}{1.5}\right)^{\frac{31-9p}{8(p-1)}} \tilde{g}^{\frac{4}{p-1}}(1.9) \chi_{\rm e,-1}^{-4} \varepsilon^{\frac{4}{p-1}}_{\rm e_r,-1} \varepsilon^{\frac{3-p}{2(p-1)}}_{\rm B_r,-2} A^{\frac{7-2p}{2(p-1)}}_{\rm W,-2} \Delta^{\frac{3(5+p)}{8(p-1)}}_{11.8}\,E^{-\frac{4-p}{2(p-1)}}_{53} t^{-\frac{23-p}{8(p-1)}}_2\hspace{0.7cm}{\rm for} \hspace{0.1cm} { 1<p<2 }\cr
7.9\times 10^{-4}\,{\rm GeV} \left(\frac{1+z}{1.5}\right)^{\frac{13}{8}}\,g^{4}(2.2)\,\chi_{\rm e,-1}^{-4}\,\epsilon^4_{\rm e_r,-1} \epsilon^{\frac12}_{\rm B_r,-2} \,A^{\frac{3}{2}}_{\rm W,-2}\,\Delta_{11.8}^{\frac{21}{8}}\, \Gamma^{4}_{2.5}\,  E^{-1}_{\rm 53}\,t^{-\frac{21}{8}}_{2}\hspace{2.75cm}{\rm for} \hspace{0.1cm} { 2<p }\cr
\end{cases}\cr
h \nu^{\rm ssc}_{\rm cut, r}&\simeq& 9.8\times 10^{-5}\,{\rm GeV} \left(\frac{1+z}{1.5}\right)^{\frac{13}{8}}\,\left(\frac{1+Y_{\rm r}}{2} \right)^{-4}  \epsilon^{-\frac72}_{\rm B_r,-2}\,A^{-\frac{9}{2}}_{\rm W,-2}\,\Delta_{11.8}^{\frac{37}{8}}\, E_{\rm 53}\, t^{-\frac{21}{8}}_{2}\,\cr
F^{\rm ssc}_{\rm max,r} &\simeq&  9.2\times 10^{-3}\,{\rm mJy}\,\left(\frac{1+z}{1.5}\right)^{\frac{19}{8}}\,g^{-1}(2.2)\,\chi_{\rm e,-1}\,  \epsilon^{\frac12}_{\rm B_r,-2} \, A^{2}_{W,-2}\, \Gamma^{-1}_{2}\,\Delta_{11.8}^{-\frac18}\, d^{-2}_{\rm z,27.9}\,E^{\frac{1}{2}}_{\rm 53}\, t^{-\frac{11}{8}}_{2}\,,
\eary
}
respectively, where the cut-off frequency $\nu^{\rm ssc}_{\rm cut, r}=\nu^{\rm ssc}_{\rm c, r}(t_{\rm x})\,\left(\frac{t}{t_{\rm x}} \right)^{-\frac{21}{8}}$ is estimated because no electrons are shocked anymore and the fluid expands adiabatically. As for the case of constant-density medium, the SSC emission generated in the reverse-shock region could decay faster due to the effect of angular time delay, and therefore, the SSC flux would evolve as $F^{\rm ssc}_{\rm \nu, r}(t > t_{\rm x})\propto t^{-(\beta+2)}$ \citep{2000ApJ...543...66P, 2003ApJ...597..455K}. 

Considering the attenuation of the KN effect, the spectral breaks in the KN regime for $\nu^{\rm ssc}_{\rm c, r}<\nu^{\rm ssc}_{\rm m, r}$ and $\nu^{\rm ssc}_{\rm m, r}<\nu^{\rm ssc}_{\rm c, r}$ for $1<p<2$ and $2 < p$ can be written as 
{\small
\bary
h \nu^{\rm ssc}_{\rm KN, m, r}&\simeq&\begin{cases}
4.2\,{\rm GeV}\,\left(\frac{1+z}{1.5}\right)^{\frac{26-15p}{16(p-1)}} \tilde{g}^{\frac{1}{p-1}}(1.9) \chi_{\rm e,-1}^{-1} \varepsilon^{\frac{1}{p-1}}_{\rm e_r,-1}\,\varepsilon^{\frac{2-p}{4(p-1)}}_{\rm B_r,-2} A^{\frac{5(2-p)}{8(p-1)}}_{\rm W,-2} \Delta^{\frac{3(p+2)}{16(p-1)}}_{11.8}\,\Gamma_{2.5}^{\frac{1}{p-1}}\,E^{-\frac{3(2-p)}{8(p-1)}}_{53} t^{-\frac{p+10}{16(p-1)}}_{2}\,\hspace{0.7cm}{\rm for} \hspace{0.1cm} { 1<p<2 }\cr
2.1\times 10\,{\rm GeV}\,\left(\frac{1+z}{1.5}\right)^{-\frac14} g(2.2)\,\chi_{e,-1}^{-1}\,\varepsilon_{\rm e_r,-1}\,\Gamma_{2.5}\,\Delta^{\frac34}_{11.8}\,t^{-\frac34}_{2}\,\hspace{5.35cm}{\rm for} \hspace{0.1cm} { 2<p }
\end{cases}\cr
h \nu^{\rm ssc}_{\rm KN, c, r}&\simeq& 2.6\times 10\,{\rm GeV}\,\left(\frac{1+z}{1.5}\right)^{-\frac32} \left(\frac{1+Y_{\rm r}}{2} \right)^{-1}\,\varepsilon^{-1}_{\rm B_r,-2}\,A^{-\frac{3}{2}}_{\rm W,-2}\,E^{\frac12}_{53}\,t^{\frac12}_{2},\,\,
\eary
}
respectively.

\subsection{Light curves in the thin-shell scenario}

The reverse shock becomes mildly relativistic in the thin-shell regime and cannot decelerate the shell (that is, $T_{90}< t_{\rm x}$). Therefore, the shock-crossing time scales as

{\small
\begin{eqnarray}\nonumber
\label{fcssc_t}
t_{\rm x}\equiv \begin{cases} 
7.9\times 10^2\,{\rm s}\left(\frac{1+z}{1.5}\right)\, n^{-\frac13}_{-1}\, E^{\frac13}_{53}\, \Gamma^{-\frac83}_{2}\hspace{0.5cm}{\rm for}\hspace{0.2cm}{\rm ISM}\,\,(\rm k=0) \cr
8.9\times 10^2\,{\rm s}\left(\frac{1+z}{1.5}\right)\, A_{\rm W,-2}^{-1}\, E_{53}\, \Gamma^{-4}_{2}\hspace{0.35cm}{\rm for}\hspace{0.16cm}{\rm wind}\,\,(\rm k=2)\,,
\end{cases}
\end{eqnarray}
}

\subsubsection{Constant-density medium ($k=0$)}

\paragraph{Light curves for $t < t_{\rm x}$.}
Before the reverse shock crosses the shell, the magnetic field, the minimum and the cooling electron Lorentz factors for $1<p<2$ ($2 < p$)  evolve as $B'_{\rm r}\propto t^0$, $\gamma_{\rm m,r}\propto t^{\frac{3}{p-1}}\,(t^3)$ and $\gamma_{\rm c,r}\propto t^{-1}$, respectively.  Moreover, the synchrotron spectral breaks and the maximum synchrotron flux in term of the observed time are $\nu^{\rm syn}_{\rm m, r}\propto t^{\frac{6}{p-1}}\,(t^6)$ and $\nu^{\rm syn}_{\rm c, r}\propto t^{-2}$ and $F^{\rm syn}_{\rm max,r }\propto t^{\frac32}$, respectively. For this time interval,   the SSC spectral breaks and the maximum SSC flux for $1<p<2$ and $2 < p$ yield

{\small
\begin{eqnarray}\label{break_thin_aft_k0}
h \nu^{\rm ssc}_{\rm m, r} &\simeq& \begin{cases}
4.2\,\times 10^{-19}\,{\rm GeV}\left(\frac{1+z}{1.5}\right)^{-\frac{11+p}{p-1}} g^{\frac{4}{p-1}}(1.9) \chi_{\rm e,-1}^{-4} \varepsilon^{\frac{4}{p-1}}_{\rm e_r,-1} \varepsilon^{\frac{3-p}{2(p-1)}}_{\rm B_r,-2} \,\Gamma_{2}^{\frac{34}{p-1}}\,n^{\frac{11-p}{2(p-1)}}_{-1}E_{53}^{-\frac{4}{p-1}}t_{2}^{\frac{12}{p-1}}\hspace{2.7cm}{\rm for} \hspace{0.1cm} { 1<p<2 }\cr
6.4\,\times 10^{-17}\,{\rm GeV}  \, \left(\frac{1+z}{1.5}\right)^{-13} g^4(2.2) \chi_{\rm e,-1}^{-4} \varepsilon^4_{\rm e_r,-1} \varepsilon^{\frac12}_{\rm B_r,-2} n^\frac92_{-1}\,\Gamma^{34}_{3}\,E^{-4}_{53}\,t^{12}_{2}, \hspace{4.75cm}{\rm for} \hspace{0.1cm} { 2<p }\cr
\end{cases}\cr
h \nu^{\rm ssc}_{\rm c, r}&\simeq& 5.2\times 10^{3}\,{\rm GeV}\, \left(\frac{1+z}{1.5} \right)^{3}\,\left(\frac{1+Y_{\rm r}}{2}\right)^{-4}\,\varepsilon^{-\frac72}_{\rm B_r,-2} n^{-\frac72}_{-1}  \Gamma^{-10}_{2} \,t^{-4}_{2},\,\cr
F^{\rm ssc}_{\rm max,r} &\simeq&  4.3\times 10^{-10}\,{\rm mJy}\,\left(\frac{1+z}{1.5} \right)^{-\frac{3}{2}}\,g^{-1}(2.2)\,\chi_{\rm e,-1}\,   \varepsilon^{\frac12}_{\rm B_r,-2}\, \Gamma^{7}_{2}\, n^2_{-1}    \,d^{-2}_{\rm z,28.3}\,E^{\frac12}_{53}\,t^{\frac {5}{2}}_{2}\,,
\eary
}
respectively.

For $\nu^{\rm ssc}_{\rm c, r}<\nu^{\rm ssc}_{\rm m, r}$ and $\nu^{\rm ssc}_{\rm m, r}<\nu^{\rm ssc}_{\rm c, r}$, the spectral breaks in the KN regime for $1<p<2$ and $2 < p$ are

{\small
\bary
h\nu^{\rm ssc}_{\rm KN, m, r}&\simeq&\begin{cases}
4.9\times 10^{-3}\,{\rm GeV}\,\left(\frac{1+z}{1.5}\right)^{-\frac{2+p}{p-1}}g^{\frac{1}{p-1}}(1.9)\chi_{e,-1}^{-1}\,\,\varepsilon^{\frac{1}{p-1}}_{\rm e_r,-1}\,\varepsilon^{\frac{2-p}{4(p-1)}}_{\rm B_r,-2}\,n^{\frac{6-p}{4(p-1)}}_{-1}\Gamma^{\frac{16+p}{2(p-1)}}_{2}E_{53}^{-\frac{1}{p-1}}t_{2}^{\frac{3}{p-1}}\hspace{2.7cm}{\rm for} \hspace{0.1cm} { 1<p<2 }\cr
1.7\times 10^{-2}\,{\rm GeV}\,\left(\frac{1+z}{1.5}\right)^{-4}g(2.2)\chi_{e,-1}^{-1}\,\varepsilon_{\rm e_r,-1}\,n_{-1}\,\Gamma^{9}_{2}\,E_{53}^{-1}\,t^{2}_{3},\hspace{6.2cm}{\rm for} \hspace{0.1cm} { 2<p }\cr
\end{cases}
\cr
h\nu^{\rm ssc}_{\rm KN, c, r}&\simeq& 1.7\times 10^{3}\,{\rm GeV}\, \left(\frac{1+Y_r}{2}\right)^{-1}\,\varepsilon^{-1}_{\rm B_r,-2}\,n^{-1}_{-1}\,\Gamma^{-2}_{2}\,t^{-1}_{2},\,\,\,\,
\eary
}
respectively.

\paragraph{Light curves for  $t_{\rm x}< t$.} 

For the spectral index of the electron population in the range of $1<p<2$ ($2 < p$), the evolution of the post-shock magnetic field is ${B'}_{\rm r} \propto t^{-\frac47}$, the breaks of the electron Lorentz factors vary as $\gamma_{\rm m, r}\propto t^{-\frac{24-7p}{35(p-1)}}\,(t^{-\frac27})$ and $\gamma_{\rm c,r}\propto  t^{\frac{19}{35}}$, the corresponding synchrotron break frequencies evolve as $\nu^{\rm syn}_{\rm m, r}\propto t^{-\frac{2(10p+7)}{35(p-1)}}\,(t^{-\frac{54}{35}})$ and $\nu^{\rm syn}_{\rm c, r}\propto t^{\frac{4}{35}}$  and the maximum flux is given by $F^{\rm syn}_{\rm max,r }\propto t^{-\frac{34}{35}}$ \citep{2000ApJ...545..807K}. In this scenario,  the spectral breaks and the maximum flux of the SSC process for $1<p<2$ and $2 < p$ can be written as

{\small
\begin{eqnarray}\label{break_thin_aft_k0}
h \nu^{\rm ssc}_{\rm m, r} &\simeq& \begin{cases}
2.0\,\times 10^{-7}\,{\rm GeV} \left(\frac{1+z}{1.5}\right)^{\frac{97-29p}{35(p-1)}}\,\tilde{g}^{\frac{4}{p-1}}(1.9)\,\chi_{\rm e,-1}^{-4}\,\varepsilon^{\frac{4}{p-1}}_{\rm e_r,-1} \varepsilon^{\frac{3-p}{2(p-1)}}_{\rm B_r,-2} \, n^{\frac{191-117p}{210(p-1)}}_{-1}\Gamma^{-\frac{286+48p}{105(p-1)}}_{2}\,  E^{\frac{2(3p+31)}{105(p-1)}}_{53}\,t^{-\frac{2(31+3p)}{35(p-1)}}_{3}\hspace{1.15cm}{\rm for} \hspace{0.1cm} { 1<p<2 }\cr
2.1\,\times 10^{-6}\,{\rm GeV} \left(\frac{1+z}{1.5}\right)^{\frac{39}{35}}\,g^4(2.2)\,\chi_{\rm e,-1}^{-4}\,\varepsilon^4_{\rm e_r,-1} \varepsilon^{\frac12}_{\rm B_r,-2}\,n^{-\frac{43}{210}}_{-1}\, \Gamma^{-\frac{382}{105}}_{2}\,  E^{\frac{74}{105}}_{53}\,t^{-\frac{74}{35}}_{3}\hspace{4.5cm}{\rm for} \hspace{0.1cm} { 2<p }\cr
\end{cases}\cr
h \nu^{\rm ssc}_{\rm cut, r}&\simeq& 9.9\times 10^{-2}\,{\rm GeV} \left(\frac{1+z}{1.5}\right)^{\frac{39}{35}}\,\left(\frac{1+Y_{\rm r}}{2}\right)^{-4}  \varepsilon^{-\frac72}_{\rm B_r,-2}\,n^{-\frac{201}{70}}_{-1} \, \Gamma^{-\frac{174}{35}}_{2}\, E^{-\frac{22}{35}}_{53}\, t^{-\frac{74}{35}}_{3}\,\cr
F^{\rm ssc}_{\rm max,r} &\simeq&  3.7\times 10^{-9}\,{\rm mJy}\, \left(\frac{1+z}{1.5}\right)^{\frac{62}{35}} \,g^{-1}(2.2)\,\chi_{\rm e,-1}\,\varepsilon^{\frac12}_{\rm B_r,-2} \, n^{\frac{191}{210}}_{-1} \,\Gamma^{-\frac{181}{105}}_{2}\,d^{-2}_{\rm z,27.9}E^{\frac{167}{105}}_{53}\, t^{-\frac{27}{35}}_{3},
\eary
}
respectively,  where the cutoff frequency $\nu^{\rm ssc}_{\rm cut, r}=\nu^{\rm ssc}_{\rm c, r}(t_{\rm x})\,\left(\frac{t}{t_{\rm x}} \right)^{-\frac{74}{35}}$ is estimated from the condition that no electrons are shocked anymore and the fluid expands adiabatically. It is worth noting that the SSC flux would evolve at high latitudes equal to the flux mentioned above in the thick-shell regime.

For this scenario, the spectral breaks in the KN regime with the cooling conditions $\nu^{\rm ssc}_{\rm c, r}<\nu^{\rm ssc}_{\rm m, r}$ and $\nu^{\rm ssc}_{\rm m, r}<\nu^{\rm ssc}_{\rm c, r}$ for $1<p<2$ and $2<p$ can be written as 
{\small
\bary
h\nu^{\rm ssc}_{\rm KN, m, r}&\simeq&\begin{cases}
4.1\,{\rm GeV}  \,\left(\frac{1+z}{1.5}\right)^{\frac{45-28p}{35(p-1)}}\tilde{g}^{\frac{1}{p-1}}(1.9)\chi_{e,-1}^{-1}\,\varepsilon^{\frac{1}{p-1}}_{\rm e_r,-1}\,\epsilon^{\frac{2-p}{4(p-1)}}_{\rm B_r,-2}\,n^{\frac{170-133p}{420(p-1)}}_{\rm -1} \,\Gamma^{-\frac{160+7p}{210(p-1)}}_{2}\,E^{\frac{10+7p}{105(p-1)}}_{53}\,t^{-\frac{10+7p}{35(p-1)}}_{3}\hspace{1cm}{\rm for} \hspace{0.1cm} { 1<p<2 }\cr
7.3\,{\rm GeV}  \,\left(\frac{1+z}{1.5}\right)^{-\frac{11}{35}}g(2.2)\chi_{e,-1}^{-1}\,\varepsilon_{\rm e_r,-1}\,n^{-\frac{8}{35}}_{\rm -1} \,\Gamma^{-\frac{29}{35}}_{2}\,E^{\frac{8}{35}}_{53}\,t^{-\frac{24}{35}}_{3},\hspace{5.2cm}{\rm for} \hspace{0.1cm} { 2<p }\cr
\end{cases}
\cr
h\nu^{\rm ssc}_{\rm KN, c, r}&\simeq& 1.3\times10^2\,{\rm GeV}  \,\left(\frac{1+z}{1.5}\right)^{-\frac{8}{7}}\, \left(\frac{1+Y_{\rm r}}{2} \right)^{-1}\,\epsilon^{-1}_{\rm B_r,-2}\,n^{-\frac{13}{21}}_{-1} \,\Gamma^{\frac{22}{21}}_{2}\,E^{-\frac{8}{21}}_{\rm  53}\,t^{\frac{1}{7}}_{3},\,\,\,
\eary
}
respectively. 
%
%
\subsubsection{Stellar-wind medium ($k=2$)}

\paragraph{Light curves for $t < t_{\rm x}$.}
Before the reverse shock crosses the shell, the magnetic field, the minimum and the cooling electron Lorentz factors for $1<p<2$ ($2 < p$) evolve as $B'_{\rm r}\propto t^{-1}$, $\gamma_{\rm m,r}\propto t^{\frac{p}{2(p-1)}}\,(t)$ and $\gamma_{\rm c,r}\propto t$, respectively.  Moreover, the synchrotron spectral breaks and the maximum synchrotron flux in term of the observed time are $\nu^{\rm syn}_{\rm m, r}\propto t^{\frac{1}{p-1}}\,(t)$ and $\nu^{\rm syn}_{\rm c, r}\propto t$ and $F^{\rm syn}_{\rm max,r }\propto t^{-\frac{1}{2}}$, respectively.  For this time interval,   the SSC spectral breaks and the maximum SSC flux for $1<p<2$ and $2 < p$ yield

{\small
\begin{eqnarray}\label{break_thin_aft_k2}
h \nu^{\rm ssc}_{\rm m, r} &\simeq& \begin{cases}
1.9\times 10^{-13}\,{\rm GeV}  \, \left(\frac{1+z}{1.5}\right)^{-\frac{2p}{p-1}} g^{\frac{4}{p-1}}(1.9) \chi_{\rm e,-1}^{-4} \varepsilon^{\frac{4}{p-1}}_{\rm e_r,-1} \varepsilon^{\frac{3-p}{2(p-1)}}_{\rm B_r,-2} \,A^{\frac{11-p}{2(p-1)}}_{\rm W,-2}\,\Gamma^{\frac{2(6+p)}{p-1}}_{2}\,E_{53}^{-\frac{4}{p-1}}\,t^{\frac{1+p}{p-1}}_{2}\hspace{3.3cm}{\rm for} \hspace{0.1cm} { 1<p<2 }\cr
5.2\times 10^{-12}\,{\rm GeV}  \, \left(\frac{1+z}{1.5}\right)^{-4} g^4(2.2) \chi_{\rm e,-1}^{-4} \varepsilon^4_{\rm e_r,-1} \varepsilon^{\frac12}_{\rm B_r,-2} A^\frac92_{\rm W,-2}\,\,\Gamma^{16}_{2}\,E^{-4}_{53}\,t^{3}_{2}\hspace{5.6cm}{\rm for} \hspace{0.1cm} { 2<p }\cr
\end{cases}\cr
h \nu^{\rm ssc}_{\rm c, r}&\simeq& 2.6\times 10^{-10}\,{\rm GeV}\, \left(\frac{1+z}{1.5} \right)^{-4}\,\left(\frac{1+Y_{\rm r}}{2}\right)^{-4}\,\varepsilon^{-\frac72}_{\rm B_r,-2} A^{-\frac72}_{\rm W,-2} \Gamma^{4}_{2} \,t^{3}_{2}\,\cr
F^{\rm ssc}_{\rm max,r} &\simeq&  8.0\times 10^{-9}\,{\rm mJy}\,\left(\frac{1+z}{1.5} \right)^{\frac{5}{2}}\,g^{-1}(2.2)\,\chi_{\rm e,-1}\,   \varepsilon^{\frac12}_{\rm B_r,-2}\,  \Gamma^{-1}_{2}\,A^2_{\rm W,-2}\,d^{-2}_{\rm z,28.3}\,E^{\frac12}_{53}\,t^{-\frac{3}{2}}_{2}\,,
\eary
}
respectively.

For $\nu^{\rm ssc}_{\rm c, r}<\nu^{\rm ssc}_{\rm m, r}$ and $\nu^{\rm ssc}_{\rm m, r}<\nu^{\rm ssc}_{\rm c, r}$, the spectral breaks in the KN regime for $1<p<2$ and $2 < p$ are

{\small
\bary
h\nu^{\rm ssc}_{\rm KN, m, r}&\simeq& \begin{cases}
4.7\times 10^{-2}\,{\rm GeV}\,\left(\frac{1+z}{1.5}\right)^{\frac{2-3p}{2(p-1)}}g^{\frac{1}{p-1}}(1.9)\chi_{e,-1}^{-1}\,\varepsilon^{\frac{1}{p-1}}_{\rm e_r,-1}\,\varepsilon^{\frac{2-p}{4(p-1)}}_{\rm B_r,-2}\,A^{\frac{6-p}{4(p-1)}}_{\rm W,-2}\Gamma^{\frac{4+3p}{2(p-1)}}_{2}\,E_{53}^{-\frac{1}{p-1}}\,t^{\frac{p}{2(p-1)}}_{2}\hspace{1.5cm}{\rm for} \hspace{0.1cm} { 1<p<2 }\cr
1.1\times 10^{-1}\,{\rm GeV}\,\left(\frac{1+z}{1.5}\right)^{-2}g(2.2)\chi_{e,-1}^{-1}\,\varepsilon_{\rm e_r,-1}\,A_{\rm W,-2}\,\Gamma^{5}_{2}\,E_{53}^{-2}\,t_{2}\hspace{5.3cm}{\rm for} \hspace{0.1cm} { 2<p }\cr
\end{cases}\cr
h \nu^{\rm ssc}_{\rm KN,c, r}&\simeq& 2.9\times 10^{-1}\,{\rm GeV}\,\left(\frac{1+z}{1.5}\right)^{-2} \left(\frac{1+Y_r}{2}\right)^{-1}\,\varepsilon^{-1}_{\rm B_r,-2}A^{-1}_{\rm W,-2}\,\Gamma^{2}_{2}\,t_{2},\,\,\,
\eary
}
respectively.

\paragraph{SSC light curves for  $t_{\rm x}<t$.} 

Once the reverse shock crosses the shell, for $k=2$ the magnetic field, the Lorentz factor for the minimum and cooling electrons, the synchrotron spectral breaks, and the maximum synchrotron flux in terms of the observed time for $1<p<2$ ($2 < p$) are ${B'}_{\rm r}\propto t^{-\frac{16}{21}}$, $\gamma_{\rm m,r}\propto t^{-\frac{22-7p}{21(p-1)}}\,(t^{-\frac{8}{21}})$ and $\gamma_{\rm c,r}\propto t^\frac67$, $\nu^{\rm syn}_{\rm m, r}\propto t^{-\frac{7+3p}{7(p-1)}}\,(t^{-\frac{13}{7}})$ and $\nu^{\rm syn}_{\rm c, r}\propto t^{\frac{13}{21}}$ and $F^{\rm syn}_{\rm max,r}\propto t^{-\frac{23}{21}}$, respectively. Therefore, the spectral breaks and the maximum flux of SSC emission for $1<p<2$ and $2 < p$ yield

{\small
\begin{eqnarray}\label{break_thin_aft_k2}
h \nu^{\rm ssc}_{\rm m, r} &\simeq& \begin{cases}
1.5\times 10^{-10}\,{\rm GeV}\, \left(\frac{1+z}{1.5}\right)^{\frac{86-26p}{21(p-1)}}\,\tilde{g}^{\frac{4}{p-1}}(1.9)\,\chi_{\rm e,-1}^{-4}\,\varepsilon^{\frac{4}{p-1}}_{\rm e_r,-1} \varepsilon^{\frac{3-p}{2(p-1)}}_{\rm B_r,-2} \,A^{\frac{59-53p}{42(p-1)}}_{\rm W,-2}\, \Gamma^{-\frac{2(46+11p)}{21(p-1)}}_{2}\,  E^{\frac{2(8p+1)}{21(p-1)}}_{53}\,t^{-\frac{5(13-p)}{21(p-1)}}_{3}\hspace{1.1cm}{\rm for} \hspace{0.1cm} { 1<p<2 }\cr
2.6\times 10^{-9}\,{\rm GeV}\, \left(\frac{1+z}{1.5}\right)^{\frac{34}{21}}\,g^4(2.2)\,\chi_{\rm e,-1}^{-4}\,\varepsilon^4_{\rm e_r,-1} \varepsilon^{\frac12}_{\rm B_r,-2}\,A^{-\frac{47}{42}}_{\rm W,-2}  \, \Gamma^{-\frac{136}{21}}_{2}\,  E^{\frac{34}{21}}_{53}\,t^{-\frac{55}{21}}_{3}\hspace{4.5cm}{\rm for} \hspace{0.1cm} { 2<p }\cr
\end{cases}\cr
h \nu^{\rm ssc}_{\rm cut, r}&\simeq& 1.3\times 10^{-7}\,{\rm GeV} \left(\frac{1+z}{1.5}\right)^{\frac{34}{21}}\,\left(\frac{1+Y_{\rm r}}{2}\right)^{-4}  \varepsilon^{-\frac72}_{\rm B_r,-2}\,A^{-\frac{383}{42}}_{\rm W,-2}  \, \Gamma^{-\frac{388}{21}}_{2}\, E^{\frac{118}{21}}_{53}\, t^{-\frac{55}{21}}_{3}\,\cr
F^{\rm ssc}_{\rm max,r} &\simeq&  8.7\times 10^{-10}\,{\rm mJy}\, \left(\frac{1+z}{1.5}\right)^{\frac{17}{7}} \,g^{-1}(2.2)\,\chi_{\rm e,-1}\,\varepsilon^{\frac12}_{\rm B_r,-2} \, A^{\frac{29}{14}}_{\rm W,-2}\, \,\Gamma^{-\frac{5}{7}}_{2}\,d^{-2}_{\rm z,27.9}E^{\frac{3}{7}}_{53}\, t^{-\frac{10}{7}}_{3},
\eary
}

respectively, where the cutoff frequency $\nu^{\rm ssc}_{\rm cut, r}=\nu^{\rm ssc}_{\rm c, r}(t_{\rm x})\,\left(\frac{t}{t_{\rm x}} \right)^{-\frac{55}{21}}$ is estimated from the condition that no electrons are shocked anymore and the fluid expands adiabatically.  It is important to emphasize that the SSC flux would undergo an evolution at high latitudes that is equivalent to the flux identified above in the thick-shell regime.  In the thin-shell case evolving in a stellar-wind model,  the spectral breaks in the KN regime for $1<p<2$ and $2 < p$ are given by
{\small
\bary
h\nu^{\rm ssc}_{\rm KN, m, r}&\simeq& \begin{cases}
0.4\,{\rm GeV}  \,\left(\frac{1+z}{1.5}\right)^{\frac{12-7p}{7(p-1)}}\tilde{g}^{\frac{1}{p-1}}(1.9)\chi_{e,-1}^{-1}\,\varepsilon^{\frac{1}{p-1}}_{\rm e_r,-1}\,\varepsilon^{\frac{2-p}{4(p-1)}}_{\rm B_r,-2}\,A^{\frac{22-21p}{28(p-1)}}_{\rm W,-2}E^{-\frac{4-7p}{14(p-1)}}_{53}\Gamma^{-\frac{12+7p}{14(p-1)}}_{2}\,t^{-\frac{5}{7(p-1)}}_{3}\hspace{1.2cm}{\rm for} \hspace{0.1cm} { 1<p<2 }\cr
0.9\,{\rm GeV}  \,\left(\frac{1+z}{1.5}\right)^{-\frac{2}{7}}g(2.2)\chi_{e,-1}^{-1}\,\varepsilon_{\rm e_r,-1}\,A^{-\frac{5}{7}}_{\rm W,-2} \,\Gamma^{-\frac{13}{7}}_{2}\,E^{\frac{5}{7}}_{53}\,t^{-\frac{5}{7}}_{3}\hspace{5.25cm}{\rm for} \hspace{0.1cm} { 2<p }\cr
\end{cases}\cr
h\nu^{\rm ssc}_{\rm KN, c, r}&\simeq& 2.5\,{\rm GeV}  \,\left(\frac{1+z}{1.5}\right)^{-\frac{32}{21}} \left(\frac{1+Y_{\rm r}}{2} \right)^{-1}\,\varepsilon^{-1}_{\rm B_r,-2}\,A^{-\frac{31}{21}}_{\rm W,-2}\,\Gamma^{\frac{2}{21}}_{2}\,E^{\frac{10}{21}}_{53}\,t^{\frac{11}{21}}_{3},\,\,\,
\eary
}

respectively.

\section{Application: Second \textit{Fermi}-LAT Catalog}\label{sec3}

We choose GRBs from the 2FLGC \citep{Ajello_2019} comprising 86 that have temporally extended emission with durations ranging from 31 to 34366 seconds \citep{2009MNRAS.400L..75K, 2010MNRAS.409..226K, 2016ApJ...818..190F}. Notably, a subset of 29 bursts exhibited photons that extend to energies higher than 10 GeV. Ten bursts display values of spectral ($\Gamma_{\rm L}\approx 2$) and temporal ($\alpha_{\rm L}\gtrsim 1.5$) indices, which are challenging to characterize using the closure relations of the conventional synchrotron forward-shock model. In addition, around fifty percent of the bursts with long-lasting LAT emission have light curves that peak just before the prompt phase is over. 

Taking into account the spectral breaks and maximum flux before and after the shock crossing time, we derive the temporal PL index of the SSC light curve in each cooling condition for $1 < p < 2$ and $2< p$ in the thick-shell scenario, as listed in Tables \ref{tableReverseBefore} and \ref{tableReverseThick}. This table shows the temporal index for both constant density and stellar wind media. Similarly, we derive the temporal index of the SSC light curve in the thin-shell scenario, as listed in Tables \ref{tableReverseBefore} and \ref{tableReverseThick}.

We employ temporal and spectral indices based on the PL function $F^{\rm ssc}_{\rm \nu} \propto t^{-\alpha} \nu^{-\beta}$, to derive before and after the shock crossing time, as listed in Tables \ref{TableCRs_before} and \ref{Table2} the SSC CRs from the reverse-shock region for thick and thin-shell scenario, respectively, with $1<p<2$ and $ 2 < p$.  The temporal index $\alpha_{\rm L}$ is derived from the 2FLGC. In contrast, the spectral index $\beta_{\rm L}$ is calculated as $\beta_{\rm L}=\Gamma_L-1$, where the photon index $\Gamma_L$ is reported in this catalog. The SSC model is considered as the reverse shock evolves in the stellar wind and interstellar environment.  

For the analysis of the CRs, we follow the strategy presented in \cite{2021PASJ...73..970D, 2021ApJS..255...13D, 2020ApJ...903...18S, 2022ApJ...934..188F, 2023MNRAS.525.1630F}. The error bars quoted in the graphs are in the 1$\sigma$ range, and we consider the error bars among the values $\alpha$ and $\beta$ dependent; thus, Figures \ref{fig1:CRs}, \ref{fig2:CRs}, \ref{fig3:CRs} and \ref{fig4:CRs} are shown with ellipses. Table~\ref{tab:CR-Results} indicates the number and proportion of GRBs described with a PL and a BPL reported in 2FLGC, satisfying each cooling condition according to the SSC reverse shock model after the shock crossing time in both the thick and thin shell regimes. We consider both a wind-like medium and 
interstellar medium (ISM) and consider both ranges of the spectral index, $1 < p < 2$ and $2 < p$. 

In general conclusion, out of the 86 GRBs considered, the CRs of the SSC mechanism of the reverse shock could at most satisfy $ 10 $ and $ 27 $ when a description of PL and BPL, respectively, is taken into account. We note that the thin shell case is preferred over a thin shell and a constant-density medium over a stellar-wind environment. The most preferred scenario of a thin shell developing in a constant ISM during the slow-cooling condition $\nu_{\rm m,r}^{\rm ssc}<\nu<\nu_{\rm cut,r}^{\rm ssc}$ for a BPL corresponds to 5. 81\% (5 GRBs). The second most preferred scenario is a five-way tie that can explain $4$ GRBs (4.65\%) from the BPL column. We note that the GRBs fitted with a BPL from the 2FLGC are the ones that present more coincidences with the reverse shock model, as compared with the PL, where the most significant number of agreements is just two (2.33\%) for a thin shell developing in a constant-density medium during the slow-cooling condition $\nu<\nu_{\rm m,r}^{\rm ssc}$. Finally, it should be noted that no case was found for the thick shell in either an ISM or a wind-like scenario for the condition $\nu<\nu_{\rm m,r}^{\rm ssc}$. As expected, the number of GRBs satisfying the same CRs when the SSC process lies in the condition  $\nu_{\rm cut,r}^{\rm ssc}<\nu$.


\subsection{Stellar wind vs ISM }

The density patterns of the external environment are considered for both short and long gamma-ray burst progenitors. Short gamma-ray bursts (GRBs) result from the mergers of binary compact objects, such as a black hole (BH) and a neutron star (NS) or two NSs \citep{1992ApJ...392L...9D, 1992Natur.357..472U, 1994MNRAS.270..480T, 2011MNRAS.413.2031M}. In contrast, long GRBs arise from the death of massive stars characterized by varying mass-loss rates and the stellar wind circumstellar medium \citep{1993ApJ...405..273W,1998ApJ...494L..45P}.  Short GRBs are predicted to occur in a medium with a low-density constant medium, whereas long GRBs linked to massive objects that die in different ways are predicted to occur in a stellar-wind environment \citep{2005ApJ...631..435R,  2006A&A...460..105V}.   It should be noted that in the case of core-collapse of massive stars, a transition from wind to interstellar medium afterglow might be thought if the wind emitted by the parent star does not exceed the deceleration ratio \citep[e.g., see][]{2017ApJ...848...15F,2019ApJ...879L..26F}.

Table \ref{tab:CR-Results} shows that a constant-density medium is preferred over a stellar-wind environment, irrespective of the evolution of the reverse shock. For instance, when the SSC mechanism lies in the cooling condition $\nu_{\rm m,r}^{\rm ssc}<\nu<\nu_{\rm cut,r}^{\rm ssc}$, 4 (2) bursts satisfy the CRs for a thick shell and 5 (2) for a thin shell evolving in ISM (wind). The number of GRBs satisfying the same CRs when the SSC process lies in the condition  $\nu_{\rm cut,r}^{\rm ssc}<\nu$ are equal.

Table \ref{table:dens_paramet} shows the SSC flux as a function of the density parameter for each cooling condition when reverse shock evolves in the thick- and thin-shell regime. It should be noted that when the condition $\nu^{\rm ssc}_{\rm c, r}< \nu$ is reached, there is no way to distinguish between a stellar wind or a homogeneous environment. Variations of density in a stellar-wind medium produce a signature more significant than ISM. Similarly, this signature is more notable in the thick than the thin-shell regime. For instance, the SSC flux under the cooling condition $\nu^{\rm ssc}_{\rm c, r}< \nu$ evolves in the thick(thin) shell case as $F^{\rm ssc}_{\rm \nu, r}\propto n^{2.9}(A_{\rm W}^{0.79})$ and $\propto n^{1.45}(A_{\rm W}^{1.4})$ in the stellar wind and ISM, respectively, for $p=2.2$. A similar analysis could consider an electron population with a hard spectral index ($1 < p <2$).


Given that such massive stars are situated inside the ISM, a transition will occur at varying distances between the stellar wind expelled by this kind of progenitors and the ISM, resulting in a discontinuity or bubble morphology \citep{1975ApJ...200L.107C, 1977ApJ...218..377W, 2006ApJ...643.1036P}. The deceleration of the outflow launched by the progenitor in the stellar wind and, subsequently, a homogeneous density allowed the discovery of this afterglow transition through multiwavelength light curves, characterized by a rise in the synchrotron light curves, as seen in some bursts \citep[GRB 030226, 050319, 081109A, 140423A, 160626B, and 190114C;][]{2003ApJ...591L..21D, 2007ApJ...664L...5K, 2009MNRAS.400.1829J, 2020ApJ...900..176L, 2017ApJ...848...15F, 2019ApJ...879L..26F}.

For example, \cite{10.1111/j.1365-2966.2009.15886.x} suggested that changes in microphysical parameters will likely occur within these bubble formations around this transition, thus interpreting the multiwavelength data collected in GRB 060206, 070311, and 070101A. \cite{2019ApJ...883..162F} developed the SSC light curves during the afterglow transition from stellar wind to ISM to elucidate the VHE observations exhibited in GRB 190114C. Although a reverse shock during the transition from wind to ISM has not yet been reported, it should be expected with a notable signature in the cooling condition $ \nu^{\rm ssc}_{\rm m,r} < \nu < \nu^{\rm ssc}_{\rm cut,r}$ (from $\propto A_{\rm W}^\frac{221 - 47p}{84}$ to $n^{\frac{425 -43p}{420}}$ for $2<p$ during the thin case), but not during the condition $\nu^{\rm ssc}_{\rm cut,r} < \nu$, neither in the thick nor the thin-shell regime.

\subsection{Maximum energy from reverse shock}

It is feasible to determine the maximum energy that the synchrotron process emits from the reverse-shock region, considering the timescales of synchrotron ($\propto\gamma_e^{-1} B'^{-2}_{\rm r}$) and acceleration ($\propto \gamma_e B'^{-1}_{\rm r}$).   Given the elementary electron charge ($q_e$), the Thompson cross-section ($\sigma_T$) and the parameter ($\xi\sim 1$) in the Bohm limit, the maximum Lorentz factor that electrons can reach is $\gamma_{\rm max, r}=\left(3q_e/\xi\sigma_T B'_{\rm r}\right)^{\frac12}$, and therefore the maximum energy emitted by the synchrotron mechanism during the thick and thin regime becomes
{\small
\begin{eqnarray}\label{break_thin_aft_k2}
h \nu^{\rm syn}_{\rm max, r} &=& \begin{cases}
2.1\,{\rm GeV} \left(\frac{1+z}{1.5}\right)^{-1}\, n^{-\frac18}_{-1} E^{\frac{1}{8}}_{53} \,t^{-\frac{7}{16}}_{\rm 2}\hspace{3.4cm}{\rm for} \hspace{0.1cm} {\rm  k=0 }\cr
1.6\,{\rm GeV} \left(\frac{1+z}{1.5}\right)^{-1}\,A^{-\frac14}_{W,-2} E^{\frac{1}{4}}_{53} \,t^{-\frac{3}{8}}_{\rm 2}\hspace{3.15cm}{\rm for} \hspace{0.1cm} {\rm  k=2 }\cr
\end{cases}
\eary
}%
and
{\small
\begin{eqnarray}\label{break_thin_aft_k2}
h \nu^{\rm syn}_{\rm max, r} &=& \begin{cases}
1.5\,{\rm GeV} \left(\frac{1+z}{1.5}\right)^{-\frac43}\, n^{-\frac19}_{-1} E^{\frac{1}{9}}_{53} \,\Gamma^{\frac19}_2 \,t^{-\frac{1}{3}}_{\rm 3}\hspace{3.2cm}{\rm for} \hspace{0.1cm} {\rm  k=0 }\cr
1.6\,{\rm GeV} \left(\frac{1+z}{1.5}\right)^{-\frac43}\,A^{-\frac13}_{W,-2} E^{\frac{1}{3}}_{53} \,\Gamma^{-\frac13}_2 \,t^{-\frac{1}{3}}_{\rm 3}\hspace{2.65cm}{\rm for} \hspace{0.1cm} {\rm  k=2 }\,,\cr
\end{cases}
\eary
}%

respectively.  Figures \ref{fig7:CRs} and \ref{fig8:CRs} show LAT-detected bursts with energies greater than $> 100\,{\rm MeV}$ together with the maximum energy radiated by the synchrotron mechanism when the reverse shock evolves in the thick- and thin-shell regime, respectively, for stellar wind and ISM environments. All photons have a probability of $>90$\% to be associated with each burst. We consider the values of equivalent kinetic energies of $E=10^{52}$ and $10^{54}$ erg, and the values of the bulk Lorentz factors $\Gamma=600$ and $100$, thick and thin-shell regime, respectively.   We note that in the case of a thick shell, all photons associated with bursts GRB 091127, 110731A, and 131108A could eventually be explained by a synchrotron from the reverse-shock region for ISM. A similar situation occurs for the thin-shell regime with GRB 110731A, 120624B, and 131108A. In general, the synchrotron reverse-shock model could explain only photons below a few GeV, so a new mechanism such as SSC from the reverse shock should be considered ($h \nu^{\rm ssc}_{\rm max, r}\gg 10\, {\rm TeV}$), as we use in this manuscript.

On the other hand, we estimate the number of photons that could be emitted by the SSC mechanism from the reverse-shock region and reach the LAT detector.   The number of photons ($N_{\gamma}$) with energy ($h\nu$) that reach the \textit{Fermi}-LAT instrument during a specific time interval ($\Delta t$) can be estimated considering the effective area of LAT ($A_{\rm eff}$) and the observed flux at the determined time. The number of photons is
{\small
\begin{eqnarray}\label{vhe-photon}
N_{\gamma}\sim 1\,{\rm ph} \left(\frac{h\nu}{2\,{\rm~GeV}}\right) \left(\frac{10\,{\rm s}}{\Delta t} \right)  \left(\frac{10^4\,{\rm cm^{2}}}{A_{\rm eff}}\right)\, \left(\frac{(\nu F_{\nu})^{\rm ssc}}{5\times 10^{-9}\,{\rm \frac{erg}{cm^{2}\,s}}}\right)\,,
\end{eqnarray}
}

where $(\nu F_{\nu})^{\rm ssc}$,  the observed SSC flux for instance in the thick-  regime evolving in the ISM and stellar wind, can be written as 

{\small
\begin{eqnarray}\label{break_thin_aft_k2}
(\nu F_{\nu})^{\rm ssc} &=& \begin{cases}
6.5\times 10^{-9}\,{\rm \frac{erg}{cm^2s}}\left(\frac{1+z}{1.5}\right)^{\frac{127+51p}{96}}\,g^{2p-3}(2.2)\,\chi_{\rm e,-1}^{3-2p}\,\varepsilon^{2(p-1)}_{\rm e_r,-1} \varepsilon^{\frac{p+1}{4}}_{\rm B_r,-2}\,n^{\frac{5+3p}{8}}_{-1}\,d_{\rm z,27.9}^{-2}  \, \Gamma_{2.5}^{2p-3}\,\Delta_{11.8}^{\frac{135p-101}{96}}\,  E^{\frac{13-p}{8}}_{53}\,t^{\frac{17-99p}{96}}_{2}\hspace{0.5cm}{\rm for} \hspace{0.1cm} {\rm k=0 }\cr
3.9\times 10^{-9}\,{\rm \frac{erg}{cm^2s}} \left(\frac{1+z}{1.5}\right)^{\frac{25+13p}{16}}\,g^{2p-3}(2.2)\,\chi_{\rm e,-1}^{3-2p}\,\varepsilon^{2(p-1)}_{\rm e_r,-1} \varepsilon^{\frac{p+1}{4}}_{\rm B_r,-2}\,A^{\frac{5+3p}{4}}_{\rm W,-2.5}\,d_{\rm z,27.9}^{-2}  \, \Gamma_{2.2}^{2p-3}\,\Delta_{11.8}^{\frac{21p-23}{16}}\,  E^{\frac{2-p}{2}}_{53}\,t^{-\frac{21p+1}{16}}_{2}\hspace{0.6cm}{\rm for} \hspace{0.1cm} {\rm  k=2 }.\cr
\end{cases}
\eary
}


Therefore, the SSC mechanism might explain the most energetic photons exhibited in the 2FLGC.

\subsection{The SSC breaks at Klein-Nishina regime}
KN effect plays a critical role in shaping SSC emission from reverse shocks. In particular, the suppression of high-energy up-scattered photons arises directly from the KN reduction in the Compton cross-section, which becomes significant when the photon energy in the electron rest frame approaches or exceeds the electron rest mass energy. This leads to a noticeable attenuation in the SSC flux at high energies. Indirectly, the KN effect also influences the overall spectral shape by modifying electron cooling rates, especially for those electrons with Lorentz factors placing their up-scattered photons in the KN regime. As a result, the SSC spectrum can exhibit additional spectral breaks, depending on the location of the synchrotron and SSC characteristic frequencies relative to the KN threshold. In appendix~\ref{app:KN_LC}, we identify the additional PLs in the SSC light curves caused by this effect for both the thick- and thin-shell regimes and for constant and wind-like media.

\subsection{High-latitude emission }

Following the peak flux ($t>t_{\rm x}$), once the SSC break has passed into the \textit{Fermi}-LAT band ($\nu^{\rm ssc}_{\rm cut,r} < \nu_{\rm LAT}$), the gamma-ray emission from the reverse-shock region ceases. However, sudden vanishing is prevented by angular-time delay (the emission generated at significant angles relative to the jet axis). Therefore the gamma-ray flux evolves as $F^{\rm ssc}_\nu\propto t^{-(\beta+2)}\nu^{-\beta}$. It is worth noting that afterglow emission with a decay index larger than 2 is difficult to reconcile with synchrotron forward-shock models.   Table \ref{tab:sample1} shows a sample of bursts modeled with a BPL, for which one of the temporal indexes, at least, is greater than 2. The evolution of the temporal index from soft to hard is usually interpreted as the contribution of two components. The first gamma-ray component with a temporal index larger than two would correspond to the SSC process from the reverse shock region, and the second one would correspond to the synchrotron or SSC mechanism from the forward-shock region. In other words, in the beginning, the emission from the reverse shock dominates, and then the emission from the forward shock begins to dominate.

Table \ref{tab:CR-Results} displays the results of Figures \ref{fig2:CRs} and \ref{fig4:CRs} and lists the coincidences of the SSC CRs from the reverse shock region and the bursts described with a BPL. This shows more coincidences when a BPL is required than when a PL is needed. A large sample of bursts with high-latitude emission could be eclipsed by other components, such as synchrotron or SSC emission from forward shock.


\subsection{GRBs with a hard spectral index ($1 < p < 2$)}

LAT-detected bursts with a hard spectral index are challenging to explain with the CR of standard synchrotron radiation. For instance,  \cite{2010MNRAS.403..926G} analyzed the high-energy emissions of 11 bursts detected by the \textit{Fermi}-LAT until October 2009. Given the LAT observations, the authors concluded that the relativistic forward shock operated in a fully radiative regime for a hard spectral index $p\sim 2$, exhibiting a temporal decay of $\sim t^{-\frac{10}{7}}$ rather than the expected $t^{-1}$ as predicted by the adiabatic synchrotron forward-shock model,  which is the best afterglow model, and that the relativistic forward shock was in the utterly radiative regime. \cite{2024MNRAS.534.3783F} derived the nine \textit{Fermi}-LAT light curves from 2FLGC that exhibited temporal and spectral PL indices with values of $\alpha\gtrsim 1.5$ and $\beta \approx 2$, respectively, and demonstrated that the closure relations of the conventional synchrotron afterglow model cannot well characterize these light curves. To interpret the bursts reported in 2FLGC, one could evolve the CRs of the synchrotron afterglow model in the radiative regime of the stellar wind and ISM scenario.

In the stellar-wind afterglow scenario, the electron Lorentz factors of the lowest energy and the cooling break evolve as $\gamma_{\rm m,f}\propto t^{-\frac{1}{4-\epsilon}}$ ($t^{\frac{2\epsilon+ p ( 3 - \epsilon) - 8}{2(4-\epsilon)(p-1)}}$) and $\gamma_{\rm c, f} \propto t^{\frac{3-\epsilon}{4-\epsilon}}$ for $2 < p$ ($1<p<2$), respectively.\footnote{The subindex ``f" refers to the quantities derived in the forward-shock region.}  The corresponding spectral breaks are  $\nu^{\rm syn}_{\rm m, f} \propto t^{\frac{\epsilon-6}{4-\epsilon}}$ ($t^{\frac{\epsilon - p - 4}{(4-\epsilon)(p-1)}}$) and  $\nu^{\rm syn}_{\rm c, f} \propto t_{4.7}^{\frac{2-\epsilon}{4-\epsilon}}$, respectively. Given the evolution of maximum flux $ F^{\rm syn}_{\rm max,f} \propto t^{-\frac{2}{4-\epsilon}}$, the synchrotron flux evolves as  $F^{\rm syn}_{\nu, f}\propto t^{\frac{2-6p-\epsilon+p\epsilon}{2(4-\epsilon)}} (t^{\frac{\epsilon-p-8}{2(4-\epsilon)}})$ for $\nu^{\rm syn}_{\rm m, f}<\nu <\nu^{\rm syn}_{\rm c, f}$, and $F^{\rm syn}_{\nu}\propto t^{\frac{4+p(\epsilon-6)-2\epsilon}{2(4-\epsilon)}} (t^{-\frac{p+6}{2(4-\epsilon)}})$ for $\{\nu^{\rm syn}_{\rm m, f},\nu^{\rm syn}_{\rm c,f}\}<\nu$.   In a homogeneous afterglow scenario, the Lorentz factors of the lowest energy and the cooling break vary as $\gamma_{\rm m}\propto t^{-\frac{3}{8-\epsilon}}$ ($t^{-\frac{3(4-p)}{2(8-\epsilon)(p-1)}}$) and $\gamma_{\rm c, f} \propto t^{\frac{1+\epsilon}{8-\epsilon}}$ for $2 < p$ ($1<p<2$), respectively.   The corresponding spectral breaks are  $\nu^{\rm syn}_{\rm m, f} \propto t^{-\frac{12}{8-\epsilon}}$ ($t^{-\frac{3(p+2)}{(8-\epsilon)(p-1)}} $) and  $\nu^{\rm syn}_{\rm c, f} \propto {\frac{2(\epsilon-2)}{8-\epsilon}}$, respectively.  Given the evolution of maximum flux  $ F^{\rm syn}_{\rm max, f} \propto t^{-\frac{3\epsilon}{8-\epsilon}}$,  the evolution of synchrotron flux becomes $F^{\rm syn}_{\nu, f}\propto t^{\frac{3(2-2p-\epsilon)}{8-\epsilon}} (t^{-\frac{3(p+2+2\epsilon)}{2(8-\epsilon)}}$ for $\nu^{\rm syn}_{\rm m, f}<\nu <\nu^{\rm syn}_{\rm c, f}$, and  $F^{\rm syn}_{\nu, f}\propto t^{\frac{2(2-3p-\epsilon)}{8-\epsilon}} (t^{-\frac{3p+10+4\epsilon}{2(8-\epsilon)}})$ for $\{\nu^{\rm syn}_{\rm m, f},\nu^{\rm syn}_{\rm c, f}\}<\nu$.   Table \ref{CR_FS_eps} shows the CR of the synchrotron forward-shock model evolving in the radiative regime in stellar wind and ISM. We derive the synchrotron process in the fast- and slow-cooling regime. 
The bursts adhere to the synchrotron closure relations of a constant environment under a cooling condition where $\nu^{\rm syn}_{\rm m, f}<\nu<\nu^{\rm syn}_{\rm c, f}$ for $\epsilon=0$, or a cooling condition where $\{\nu^{\rm syn}_{\rm m, f}, \nu^{\rm syn}_{\rm c, f}\}<\nu$ for $\epsilon=1$. For instance, with a value of $p=2.8$ and a cooling condition of $\nu^{\rm syn}_{\rm m, f}<\nu<\nu^{\rm syn}_{\rm c, f}$, the associated spectral and temporal PL indices are $\beta=0.9$ and $\alpha=1.35$, respectively. A comparable result may be achieved by requiring a value of $p=1.8$ and a cooling condition of $\{\nu^{\rm syn}_{\rm m, f}, \nu^{\rm syn}_{\rm c,f}\}<\nu$ with $\epsilon=0.92$. It is important to observe that if $h\nu\approx 100\,{\rm MeV}$, the initial condition $\nu^{\rm syn}_{\rm m,f}<\nu<\nu^{\rm syn}_{\rm c, f}$ may require an unusual value of the magnetic microphysical parameter $\varepsilon_{B_f} \lesssim 10^{-6}$ \citep{2009MNRAS.400L..75K,2010MNRAS.409..226K} while the latter results in a typical value of the microphysical parameter \citep[e.g., $3\times10^{-5}\lesssim\varepsilon_{B_f}\lesssim 0.3$;][]{2014ApJ...785...29S}.
 
The SSC mechanism from the reverse-shock region could explain the evolution of the spectral and temporal features. For example, given $2 < p$, the observed flux evolves with a temporal index of $F^{\rm ssc}_{\nu, r}\sim t^{-(1.8 - 2.0)}$ for both stellar wind and ISM. Although the CRs of the synchrotron afterglow model in the radiative regime could explain LAT-detected bursts with a hard spectral index, they cannot explain photons above hundreds of MeV. However, the CRs of the SSC mechanism from the reverse-shock region can explain naturally the hard spectral index and photons above hundreds of MeV.

\subsubsection{GRBs with an atypical spectral index $\beta_{\rm L}\approx 0$}

Table \ref{tab:PLsample} lists a sample of bursts included in 2FLGC, which presents an atypical spectral index ($\beta\approx 0$). This sample can hardly be described with the standard synchrotron afterglow model due to the \textit{Fermi}-LAT band being below the two spectral breaks ($\nu_{\rm LAT} < \nu^{\rm syn}_{\rm m,f}< \nu^{\rm syn}_{\rm c,f}$ or $\nu_{\rm LAT} < \nu^{\rm syn}_{\rm c,f}< \nu^{\rm syn}_{\rm m,f}$) \citep[e.g., see][]{1998ApJ...497L..17S}. However, the CRs in Table \ref{Table2} show that these bursts could be described by SSC when this process is under the cooling condition $\nu_{\rm LAT} < \nu^{\rm ssc}_{\rm m,r}< \nu^{\rm ssc}_{\rm cut,r}$. For this cooling condition, we show the parameter space in Figure \ref{fig:param_space}  when the outflow is decelerated in a homogeneous environment (top) and a circumburst medium formed by stellar wind (bottom). We consider the evolution of the reverse shock in the thick-shell regime on the left-hand panels, whereas, on the right-hand panels, the reverse shock in the thin-shell case is considered. This figure shows that the bulk Lorentz factor lies in the range $500 <\Gamma<1000$ for the evolution of the reverse shock in the thick-shell regime and case and  $\sim 100$ for the thin-shell regime. As expected, the bulk Lorentz factors are greater in the thick regime than in the thin regime. The scenario with the most extensive set of parameters corresponds to the thick-shell regime in a stellar wind, and the lowest set of parameters is for the thin-shell case in a stellar wind.

When comparing the two types of external environment, we note that the thin-shell scenario is more likely to account for CRs in the homogeneous medium case. Contrastingly, the thick-shell case prefers a stellar-wind medium. This distinction arises because, in the thin-shell regime, the reverse shock remains mildly relativistic, leading to a longer shock-crossing time and a slower evolution of the SSC emission, which aligns well with the CRs observed in a constant-density medium. In contrast, the thick-shell case involves a more substantial reverse shock, which decelerates the ejecta more efficiently. It favors a stratified stellar-wind environment where the density decreases with radius. This preference suggests that the thick-shell regime is more likely to produce SSC-dominated emission for GRBs occurring in wind-like environments, typically associated with long GRBs from massive stellar progenitors.

\section{LAT-detected bursts and Optical correlations}

\subsection{A sample of Bursts}

Analyses have been conducted on the correlations between optical and gamma-ray emissions in GRBs to elucidate whether these emissions arise from internal and/or external shocks.   Among the bursts presented in 2FLGC, GRB 160625B and GRB 180720B exhibited early optical flashes in temporal correlations with the LAT emission.

\subsubsection{GRB 160625B}%

\textit{Fermi}-GBM detected and localized GRB 160625B on June 25, 2016 at 22:40:16.28 UT \citep{2016GCN..19581...1B}. Shortly after, \textit{Fermi}-LAT also triggered this burst at 22:43:24.82 UT \citep{2016GCN..19586...1D}. The GBM instrument measured a prompt emission duration of $T_{90}=453.38\,{\rm s}$, with a corresponding fluence of $(2.5\pm 0.3)\times 10^{-5}\,{\rm erg\,cm^{-2}}$ and an isotropic-equivalent energy of $E_{\rm \gamma, iso}=(1.5\pm 0.1)\times 10^{53}\, {\rm erg}$ \citep{Ajello_2019}. The Swift/XRT instrument observed GRB 160625B in PC mode over a time span of $9.8\times10^4$ to $4.1\times 10^6\,{\rm s}$. Moreover, a series of optical observations were conducted employing ground-based telescopes, locating this burst with a redshift of $z=1.406$ \citep{2016arXiv161203089Z, 2016GCN..19600...1X, 2017Natur.547..425T}.

\subsubsection{GRB 180720B}

GRB 180720B was first identified by \textit{Fermi}-GBM on 20 July 2018 at $T_0=$ 14:21:39.65 Universal Time (UT) \citep{Roberts_2018GCN.22981....1B} and was quickly followed by Swift-BAT \citep{Siegel_2018GCN.22981....1B}. This event was classified as a long GRB due to its duration of $\sim 48.9 \pm 0.4$ s; furthermore, the estimated isotropic energy output in the 50-300 keV band was $E_{\rm \gamma,iso} = (6.0 \pm 0.1) \times 10^{53} \, \rm erg$. Furthermore, this GRB was also observed by \textit{Fermi}-LAT during the initial 700 s, collecting an energetic photon of 5 GeV at 137 s after the trigger time \citep{Bissaldi_2018GCN.22980....1B}. Due to the prompt episode observed by the GBM, this burst was classified as very luminous, ranking among the top seven brightest occurrences to date \citep{2020A&A...636A..55R}. Swift-XRT also detected a luminous afterglow 11 hours after trigger in the 0.3-10 keV range, which remained observable until 30 days. The redshift was determined to be $z=0.654$ by detecting many absorption signatures, including Ca II, Mg II, and Fe II \citep{Vreeswijk_2018GCN.22996....1V}. Very-high-energy gamma-ray emission was detected 
by the HESS collaboration  at 10.1 hours after trigger time \citep{Abdalla_2019Natur.575..464A}. 

\subsection{Analysis}

We derive and show in Table \ref{Table5} the CRs of the synchrotron afterglow model from the reverse shock after the shock crossing time evolving in stellar wind and ISM.    This table shows the thick- and thin-shell regime for the reverse shock in the cooling conditions  $\nu^{\rm ssc}_{\rm m,r} < \nu < \nu^{\rm ssc}_{\rm cut,r}$ and $ \nu^{\rm ssc}_{\rm cut,r} < \nu $, and the electron distribution with spectral PL indexes in ranges of $1<p<2$ and $2<p$. Figure \ref{fig5:CRs} shows the spectral and temporal indexes of GRB 160625B and GRB 180720B in the LAT and optical bands together with the CRs of SSC and synchrotron flux from the reverse-shock region in the cooling condition $ \nu^{\rm ssc}_{\rm m,r} < \nu < \nu^{\rm ssc}_{\rm cut,r}$. Figure \ref{fig6:CRs} is the same as Figure \ref{fig5:CRs}, but for the cooling conditions $\nu^{\rm syn}_{\rm m,r} < \nu < \nu^{\rm syn}_{\rm cut,r}$ (right) and $\nu^{\rm ssc}_{\rm m,r} < \nu < \nu^{\rm ssc}_{\rm cut,r}$ (left). We note that the only cases that satisfy the CRs for GRB 180720B and GRB 160625B are the thick- and thin-shell regime evolving in a constant-density medium and the cooling conditions $ \nu^{\rm ssc}_{\rm m,r} < \nu < \nu^{\rm ssc}_{\rm cut,r}$ and the cooling condition $ \nu^{\rm syn}_{\rm m,r} < \nu < \nu^{\rm syn}_{\rm cut,r}$.

Figure \ref{fig:best_fit} shows the early optical and LAT observations of GRB 160625B (left) and GRB 180720B (right), together with the best-fit curves obtained from the synchrotron and SSC processes. In addition, this figure shows the radio observations with the contribution of the reverse shock using the best-fit values found from the early optical and LAT observations. We consider the evolution of the reverse shock in the thick-shell regime in a constant-density medium. The early optical and LAT observations were modeled using the Chi-square ($\chi^2$) minimization method implemented in the Non-Linear Least-Squares Minimization and Curve-Fitting (LMFIT) for Python software \citep{newville_2025_15014437}. This figure illustrates that optical flux is temporally correlated with the LAT emission, suggesting that the GeV and optical fluxes were generated from the same accelerated region and electron population. Therefore, we conclude that LAT emission may have originated during the early afterglow and that the synchrotron emission from the reverse shock also contributes to the radio observations, as suggested by \cite{2020ApJ...904..166C} and
\cite{2020MNRAS.496.3326R} for GRB 160625B and GRB 180720B, respectively.

\section{Summary}\label{sec4}
We have derived the CRs during the adiabatic regime of the SSC mechanism emitted and scattered by a relativistic electron population accelerated in the reverse shock region with a hard ($1<p<2$) and a soft ($2<p$) spectral index. We have considered the evolution of the reverse shock after the shock-crossing time in the thick- and thin-shell case when the outflow is decelerated in a stratified environment formed by the stellar wind launched by the progenitor and a homogeneous medium. Furthermore, we incorporated the effect of EBL absorption as proposed in \cite{2017A&A...603A..34F} and estimated the spectral breaks that occurred in the KN regime \citep{2009ApJ...703..675N, 2010ApJ...712.1232W}.  We conducted an analysis that compared the CRs of the SSC reverse-shock model in each cooling condition with the description of the spectral and temporal characteristics of bursts documented in 2FLGC.

We have demonstrated that the CRs of the SSC mechanism resulting from the reverse shock could only fulfill a maximum of ten and twenty-seven of the 86 GRBs being considered, respectively, when a description of PL and BPL is required. It is important to note that the thin shell case is preferred over the thick shell scenario and that a constant-density medium environment is preferable over an environment with stellar wind.    In the case of a BPL, the most favorable scenario is the development of a thin shell in a constant ISM under the slow-cooling condition $\nu_{\rm m,r}^{\rm ssc}<\nu < \nu_{\rm cut,r}^{\rm ssc}$. This scenario corresponds to 5.81 percent, which is equivalent to five GRBs.

The second favored scenario is a five-way tie, accounting for $4$ GRBs (4.65\%) from the BPL column. We observed that the GRBs fitted with a BPL from the 2FLGC exhibited more coincidences with the reverse shock model than the PL. In contrast, the PL had the lowest number of agreements, with only two (2.33\%) for a thin shell that develops in a constant-density medium during the slow-cooling condition $\nu<\nu_{\rm m,r}^{\rm ssc}$. In the end, it is essential to mention that no instance of the thick shell was identified in either an ISM or a wind-like scenario for the condition $\nu<\nu_{\rm m,r}^{\rm ssc}$. Regardless of the evolution of the reverse shock, a constant-density medium is preferred over a stellar-wind environment.  We have provided the SSC flux as a function of the density parameter for each cooling condition as reverse shock progresses in both the thick and thin shell regimes. It is essential to recognize that when the criterion $\nu^{\rm ssc}_{\rm c, r}< \nu$ has been satisfied, it becomes impossible to differentiate between a stellar wind and a homogeneous environment. Density variations in a stellar-wind medium yield a more pronounced signal than that of the ISM. This signal is more pronounced in the thick-shell regime than in the thin-shell regime.

Taking into account the bursts in 2FLGC with high-energy photons above $\geq$ 1 GeV,  we have obtained all photons with a probability of $>90$\% to be associated with each burst and derived the maximum energy radiated by the synchrotron mechanism when the reverse shock evolves in the thick and thin-shell regime, respectively, for stellar wind and ISM environments. We have shown that the maximum synchrotron energy radiated from the reverse-shock scenario could explain some of the GRBs above 100 GeV, except GRB 091127, 110731A, 120624B, and 131108A, so that scattered photons by the SSC process must be required.

It is worth noting that when a superposition of emissions from the forward and reverse shock regions forms, the transient peaks in the LAT light curves may be partially eclipsed by the temporarily extended emission. In the previous case, the CRs of the SSC reverse-shock model are expected to satisfy only some segments of these. This would correspond to those LAT-detected bursts described by a BPL.

Bursts displaying temporal and spectral indices of $\gtrsim 1.5$ and $\approx 2$, respectively, are difficult to characterize using the CRs of the standard synchrotron afterglow model. We have derived the CRs of the synchrotron afterglow model during the radiative regime and shown that although they can describe the evolution of spectral and temporal indexes of most of the bursts, they can hardly explain those with energetic photons higher than 1 GeV. However, the SSC process from the reverse-shock region can naturally explain these bursts. 

Bursts exhibiting an unusual spectral index ($\beta_{\rm LAT}\approx 0$) are challenging to characterize using the synchrotron afterglow approach, as the \textit{Fermi}-LAT band must be situated below both spectral breaks ($\nu_{\rm LAT} < \nu^{\rm syn}_{\rm m,f}< \nu^{\rm syn}_{\rm c,f}$ or $\nu_{\rm LAT} < \nu^{\rm syn}_{\rm c,f}< \nu^{\rm syn}_{\rm m,f}$).   In contrast, we have demonstrated that these bursts can be characterized by the SSC reverse-shock model when this process is in this cooling regime $\nu_{\rm LAT} < \nu^{\rm ssc}_{\rm m,r}< \nu^{\rm ssc}_{\rm c,r}$ or $\nu_{\rm LAT} < \nu^{\rm ssc}_{\rm c,f}< \nu^{\rm ssc}_{\rm m,r}$.

In order to describe GRB 160625B and GRB 180720B, which exhibited early optical and gamma-ray photons in temporal correlations, we derived the CRs of the synchrotron afterglow model from the reverse shock evolving in stellar and ISM. We have considered that the reverse shock evolves in a thick- and thin-shell regime and the electron distribution with spectral PL indexes in ranges of $1<p<2$ and $2<p$. We showed that the same electron population may have generated the GeV and optical fluxes during the forward shock.

\section*{Acknowledgements}

We are deeply appreciative of the anonymous referee's comprehensive evaluation of the paper and their insightful recommendations, which significantly improved the manuscript's clarity. We greatly appreciate the useful discussions with Peter Veres, Rodolofo Barniol-Duran and Tanmoy Laskar. NF is grateful to UNAM-DGAPA-PAPIIT for the funding provided by grant IN112525. BBK is supported by IBS under the project code IBS-R018-D3. AG was supported by Universidad Nacional Autónoma de México Postdoctoral Program (POSDOC). M.G.D. acknowledges funding from the AAS Chretienne Fellowship and the MINIATURA2 grant. 

\section*{Data Availability}

There are no new data associated with this article.



\bibliographystyle{mnras}
\bibliography{RS_CRs} 



\clearpage
\newpage
\appendix

\section{Derivation of SSC emission}

Table \ref{tab:sample} displays the summary of notation list of all parameters, the physical meaning, units and normalization convention used.

\subsection{Thick-shell regime}
\subsubsection{Constant-density medium ($k=0$)}
\paragraph{Synchrotron spectral breaks for $t<t_{\rm x}$.}

The minimum and the cooling Lorentz factors for $1<p<2$ and $2 < p$ are 

{\small
\begin{eqnarray}
\gamma_{\rm m, r} &\simeq& \begin{cases}
9.9\times 10\,\left(\frac{1+z}{1.5}\right)^{-\frac{p}{8(p-1)}} g^{\frac{1}{p-1}}(1.9) \chi_{\rm e,-1}^{-1} \varepsilon^{\frac{1}{p-1}}_{\rm e_r,-1} \varepsilon^{\frac{2-p}{4(p-1)}}_{\rm B_r,-2}\,\Gamma_{2.5}^{\frac{1}{p-1}}\, n^{\frac{8-3p}{16(p-1)}}_{-1}\Delta^{\frac{p}{16(p-1)}}_{11.8}\,E^{-\frac{p}{16(p-1)}}_{53} t^{\frac{p}{8(p-1)}}_1\hspace{1.4cm}{\rm for} \hspace{0.1cm} { 1<p<2 }\cr
5.2\times 10^2\,\left(\frac{1+z}{1.5}\right)^{-\frac14} g(2.2) \chi_{\rm e,-1}^{-1} \varepsilon_{\rm e_r,-1}  \Gamma_{2.5} n^{\frac18}_{-1}\Delta^{\frac{1}{8}}_{11.8}\,E^{-\frac{1}{8}}_{53} t^{\frac{1}{4}}_1 \hspace{5.8cm}{\rm for} \hspace{0.1cm} {2<p }
\end{cases}\cr
\gamma_{\rm c, r}&\simeq& 1.2\times 10^4\,\left(\frac{1+z}{1.5}\right)^{\frac14}  \left(\frac{1+Y_{\rm r}}{2} \right)^{-1}\varepsilon^{-1}_{\rm B_r,-2} n^{-\frac58}_{-1}\Delta^{\frac{3}{8}}_{11.8} E^{-\frac{3}{8}}_{53}\,t^{-\frac14}_1\,,
\eary
}

respectively. The characteristic and cooling spectral breaks of synchrotron emission for $1<p<2$ and $2 < p$ are  

{\small
\begin{eqnarray}
h \nu^{\rm syn}_{\rm m, r} &\simeq& \begin{cases}
7.4\times 10^{-2}\,{\rm eV}\,\left(\frac{1+z}{1.5}\right)^{\frac{2-3p}{4(p-1)}} g^{\frac{2}{p-1}}(1.9) \chi_{\rm e,-1}^{-2} \varepsilon^{\frac{2}{p-1}}_{\rm e_r,-1} \varepsilon^{\frac{1}{2(p-1)}}_{\rm B_r,-2}\,\Gamma_{2.5}^{\frac{2}{p-1}}\, n^{\frac{6-p}{8(p-1)}}_{-1}\Delta^{\frac{2-p}{8(p-1)}}_{11.8}\,E^{-\frac{2-p}{8(p-1)}}_{53} t^{\frac{2-p}{4(p-1)}}_1\hspace{1cm}{\rm for} \hspace{0.1cm} { 1<p<2 }\cr
2.0\,{\rm eV}\,\left(\frac{1+z}{1.5}\right)^{-1} g^2(2.2) \chi_{\rm e,-1}^{-2} \varepsilon^2_{\rm e_r,-1} \varepsilon^{\frac12}_{\rm B_r,-2} \Gamma^2_{2.5} n^{\frac12}_{-1} \hspace{6.8cm}{\rm for} \hspace{0.1cm} {2<p }
\end{cases}\cr
h\nu^{\rm syn}_{\rm c, r}&\simeq& 1.1\times 10^3\,{\rm eV}\,\left(\frac{1+Y_{\rm r}}{2} \right)^{-2}\varepsilon^{-\frac32}_{\rm B_r,-2} n^{-1}_{-1}\Delta^{\frac{1}{2}}_{11.8} E^{-\frac{1}{2}}_{53}\,t^{-1}_1,\cr
F^{\rm syn}_{\rm max,r} &\simeq& 3.5\times 10^3\,{\rm mJy} \,\left(\frac{1+z}{1.5}\right)^{\frac12}\chi_{\rm e,-1} \varepsilon^{\frac12}_{\rm B_r,-2}\,n^{\frac{1}{4}}_{-1} \, d^{-2}_{\rm z,28.3}\,\Gamma^{-1}_{2.5}\,\Delta^{-\frac{5}{4}}_{11.8}\,E^{\frac{5}{4}}_{53}\, t^{\frac12}_1\,,\,\,\,\,\,\,\,\,
\eary
}
respectively.

\paragraph{Synchrotron spectral breaks for $t_{\rm x} < t$.} The minimum and the cooling Lorentz factors for $1<p<2$ and $2 < p$ are

{\small
\begin{eqnarray}
\gamma_{\rm m, r} &\simeq& \begin{cases}
9.1\times 10\,\left(\frac{1+z}{1.5}\right)^{\frac{68-21p}{96(p-1)}} g^{\frac{1}{p-1}}(1.9) \chi_{\rm e,-1}^{-1} \varepsilon^{\frac{1}{p-1}}_{\rm e_r,-1} \varepsilon^{\frac{2-p}{4(p-1)}}_{\rm B_r,-2}\,\Gamma_{2.5}^{\frac{1}{p-1}}\, n^{\frac{8-3p}{16(p-1)}}_{-1}\Delta^{\frac{68-3p}{96(p-1)}}_{11.8}\,E^{-\frac{p}{16(p-1)}}_{53} t^{-\frac{68-21p}{96(p-1)}}_2\hspace{1cm}{\rm for} \hspace{0.1cm} { 1<p<2 }\cr
5.0\times 10^2\,\left(\frac{1+z}{1.5}\right)^{\frac{13}{48}} g(2.2) \chi_{\rm e,-1}^{-1} \varepsilon_{\rm e_r,-1} \Gamma_{2} n^{\frac18}_{-1} \Delta^{\frac{31}{48}}_{11.8} E^{-\frac{1}{8}}_{53}t^{-\frac{13}{48}}_2\hspace{5.65cm}{\rm for} \hspace{0.1cm} {2<p }
\end{cases}\cr
\gamma_{\rm c, r}&\simeq&1.7\times 10^{4}\,\left(\frac{1+z}{1.5}\right)^{-\frac{25}{48}}  \left(\frac{1+Y_{\rm r}}{2} \right)^{-1}\varepsilon^{-1}_{\rm B_r,-2} n^{-\frac58}_{-1}\Delta^{-\frac{19}{48}}_{11.8} E^{-\frac{3}{8}}_{53}\,t^{\frac{25}{48}}_2,
\eary
}
respectively. The characteristic and cooling spectral breaks of the synchrotron model for $1<p<2$ and $2 < p$ are

{\small
\begin{eqnarray}
h \nu^{\rm syn}_{\rm m, r} &\simeq& \begin{cases}
1.2\times 10^{-2}\,{\rm eV}\,\left(\frac{1+z}{1.5}\right)^{\frac{69-22p}{48(p-1)}} g^{\frac{2}{p-1}}(1.9) \chi_{\rm e,-1}^{-2} \varepsilon^{\frac{2}{p-1}}_{\rm e_r,-1} \varepsilon^{\frac{1}{2(p-1)}}_{\rm B_r,-2}\,\Gamma_{2.5}^{\frac{2}{p-1}}\, n^{\frac{6-p}{8(p-1)}}_{-1}\Delta^{\frac{57+8p}{48(p-1)}}_{11.8}\,E^{-\frac{2-p}{8(p-1)}}_{53} t^{-\frac{21+26p}{48(p-1)}}_2\hspace{1cm}{\rm for} \hspace{0.1cm} { 1<p<2 }\cr
3.5\times 10^{-1}\,{\rm eV}\,\left(\frac{1+z}{1.5}\right)^{\frac{25}{48}} g^2(2.2) \chi_{\rm e,-1}^{-2} \varepsilon^2_{\rm e_r,-1} \varepsilon^{\frac12}_{\rm B_r,-2} \Gamma^2_{2} n^{\frac12}_{-1} \Delta^{\frac{73}{48}}_{11.8} t^{-\frac{73}{48}}_2\hspace{5.3cm}{\rm for} \hspace{0.1cm} {2<p }
\end{cases}\cr
h\nu^{\rm syn}_{\rm cut, r}&\simeq& 6.8\times 10\,{\rm eV}\,\left(\frac{1+z}{1.5}\right)^{\frac{25}{48}}  \left(\frac{1+Y_{\rm r}}{2} \right)^{-2}\varepsilon^{-\frac32}_{\rm B_r,-2} n^{-1}_{-1}\Delta^{\frac{49}{48}}_{11.8} E^{-\frac{1}{2}}_{53}\,t^{-\frac{73}{48}}_2,\cr
F^{\rm syn}_{\rm max,r} &\simeq& 2.0\times 10^3\,{\rm mJy} \,\left(\frac{1+z}{1.5}\right)^{\frac{95}{48}}\chi_{\rm e,-1} \varepsilon^{\frac12}_{\rm B_r,-2}\,n^{\frac{1}{4}}_{-1} \, d^{-2}_{\rm z,28.3}\,\Gamma^{-1}_{2.5}\,\Delta^{\frac{11}{48}}_{11.8}\,E^{\frac{5}{4}}_{53}\, t^{-\frac{47}{48}}_{2}\,,\,\,\,\,\,\,\,\,
\eary
}
respectively.
\subsubsection{Stellar-wind medium ($k=2$)}
\paragraph{Synchrotron spectral breaks for $t<t_{\rm x}$.}The minimum andling Lorentz factors for $1<p<2$ and $2 < p$ are

{\small
\begin{eqnarray}
\gamma_{\rm m, r} &\simeq& \begin{cases}
2.4\times 10^2\,\left(\frac{1+z}{1.5}\right)^{\frac{2-p}{2(p-1)}} g^{\frac{1}{p-1}}(1.9) \chi_{\rm e,-1}^{-1} \varepsilon^{\frac{1}{p-1}}_{\rm e_r,-1} \varepsilon^{\frac{2-p}{4(p-1)}}_{\rm B_r,-2}\,\Gamma_{2.5}^{\frac{1}{p-1}}\, A^{\frac{8-3p}{8(p-1)}}_{\rm W, -2}\Delta^{\frac{4-p}{8(p-1)}}_{11.8}\,E^{-\frac{4-p}{8(p-1)}}_{53} t^{\frac{2-p}{2(p-1)}}_1\hspace{1.4cm}{\rm for} \hspace{0.1cm} { 1<p<2 }\cr
9.9\times 10^2\,g(2.2) \chi_{\rm e,-1}^{-1} \varepsilon_{\rm e_r,-1} \Gamma_{2.5} A_{\rm W,-2}^{\frac14} \Delta^{\frac{1}{4}}_{11.8} E^{-\frac{1}{4}}_{53} \hspace{6.75cm}{\rm for} \hspace{0.1cm} {2<p }
\end{cases}\cr
\gamma_{\rm c, r}&\simeq& 9.2\times 10\,\left(\frac{1+z}{1.5}\right)^{-1}  \left(\frac{1+Y_{\rm r}}{2} \right)^{-1}\varepsilon^{-1}_{\rm B_r,-2} A_{\rm W,-2}^{-\frac54}\Delta^{-\frac{1}{4}}_{11.8} E^{\frac{1}{4}}_{53}\,t_1,
\eary
}

respectively. The characteristic and cooling spectral breaks of synchrotron emission for $1<p<2$ and $2 < p$ are  

{\small
\begin{eqnarray}
h \nu^{\rm syn}_{\rm m, r} &\simeq& \begin{cases}
3.0\,{\rm eV}\,\left(\frac{1+z}{1.5}\right)^{\frac{2-p}{p-1}} g^{\frac{2}{p-1}}(1.9) \chi_{\rm e,-1}^{-2} \varepsilon^{\frac{2}{p-1}}_{\rm e_r,-1} \varepsilon^{\frac{1}{2(p-1)}}_{\rm B_r,-2}\,\Gamma_{2.5}^{\frac{2}{p-1}}\, A_{\rm W,-2}^{\frac{6-p}{4(p-1)}}\Delta^{\frac{4-p}{4(p-1)}}_{11.8}\,E^{-\frac{4-p}{4(p-1)}}_{53} t^{-\frac{1}{p-1}}_1\hspace{1.7cm}{\rm for} \hspace{0.1cm} { 1<p<2 }\cr
5.3\times 10\,{\rm eV}\,g^2(2.2) \chi_{\rm e,-1}^{-2} \varepsilon^2_{\rm e_r,-1} \varepsilon^{\frac12}_{\rm B_r,-2} \Gamma^2_{2.5} A_{\rm W,-2} \Delta^{\frac{1}{2}}_{11.8} E^{-\frac{1}{2}}_{53}\,t^{-1}_1 \hspace{4.65cm}{\rm for} \hspace{0.1cm} {2<p }
\end{cases}\cr
h\nu^{\rm syn}_{\rm c, r}&\simeq& 4.6\times 10^{-1}\,{\rm eV}\,\left(\frac{1+z}{1.5}\right)^{-2}  \left(\frac{1+Y_{\rm r}}{2} \right)^{-2}\varepsilon^{-\frac32}_{\rm B_r,-2} A_{\rm W,-2}^{-2}\Delta^{-\frac{1}{2}}_{11.8} E^{\frac{1}{2}}_{53}\,t_1,\cr
F^{\rm syn}_{\rm max,r} &\simeq& 2.6\times 10^4\,{\rm mJy} \,\left(\frac{1+z}{1.5}\right) \chi_{\rm e,-1}\varepsilon^{\frac12}_{\rm B_r,-2}\, A_{\rm W,-2}^{\frac{1}{2}} \, d^{-2}_{\rm z,28.3}\,\Gamma^{-1}_{2.5}\,\Delta^{-1}_{11.8}\,E_{53},\,\,\,\,\,\,\,\,
\eary
}
respectively.
\paragraph{Synchrotron spectral breaks for $t_{\rm x} < t$.} The minimum and and Lorentz factors for $1<p<2$ and $2 < p$ are

{\small
\begin{eqnarray}
\gamma_{\rm m, r} &\simeq& \begin{cases}
1.3\times 10^2\,\left(\frac{1+z}{1.5}\right)^{\frac{16-5p}{16(p-1)}} g^{\frac{1}{p-1}}(1.9) \chi_{\rm e,-1}^{-1} \varepsilon^{\frac{1}{p-1}}_{\rm e_r,-1} \varepsilon^{\frac{2-p}{4(p-1)}}_{\rm B_r,-2}\,\Gamma_{2.5}^{\frac{1}{p-1}}\, A_{\rm W,-2}^{\frac{8-3p}{8(p-1)}}\Delta^{\frac{8+p}{16(p-1)}}_{11.8}\,E^{-\frac{4-p}{8(p-1)}}_{53} t^{-\frac{16-5p}{16(p-1)}}_2\hspace{1.4cm}{\rm for} \hspace{0.1cm} { 1<p<2 }\cr
6.4\times 10^2\,\left(\frac{1+z}{1.5}\right)^{\frac38} g(2.2) \chi_{\rm e,-1}^{-1} \varepsilon_{\rm e_r,-1} \Gamma_{2.5} A_{\rm W,-2}^{\frac14} \Delta^{\frac{5}{8}}_{11.8} E^{-\frac{1}{4}}_{53}\,t^{-\frac{3}{8}}_2 \hspace{5.65cm}{\rm for} \hspace{0.1cm} {2<p }
\end{cases}\cr
\gamma_{\rm c, r}&\simeq& 8.1\times 10^2\,\left(\frac{1+z}{1.5}\right)^{-\frac{7}{8}}  \left(\frac{1+Y_{\rm r}}{2} \right)^{-1}\varepsilon^{-1}_{\rm B_r,-2} A_{\rm W,-2}^{-\frac54}\Delta^{-\frac{1}{8}}_{11.8} E^{\frac{1}{4}}_{53}\,t^{\frac{7}{8}}_2,
\eary
}

respectively. The characteristic and cooling spectral breaks of synchrotron radiation for $1<p<2$ and $2 < p$ are  

{\small
\begin{eqnarray}
h \nu^{\rm syn}_{\rm m, r} &\simeq& \begin{cases}
8.0\times 10^{-2}\,{\rm eV}\,\left(\frac{1+z}{1.5}\right)^{\frac{15-4p}{8(p-1)}} g^{\frac{2}{p-1}}(1.9) \chi_{\rm e,-1}^{-2} \varepsilon^{\frac{2}{p-1}}_{\rm e_r,-1} \varepsilon^{\frac{1}{2(p-1)}}_{\rm B_r,-2}\,\Gamma_{2.5}^{\frac{2}{p-1}}\, A_{\rm W,-2}^{\frac{6-p}{4(p-1)}}\Delta^{\frac{7+2p}{8(p-1)}}_{11.8}\,E^{-\frac{4-p}{4(p-1)}}_{53} t^{-\frac{7+4p}{8(p-1)}}_2\hspace{1.1cm}{\rm for} \hspace{0.1cm} { 1<p<2 }\cr

1.9\,{\rm eV}\,\left(\frac{1+z}{1.5}\right)^{\frac78} g^2(2.2) \chi_{\rm e,-1}^{-2} \varepsilon^2_{\rm e_r,-1} \varepsilon^{\frac12}_{\rm B_r,-2} \Gamma^2_{2.5} A_{\rm W,-2} \Delta^{\frac{11}{8}}_{11.8} E^{-\frac{1}{2}}_{53}\,t^{-\frac{15}{8}}_2 \hspace{4.95cm}{\rm for} \hspace{0.1cm} {2<p }
\end{cases}\cr
h\nu^{\rm syn}_{\rm cut, r}&\simeq& 3.6\times 10^{-1}\,{\rm eV}\,\left(\frac{1+z}{1.5}\right)^{\frac{7}{8}}  \left(\frac{1+Y_{\rm r}}{2} \right)^{-2}\varepsilon^{-\frac32}_{\rm B_r,-2} A_{\rm W,-2}^{-2}\Delta^{\frac{19}{8}}_{11.8} E^{\frac{1}{2}}_{53}\,t^{-\frac{15}{8}}_2,\cr
F^{\rm syn}_{\rm max,r} &\simeq& 7.1\times 10^{3}\,{\rm mJy} \,\left(\frac{1+z}{1.5}\right)^{\frac{17}{8}} \chi_{\rm e,-1}\varepsilon^{\frac12}_{\rm B_r,-2}\, A_{\rm W,-2}^{\frac{1}{2}} \, d^{-2}_{\rm z,28.3}\,\Gamma^{-1}_{2.5}\,\Delta^{\frac18}_{11.8}\, E_{53}\,t^{-\frac{9}{8}}_2   \,,\,\,\,\,\,\,\,\,
\eary
}
respectively.

\subsection{Thin-shell regime}
\subsubsection{Constant-density medium ($k=0$)}
\paragraph{Synchrotron spectral breaks for $t<t_{\rm x}$.} The minimum and the cooling Lorentz factors for $1<p<2$ and $2 < p$ are

{\small
\begin{eqnarray}
\gamma_{\rm m, r} &\simeq& \begin{cases}
1.4\times 10^{-1}\,\left(\frac{1+z}{1.5}\right)^{-\frac{3}{p-1}} g^{\frac{1}{p-1}}(1.9) \chi_{\rm e,-1}^{-1} \varepsilon^{\frac{1}{p-1}}_{\rm e_r,-1} \varepsilon^{\frac{2-p}{4(p-1)}}_{\rm B_r,-2}\,\Gamma_{2}^{\frac{18-p}{2(p-1)}}\, n^{\frac{6-p}{4(p-1)}}_{-1}\,E^{-\frac{1}{p-1}}_{53} t^{\frac{3}{p-1}}_2\hspace{1.4cm}{\rm for} \hspace{0.1cm} { 1<p<2 }\cr
5.1\times 10^{-1}\,\left(\frac{1+z}{1.5}\right)^{-3} g(2.2) \chi_{\rm e,-1}^{-1} \varepsilon_{\rm e_r,-1} \Gamma^{8}_{2} n_{-1} E^{-1}_{53}\,t^{3}_2 \hspace{5.1cm}{\rm for} \hspace{0.1cm} {2<p }
\end{cases}\cr
\gamma_{\rm c, r}&\simeq& 4.8\times 10^4\,\left(\frac{1+z}{1.5}\right)  \left(\frac{1+Y_{\rm r}}{2} \right)^{-1}\varepsilon^{-1}_{\rm B_r,-2} n^{-1}_{-1}\, \Gamma^{-3}_{2} \,t^{-1}_2,
\eary
}

respectively. The characteristic and cooling spectral breaks of synchrotron emission for $1<p<2$ and $2 < p$ are

{\small
\begin{eqnarray}
h \nu^{\rm syn}_{\rm m, r} &\simeq& \begin{cases}
2.0\times 10^{-8}\,{\rm eV}\,\left(\frac{1+z}{1.5}\right)^{-\frac{5+p}{p-1}} g^{\frac{2}{p-1}}(1.9) \chi_{\rm e,-1}^{-2} \varepsilon^{\frac{2}{p-1}}_{\rm e_r,-1} \varepsilon^{\frac{1}{2(p-1)}}_{\rm B_r,-2}\,\Gamma_{2}^{\frac{16+p}{p-1}}\, n^{\frac{5}{2(p-1)}}_{-1}\,E^{-\frac{2}{p-1}}_{53} t^{\frac{6}{p-1}}_2\hspace{1.4cm}{\rm for} \hspace{0.1cm} { 1<p<2 }\cr
2.5\times 10^{-7}\,{\rm eV}\,\left(\frac{1+z}{1.5}\right)^{-7} g^2(2.2) \chi_{\rm e,-1}^{-2} \varepsilon^2_{\rm e_r,-1} \varepsilon^{\frac12}_{\rm B_r,-2} \Gamma^{18}_{2} n_{-1}^{\frac52} E^{-2}_{53}\,t^{6}_2 \hspace{3.85cm}{\rm for} \hspace{0.1cm} {2<p }
\end{cases}\cr
h\nu^{\rm syn}_{\rm c, r}&\simeq& 2.2\times 10^3\,{\rm eV}\,\left(\frac{1+z}{1.5}\right)  \left(\frac{1+Y_{\rm r}}{2} \right)^{-2}\varepsilon^{-\frac32}_{\rm B_r,-2} n_{-1}^{-\frac32}\, \Gamma^{-4}_{2} \,t^{-2}_2,\cr
F^{\rm syn}_{\rm max,r} &\simeq& 1.3\times 10^{2}\,{\rm mJy} \,\left(\frac{1+z}{1.5}\right)^{-\frac12} \chi_{\rm e,-1}\varepsilon^{\frac12}_{\rm B_r,-2}\,n_{-1} \, d^{-2}_{\rm z,28.3}\,\Gamma^{5}_{2}\, E^{\frac12}_{53}\,t^{\frac32}_2\,,\,\,\,\,\,\,\,\,
\eary
}

respectively.

\paragraph{Synchrotron spectral breaks for $t_{\rm x}<t$.} The minimum and the cooling Lorentz factors for $1<p<2$ and $2 < p$ are

{\small
\begin{eqnarray}
\gamma_{\rm m, r} &\simeq& \begin{cases}
1.3\times 10^2\,\left(\frac{1+z}{1.5}\right)^{\frac{24-7p}{35(p-1)}} g^{\frac{1}{p-1}}(1.9) \chi_{\rm e,-1}^{-1} \varepsilon^{\frac{1}{p-1}}_{\rm e_r,-1} \varepsilon^{\frac{2-p}{4(p-1)}}_{\rm B_r,-2}\,\Gamma_{2}^{-\frac{174-7p}{210(p-1)}}\, n^{\frac{114-77p}{420(p-1)}}_{-1}\,E^{\frac{24-7p}{105(p-1)}}_{53} t^{-\frac{24-7p}{35(p-1)}}_3\hspace{1.2cm}{\rm for} \hspace{0.1cm} { 1<p<2 }\cr
2.3\times 10^2\,\left(\frac{1+z}{1.5}\right)^{\frac27} g(2.2) \chi_{\rm e,-1}^{-1} \varepsilon_{\rm e_r,-1} \Gamma^{-\frac{16}{21}}_{2} n_{-1}^{-\frac{2}{21}} E^{\frac{2}{21}}_{53}\,t^{-\frac{2}{7}}_3 \hspace{5.8cm}{\rm for} \hspace{0.1cm} {2<p }
\end{cases}\cr
\gamma_{\rm c, r}&\simeq& 4.0\times 10^3\,\left(\frac{1+z}{1.5}\right)^{-\frac{19}{35}}  \left(\frac{1+Y_{\rm r}}{2} \right)^{-1}\varepsilon^{-1}_{\rm B_r,-2} n^{-\frac{17}{35}}_{-1}\, \Gamma^{\frac{39}{35}}_{2}E^{-\frac{18}{35}}_{53} \,t^{\frac{19}{35}}_3,
\eary
}
respectively. The characteristic and cooling spectral breaks of synchrotron model for $1<p<2$ and $2 < p$ are

{\small
\begin{eqnarray}
h \nu^{\rm syn}_{\rm m, r} &\simeq& \begin{cases}
1.2\times 10^{-2}\,{\rm eV}\,\left(\frac{1+z}{1.5}\right)^{\frac{49-15p}{35(p-1)}} g^{\frac{2}{p-1}}(1.9) \chi_{\rm e,-1}^{-2} \varepsilon^{\frac{2}{p-1}}_{\rm e_r,-1} \varepsilon^{\frac{1}{2(p-1)}}_{\rm B_r,-2}\,\Gamma_{2}^{-\frac{112+55p}{105(p-1)}}\, n^{\frac{77-40p}{210(p-1)}}_{-1}\,E^{\frac{2(7+10p)}{105(p-1)}}_{53} t^{-\frac{2(7+10p)}{35(p-1)}}_3\hspace{1.4cm}{\rm for} \hspace{0.1cm} { 1<p<2 }\cr
3.9\times 10^{-2}\,{\rm eV}\,\left(\frac{1+z}{1.5}\right)^{\frac{19}{35}} g^2(2.2) \chi_{\rm e,-1}^{-2} \varepsilon^2_{\rm e_r,-1} \varepsilon^{\frac12}_{\rm B_r,-2} \Gamma^{-\frac{74}{35}}_{2} n_{-1}^{-\frac{1}{70}} E^{\frac{18}{35}}_{53}\,t^{-\frac{54}{35}}_3 \hspace{5.1cm}{\rm for} \hspace{0.1cm} {2<p }
\end{cases}\cr
h\nu^{\rm syn}_{\rm cut, r}&\simeq& 6.2\,{\rm eV}\,\left(\frac{1+z}{1.5}\right)^{\frac{19}{35}}  \left(\frac{1+Y_{\rm r}}{2} \right)^{-2}\varepsilon^{-\frac32}_{\rm B_r,-2} n^{-\frac{283}{210}}_{-1}\, \Gamma^{-\frac{292}{105}}_{2}E^{-\frac{16}{105}}_{53} \,t^{-\frac{54}{35}}_3,\cr
F^{\rm syn}_{\rm max,r} &\simeq& 2.3\times 10^{3}\,{\rm mJy} \,\left(\frac{1+z}{1.5}\right)^{\frac{69}{35}} \chi_{\rm e,-1}\varepsilon^{\frac12}_{\rm B_r,-2}\,n_{-1}^{\frac{37}{210}} \, d^{-2}_{\rm z,28.3}\,\Gamma^{-\frac{167}{105}}_{2}\, E^{\frac{139}{105}}_{53}\,t^{-\frac{34}{35}}_3\,,\,\,\,\,\,\,\,\,
\eary
}

respectively.

\subsubsection{Stellar-wind medium ($k=2$)}
\paragraph{Synchrotron spectral breaks for $t<t_{\rm x}$.} The minimum and the cooling Lorentz factors for $1<p<2$ and $2 < p$ are

{\small
\begin{eqnarray}
\gamma_{\rm m, r} &\simeq& \begin{cases}
1.4\,\left(\frac{1+z}{1.5}\right)^{-\frac{p}{2(p-1)}} g^{\frac{1}{p-1}}(1.9) \chi_{\rm e,-1}^{-1} \varepsilon^{\frac{1}{p-1}}_{\rm e_r,-1} \varepsilon^{\frac{2-p}{4(p-1)}}_{\rm B_r,-2}\,\Gamma_{2}^{\frac{6+p}{2(p-1)}}\, A^{\frac{6-p}{4(p-1)}}_{\rm W,-2}\,E^{-\frac{1}{p-1}}_{53} t^{\frac{p}{2(p-1)}}_2\hspace{1.4cm}{\rm for} \hspace{0.1cm} { 1<p<2 }\cr
3.2\,\left(\frac{1+z}{1.5}\right)^{-1}g(2.2) \chi_{\rm e,-1}^{-1} \varepsilon_{\rm e_r,-1}  \Gamma^4_{2} A_{\rm W,-2} E^{-1}_{53}\,t_2 \hspace{5.3cm}{\rm for} \hspace{0.1cm} {2<p }
\end{cases}\cr
\gamma_{\rm c, r}&\simeq& 8.5\,\left(\frac{1+z}{1.5}\right)^{-1}  \left(\frac{1+Y_{\rm r}}{2} \right)^{-1}\varepsilon^{-1}_{\rm B_r,-2} A_{\rm W,-2}^{-1}  \Gamma_{2} \,t_2,
\eary
}

respectively. The characteristic and cooling spectral breaks of synchrotron emission for $1<p<2$ and $2 < p$ are  

{\small
\begin{eqnarray}
h \nu^{\rm syn}_{\rm m, r} &\simeq& \begin{cases}
9.8\times 10^{-5}\,{\rm eV}\,\left(\frac{1+z}{1.5}\right)^{-\frac{p}{p-1}} g^{\frac{2}{p-1}}(1.9) \chi_{\rm e,-1}^{-2} \varepsilon^{\frac{2}{p-1}}_{\rm e_r,-1} \varepsilon^{\frac{1}{2(p-1)}}_{\rm B_r,-2}\,\Gamma_{2}^{\frac{6+p}{p-1}}\, A^{\frac{5}{2(p-1)}}_{\rm W,-2}\,E^{-\frac{2}{p-1}}_{53} t^{\frac{1}{p-1}}_2\hspace{1.4cm}{\rm for} \hspace{0.1cm} { 1<p<2 }\cr
5.1\times 10^{-4}\,{\rm eV}\,\left(\frac{1+z}{1.5}\right)^{-2}g^2(2.2) \chi_{\rm e,-1}^{-2} \varepsilon^2_{\rm e_r,-1} \varepsilon^{\frac12}_{\rm B_r,-2} \Gamma^8_{2} A_{\rm W,-2}^{\frac52} E^{-2}_{53}\,t_2 \hspace{3.85cm}{\rm for} \hspace{0.1cm} {2<p }
\end{cases}\cr
h\nu^{\rm syn}_{\rm c, r}&\simeq& 3.7\times 10^{-3}\,{\rm eV}\,\left(\frac{1+z}{1.5}\right)^{-2}  \left(\frac{1+Y_{\rm r}}{2} \right)^{-2}\varepsilon^{-\frac32}_{\rm B_r,-2} A_{\rm W,-2}^{-\frac32}  \Gamma^2_{2} \,t_2,\cr
F^{\rm syn}_{\rm max,r} &\simeq& 5.4\times 10^4\,{\rm mJy} \,\left(\frac{1+z}{1.5}\right)^{\frac32} \chi_{\rm e,-1}\varepsilon^{\frac12}_{\rm B_r,-2}\, A_{\rm W,-2} \, d^{-2}_{\rm z,28.3}\,\Gamma_{2}\, E^{\frac12}_{53}   t^{-\frac12}_{2}\,,\,\,\,\,\,\,\,\,
\eary
}
respectively.
\paragraph{Synchrotron spectral breaks for $t_{\rm x} < t$.} The minimum and the cooling Lorentz factors for $1<p<2$ and $2 < p$ are

{\small
\begin{eqnarray}
\gamma_{\rm m, r} &\simeq& \begin{cases}
0.6\,{\rm GeV}\,\left(\frac{1+z}{1.5}\right)^{\frac{22-7p}{21(p-1)}} g^{\frac{1}{p-1}}(1.9) \chi_{\rm e,-1}^{-1} \varepsilon^{\frac{1}{p-1}}_{\rm e_r,-1} \varepsilon^{\frac{2-p}{4(p-1)}}_{\rm B_r,-2}\,\Gamma_{2}^{-\frac{50+7p}{42(p-1)}}\, A_{\rm W,-2}^{\frac{38-35p}{84(p-1)}}\,E^{\frac{2+7p}{42(p-1)}}_{53} t^{-\frac{22-7p}{21(p-1)}}_3\hspace{1cm}{\rm for} \hspace{0.1cm} { 1<p<2 }\cr
0.6\,{\rm GeV}\,\left(\frac{1+z}{1.5}\right)^{\frac{8}{21}} g(2.2) \chi_{\rm e,-1}^{-1} \varepsilon_{\rm e_r,-1} \Gamma^{-\frac{32}{21}}_{2} A^{-\frac{8}{21}}_{\rm W} E^{\frac{8}{21}}_{53}\,t^{-\frac{8}{21}}_3 \hspace{5cm}{\rm for} \hspace{0.1cm} {2<p }
\end{cases}\cr
\gamma_{\rm c, r}&\simeq& 0.7\,{\rm eV}\,\left(\frac{1+z}{1.5}\right)^{-\frac{6}{7}}  \left(\frac{1+Y_{\rm r}}{2} \right)^{-1}\varepsilon^{-1}_{\rm B_r,-2} A_{\rm W,-2}^{-\frac{8}{7}}\Gamma^{\frac{3}{7}}_{2} E^{\frac{1}{7}}_{53}\,t^{\frac{6}{7}}_3,
\eary
}

respectively. The characteristic and cooling spectral breaks  of synchrotron radiation for $1<p<2$ and $2 < p$ are

{\small
\begin{eqnarray}
h \nu^{\rm syn}_{\rm m, r} &\simeq& \begin{cases}
0.6\,{\rm GeV}\,\left(\frac{1+z}{1.5}\right)^{\frac{2(7-2p)}{7(p-1)}} g^{\frac{2}{p-1}}(1.9) \chi_{\rm e,-1}^{-2} \varepsilon^{\frac{2}{p-1}}_{\rm e_r,-1} \varepsilon^{\frac{1}{2(p-1)}}_{\rm B_r,-2}\,\Gamma_{2}^{-\frac{14+5p}{7(p-1)}}\, A_{\rm W,-2}^{\frac{7-6p}{14(p-1)}}\,E^{\frac{3p}{7(p-1)}}_{53} t^{-\frac{7+3p}{7(p-1)}}_3\hspace{1.4cm}{\rm for} \hspace{0.1cm} { 1<p<2 }\cr
0.6\,{\rm GeV}\,\left(\frac{1+z}{1.5}\right)^{\frac67} g^2(2.2) \chi_{\rm e,-1}^{-2} \varepsilon^2_{\rm e_r,-1} \varepsilon^{\frac12}_{\rm B_r,-2} \Gamma^{-\frac{24}{7}}_{2} A^{-\frac{5}{14}}_{\rm W,-2} E^{\frac{6}{7}}_{53}\,t^{-\frac{13}{7}}_3 \hspace{4.65cm}{\rm for} \hspace{0.1cm} {2<p }
\end{cases}\cr
h\nu^{\rm syn}_{\rm cut, r}&\simeq& 0.7\,{\rm eV}\,\left(\frac{1+z}{1.5}\right)^{\frac{6}{7}}  \left(\frac{1+Y_{\rm r}}{2} \right)^{-2}\varepsilon^{-\frac32}_{\rm B_r,-2} A_{\rm W,-2}^{-\frac{61}{14}}\Gamma^{-\frac{66}{7}}_{2} E^{\frac{20}{7}}_{53}\,t^{-\frac{13}{7}}_3,\cr
F^{\rm syn}_{\rm max,r} &\simeq& 6.8\,{\rm mJy} \,\left(\frac{1+z}{1.5}\right)^{\frac{44}{21}} \chi_{\rm e,-1}\varepsilon^{\frac12}_{\rm B_r,-2}\, A_{\rm W,-2}^{\frac{17}{42}} \, d^{-2}_{\rm z,28.3}\,\Gamma^{-\frac{29}{21}}_{2}\, E^{\frac{23}{21}}_{53}\,t^{-\frac{23}{21}}_3   \,,\,\,\,\,\,\,\,\,
\eary
}

respectively.

\newpage

\section{Klein-Nishina Light Curves}\label{app:KN_LC}
Based on \citeauthor{2009ApJ...703..675N}'s definition of a new spectral break ($h\nu^{\rm ssc}_{\rm KN, 0, r} \simeq\frac{\gamma_3\gamma_{0,r}}{(1+z)}\,m_e c^2$), we identify additional PLs in the SSC light curves caused by KN effects when jet decelerates in the constant-density medium and stellar wind, and the reverse-shock evolves in thick- and thin-shell regime. The definition of $\gamma_{\rm 0,r}$ in the fast- and slow-cooling regimes is explicitly given through Eqs. 20 and 49 of \cite{2009ApJ...703..675N}, respectively, and $\gamma_3$ is defined in \cite{2024MNRAS.527.1674F} for each case.

\subsection{Thick-shell regime}

\subsubsection{Constant-density medium ($k=0$)}

Additional PLs of the SSC light curves for the fast (weak) and slow-cooling regimes are

{\small
\begin{eqnarray}
F^{\rm ssc}_{\rm \nu, r} \propto \begin{cases} 
 t^{\frac{p}{4}}\, \nu_{\rm }^{-(p-1)},\hspace{1.76cm} \nu^{\rm ssc}_{\rm KN, m, r}<\nu<\nu^{\rm ssc}_{\rm KN, 0, r}, \cr
 t^{\frac{p}{4}}\, \nu_{\rm }^{-(p-\frac12)},\hspace{1.69cm} \nu^{\rm ssc}_{\rm KN, 0, r}<\nu<\nu^{\rm ssc}_{\rm KN, c, r}, \cr
 t^{-\frac{5-3p}{12}}\,\nu_{\rm }^{-(p+\frac13)},\,\,\,\,\, \hspace{0.9cm} \nu^{\rm ssc}_{\rm KN, c, r}<\nu\,, \cr
\end{cases}
\end{eqnarray}
}

and

{\small
\begin{eqnarray}
F^{\rm ssc}_{\rm \nu, r} \propto \begin{cases} 
t^{\frac{1}{2}}\, \nu_{\rm }^{-(p-1)},\hspace{1.5cm} \nu^{\rm ssc}_{\rm KN, c, r}<\nu<\nu^{\rm ssc}_{\rm KN, 0, r}, \cr
t^{\frac{1}{4}}\, \nu_{\rm }^{-\frac{p+1}{2}},\hspace{1.69cm} \nu^{\rm ssc}_{\rm KN, 0, r}<\nu<\nu^{\rm ssc}_{\rm m, r}, \cr
t^{\frac{1}{4}}\,\nu_{\rm }^{-(p+\frac{1}{3})},\,\,\,\,\,\hspace{1.2cm}\nu^{\rm ssc}_{\rm m, r}<\nu\,, \cr
\end{cases}
\end{eqnarray}
}
respectively.

\subsubsection{Stellar-wind medium ($k=2$)}

For the fast (weak) and slow cooling, the new PLs of the SSC emission are

{\small
\begin{eqnarray}
F^{\rm ssc}_{\rm \nu, r} \propto \begin{cases} 
 t^{\frac{2 - p}{2}}\, \nu_{\rm }^{-(p-1)},\hspace{1.76cm} \nu^{\rm ssc}_{\rm KN, m, r}<\nu<\nu^{\rm ssc}_{\rm KN, 0, r}, \cr
 t^{\frac{2 - p}{2}}\, \nu_{\rm }^{-(p-\frac12)},\hspace{1.69cm} \nu^{\rm ssc}_{\rm KN, 0, r}<\nu<\nu^{\rm ssc}_{\rm KN, c, r}, \cr
 t^{-\frac{3p-11}{6}}\,\nu_{\rm }^{-(p+\frac13)},\,\,\,\,\, \hspace{1.1cm} \nu^{\rm ssc}_{\rm KN, c, r}<\nu\,, \cr
\end{cases}
\end{eqnarray}
}

and

{\small
\begin{eqnarray}
F^{\rm ssc}_{\rm \nu, r} \propto \begin{cases} 
t^{0}\, \nu_{\rm }^{-(p-1)},\hspace{1.5cm} \nu^{\rm ssc}_{\rm KN, c, r}<\nu<\nu^{\rm ssc}_{\rm KN, 0, r}, \cr
t^{\frac{1}{2}}\, \nu_{\rm }^{-\frac{p+1}{2}},\hspace{1.69cm} \nu^{\rm ssc}_{\rm KN, 0, r}<\nu<\nu^{\rm ssc}_{\rm m, r}, \cr
t^{\frac{1}{2}}\,\nu_{\rm }^{-(p+\frac{1}{3})},\,\,\,\,\,\hspace{1.2cm}\nu^{\rm ssc}_{\rm m, r}<\nu\,, \cr
\end{cases}
\end{eqnarray}
}

respectively.

\subsection{Thin-shell regime}

\subsubsection{Constant-density medium ($k=0$)}

For the fast (weak) and slow-cooling domain, additional PLs of the SSC light curves are 

{\small
\begin{eqnarray}
F^{\rm ssc}_{\rm \nu, r} \propto \begin{cases} 
 t^{-\frac{17-15p}{2}}\, \nu_{\rm }^{-(p-1)},\hspace{1.5cm} \nu^{\rm ssc}_{\rm KN, m, r}<\nu<\nu^{\rm ssc}_{\rm KN, 0, r}, \cr
 t^{-\frac{14-15p}{2}}\, \nu_{\rm }^{-(p-\frac12)},\hspace{1.4cm} \nu^{\rm ssc}_{\rm KN, 0, r}<\nu<\nu^{\rm ssc}_{\rm KN, c, r}, \cr
 t^{-\frac{47-45p}{6}}\,\nu_{\rm }^{-(p+\frac13)},\,\,\,\,\, \hspace{1.22cm} \nu^{\rm ssc}_{\rm KN, c, r}<\nu\,, \cr
\end{cases}
\end{eqnarray}
}

and

{\small
\begin{eqnarray}
F^{\rm ssc}_{\rm \nu, r} \propto \begin{cases} 
t^{-\frac{9-11p}{2}}\, \nu_{\rm }^{-(p-1)},\hspace{1.5cm} \nu^{\rm ssc}_{\rm KN, c, r}<\nu<\nu^{\rm ssc}_{\rm KN, 0, r}, \cr
t^{2(2p-1)}\, \nu_{\rm }^{-\frac{p+1}{2}},\hspace{1.7cm} \nu^{\rm ssc}_{\rm KN, 0, r}<\nu<\nu^{\rm ssc}_{\rm m, r}, \cr
t^{-\frac{5-11p}{2}}\,\nu_{\rm }^{-(p+\frac{1}{3})},\,\,\,\,\,\hspace{1.25cm}\nu^{\rm ssc}_{\rm m, r}<\nu\,, \cr
\end{cases}
\end{eqnarray}
}

respectively.

\subsubsection{Stellar-wind medium ($k=2$)}

Additional PLs of the SSC light curves for the fast (weak) and slow-cooling regimes are

{\small
\begin{eqnarray}
F^{\rm ssc}_{\rm \nu, r} \propto \begin{cases} 
 t^{-\frac{5+6p}{2}}\, \nu_{\rm }^{-(p-1)},\hspace{1.5cm} \nu^{\rm ssc}_{\rm KN, m, r}<\nu<\nu^{\rm ssc}_{\rm KN, 0, r}, \cr
 t^{2(p-1)}\, \nu_{\rm }^{-(p-\frac12)},\hspace{1.4cm} \nu^{\rm ssc}_{\rm KN, 0, r}<\nu<\nu^{\rm ssc}_{\rm KN, c, r}, \cr
 t^{-\frac{7-12p}{6}}\,\nu_{\rm }^{-(p+\frac13)},\,\,\,\,\, \hspace{1.22cm} \nu^{\rm ssc}_{\rm KN, c, r}<\nu\,, \cr
\end{cases}
\end{eqnarray}
}

and

{\small
\begin{eqnarray}
F^{\rm ssc}_{\rm \nu, r} \propto \begin{cases} 
t^{-\frac{5-4p}{2}}\, \nu_{\rm }^{-(p-1)},\hspace{1.5cm} \nu^{\rm ssc}_{\rm KN, c, r}<\nu<\nu^{\rm ssc}_{\rm KN, 0, r}, \cr
t^{-\frac{2-3p}{2}}\, \nu_{\rm }^{-\frac{p+1}{2}},\hspace{1.7cm} \nu^{\rm ssc}_{\rm KN, 0, r}<\nu<\nu^{\rm ssc}_{\rm m, r}, \cr
t^{-\frac{7-12p}{6}}\,\nu_{\rm }^{-(p+\frac{1}{3})},\,\,\,\,\,\hspace{1.15cm}\nu^{\rm ssc}_{\rm m, r}<\nu\,, \cr
\end{cases}
\end{eqnarray}
}

respectively.

\newpage
\begin{table}
\centering \renewcommand{\arraystretch}{1.6}\addtolength{\tabcolsep}{2pt}
\caption{Evolution of the SSC light curves ($F^{\rm ssc}_{\rm \nu, r}\propto t^{-\alpha}\nu^{-\beta}$) from the reverse shock in the thick- and thin-shell regime before shock crossing time.}
\label{tableReverseBefore}
\begin{tabular}{cccccc}
\hline
\hline
     &               & ISM     & ISM      & wind    & wind      \\

     &               & $1<p<2$     & $2<p$       & $1<p<2$     & $2<p$       \\
\hline

 & $\beta$    & $\alpha$    & $\alpha$    & $\alpha$    & $\alpha$    \\
\hline
Thick shell &    &    &    &    &    \\
\hline
$ \nu < \nu^{\rm ssc}_{\rm m,r}$    &    $-\frac13$         &     $\frac{7-6p}{6(p-1)}$        &      $-\frac{5}{6}$       &       $-\frac{2(3-2p)}{3(p-1)}$      &      $\frac{2}{3}$       \\
$ \nu^{\rm ssc}_{\rm m,r} < \nu < \nu^{\rm ssc}_{\rm c,r}$    &    $\frac{p-1}{2}$           &       $-\frac{5}{4}$      &       $-\frac{3+p}{4}$      &      $\frac{5-p}{2}$       &      $\frac{p+1}{2}$       \\

$ \nu^{\rm ssc}_{\rm c,r} < \nu $   &    $\frac{p}{2}$           &       $-\frac{1}{2}$      &       $-\frac{p}{4}$      &      $\frac{2-p}{2}$       &      $\frac{p-2}{2}$       \\


\hline
Thin shell &  &     &     &    &    \\
\hline
$ \nu < \nu^{\rm ssc}_{\rm m,r}$    &    $-\frac13$         &     $\frac{13-5p}{2(p-1)}$        &      $\frac{3}{2}$       &       $-\frac{7-11p}{6(p-1)}$      &      $\frac{5}{2}$       \\
$ \nu^{\rm ssc}_{\rm c,r} < \nu < \nu^{\rm ssc}_{\rm c,r}$    &    $\frac{p-1}{2}$          &       $-\frac{17}{2}$      &       $\frac{7-12p}{2}$      &      $\frac{2-p}{2}$       &      $\frac{3(2-p)}{2}$       \\
$ \nu^{\rm ssc}_{\rm c,r} < \nu $    &    $\frac{p}{2}$          &       $-\frac{13}{2}$      &       $\frac{11-12p}{2}$      &      $-\frac{p+1}{2}$       &      $\frac{3(1-p)}{2}$       \\


\hline
\end{tabular}
\end{table}

\begin{table}
\centering \renewcommand{\arraystretch}{1.85}\addtolength{\tabcolsep}{1.5pt}
\caption{Closure relations of SSC ($F^{\rm ssc}_{\rm \nu,r}\propto t^{-\alpha(\beta)}\nu^{-\beta}$)  from the reverse shock afterglow model evolving in a thick- and thin-shell case before shock crossing time.}
\label{TableCRs_before}
\begin{tabular}{c c c  c c c}
 \hline \hline
&\hspace{0.5cm}     &\hspace{0.5cm}   ISM &\hspace{0.2cm}   ISM  & \hspace{0.5cm}   wind   & \hspace{0.5cm}  wind \\ 
&\hspace{0.5cm}     &\hspace{0.5cm} $1 < p <2$   &\hspace{0.5cm}   $2 < p$  & \hspace{0.2cm}    $1 < p <2$   & \hspace{0.5cm}  $2 < p$ \\
\hline
 &\hspace{0.5cm} $\beta$    &\hspace{0.5cm} $\alpha(\beta)$   &\hspace{0.5cm}   $\alpha(\beta)$  & \hspace{0.2cm}    $\alpha(\beta)$   & \hspace{0.5cm}  $\alpha(\beta)$    \\ \hline

Thick shell &\hspace{0.5cm}     &\hspace{0.2cm}    &\hspace{0.5cm}     & \hspace{0.5cm}       & \hspace{0.5cm}  \\ \hline 	
 $ \nu < \nu^{\rm ssc}_{\rm m, r}$   	        & \hspace{0.5cm}  $-\frac{1}{3}$           &\hspace{0.5cm}  $-\frac{(7-6p)\beta}{2(p-1)}$        &      $\frac{5\beta}{2}$       &       $\frac{2(3-2p)\beta}{p-1}$      &      $-2\beta$	\\	
 $ \nu^{\rm ssc}_{\rm m, r} < \nu < \nu^{\rm ssc}_{\rm c, r}$   	                & \hspace{0.5cm}  $\frac{p-1}{2}$  &\hspace{0.5cm}  $-$	                    &\hspace{0.2cm}  $-\frac{2+\beta}{2}$ &\hspace{0.5cm}  $2-\beta$ &\hspace{0.5cm}  $\beta+1$ \\ 	
 $\nu^{\rm ssc}_{\rm c,r} < \nu$   	                                 & \hspace{0.5cm}  $\frac{p}{2}$     &\hspace{0.5cm}  $-$	                    &\hspace{0.2cm}  $-\frac{\beta}{2}$ &\hspace{0.5cm}  $1-\beta$ &\hspace{0.5cm}  $\beta-1$\\ 
\hline	
Thin shell &\hspace{0.5cm}     &\hspace{0.5cm}    &\hspace{0.5cm}     & \hspace{0.5cm}       & \hspace{0.5cm}  \\ \hline 	
 $ \nu < \nu^{\rm ssc}_{\rm m, r}$   	        & \hspace{0.5cm}  $-\frac{1}{3}$           &\hspace{0.5cm}  $-\frac{3(13-5p)\beta}{2(p-1)}$        &      $-\frac{9\beta}{2}$       &       $\frac{(7-11p)\beta}{2(p-1)}$      &      $-\frac{15\beta}{2}$	\\	
 $ \nu^{\rm ssc}_{\rm m, r} < \nu < \nu^{\rm ssc}_{\rm c, r}$   	                & \hspace{0.5cm}  $\frac{p-1}{2}$  &\hspace{0.5cm}  $-$	                    &\hspace{0.5cm}  $-\frac{5+24\beta}{2}$ &\hspace{0.5cm}  $\frac{1-2\beta}{2}$ &\hspace{0.5cm}  $\frac{3(1-2\beta)}{2}$ \\ 	
 $\nu^{\rm ssc}_{\rm c,r} < \nu$   	                                 & \hspace{0.5cm}  $\frac{p}{2}$     &\hspace{0.5cm}  $-$	                    &\hspace{0.5cm}  $\frac{11-24\beta}{2}$ &\hspace{0.5cm}  $-\frac{1+2\beta}{2}$ &\hspace{0.5cm}  $\frac{3(1-2\beta)}{2}$\\

%
%
\hline
\end{tabular}
\end{table}

\begin{table}
\centering \renewcommand{\arraystretch}{1.6}\addtolength{\tabcolsep}{2pt}
\caption{Evolution of the SSC light curves ($F^{\rm ssc}_{\rm \nu, r}\propto t^{-\alpha}\nu^{-\beta}$) from the reverse shock in the thick- and thin-shell regime after shock crossing time.}
\label{tableReverseThick}
\begin{tabular}{cccccc}
\hline
\hline
     &               & ISM     & ISM      & wind    & wind      \\

     &               & $1<p<2$     & $2<p$       & $1<p<2$     & $2<p$       \\
\hline

 & $\beta$    & $\alpha$    & $\alpha$    & $\alpha$    & $\alpha$    \\
\hline
Thick shell &    &    &    &    &    \\
\hline
$ \nu < \nu^{\rm ssc}_{\rm m,r}$    &    $-\frac13$         &     $-\frac{106-59p}{72(p-1)}$        &      $\frac{1}{6}$       &       $-\frac{28-17p}{12(p-1)}$      &      $\frac{1}{2}$       \\
$ \nu^{\rm ssc}_{\rm m,r} < \nu < \nu^{\rm ssc}_{\rm cut,r}$    &    $\frac{p-1}{2}$           &       $\frac{5p+171}{96}$      &       $\frac{99p-17}{96}$      &      $\frac{45-p}{16}$       &      $\frac{21p+1}{16}$       \\

\hline
Thin shell &  &     &     &    &    \\
\hline
$ \nu < \nu^{\rm ssc}_{\rm m,r}$    &    $-\frac13$         &     $-\frac{143-75p}{105(p-1)}$        &      $\frac{1}{15}$       &       $-\frac{5(31-19p)}{63(p-1)}$      &      $\frac{5}{9}$       \\
$ \nu^{\rm ssc}_{\rm m,r} < \nu < \nu^{\rm ssc}_{\rm cut,r}$    &    $\frac{p-1}{2}$          &       $\frac{58+3p}{35}$      &       $\frac{37p-10}{35}$      &      $\frac{5(25-p)}{42}$       &      $\frac{5(11p+1)}{42}$       \\

\hline
\end{tabular}
\end{table}

\begin{table}
\centering \renewcommand{\arraystretch}{1.85}\addtolength{\tabcolsep}{1.5pt}
\caption{Closure relations of SSC ($F^{\rm ssc}_{\rm \nu,r}\propto t^{-\alpha(\beta)}\nu^{-\beta}$)  from the reverse shock afterglow model evolving in a thick- and thin-shell case after shock crossing time.}
\label{Table2}
\begin{tabular}{c c c  c c c}
 \hline \hline
&\hspace{0.5cm}     &\hspace{0.5cm}   ISM &\hspace{0.2cm}   ISM  & \hspace{0.5cm}   wind   & \hspace{0.5cm}  wind \\ 
&\hspace{0.5cm}     &\hspace{0.5cm} $1 < p <2$   &\hspace{0.5cm}   $2 < p$  & \hspace{0.2cm}    $1 < p <2$   & \hspace{0.5cm}  $2 < p$ \\
\hline
 &\hspace{0.5cm} $\beta$    &\hspace{0.5cm} $\alpha(\beta)$   &\hspace{0.5cm}   $\alpha(\beta)$  & \hspace{0.2cm}    $\alpha(\beta)$   & \hspace{0.5cm}  $\alpha(\beta)$    \\ \hline

Thick shell &\hspace{0.5cm}     &\hspace{0.2cm}    &\hspace{0.5cm}     & \hspace{0.5cm}       & \hspace{0.5cm}  \\ \hline 	
 $ \nu < \nu^{\rm ssc}_{\rm m, r}$   	        & \hspace{0.5cm}  $-\frac{1}{3}$           &\hspace{0.5cm}  $\frac{(106-59p)\beta}{24(p-1)}$	                    &\hspace{0.2cm}  $-\frac{\beta}{2}$ &\hspace{0.5cm}  $\frac{(28-17p)\beta}{4(p-1)}$ &\hspace{0.5cm}  $-\frac{3\beta}{2}$	\\	
 $ \nu^{\rm ssc}_{\rm m, r} < \nu < \nu^{\rm ssc}_{\rm cut, r}$   	                & \hspace{0.5cm}  $\frac{p-1}{2}$  &\hspace{0.5cm}  $\frac{5\beta+88}{48}$	                    &\hspace{0.2cm}  $\frac{99\beta+41}{48}$ &\hspace{0.5cm}  $\frac{22-\beta}{8}$ &\hspace{0.5cm}  $\frac{21\beta+11}{8}$ \\ 	
 $\nu^{\rm ssc}_{\rm cut,r} < \nu$   	                                 & \hspace{0.5cm}  $\frac{p}{2}$     &\hspace{0.5cm}  $\beta+2$	                    &\hspace{0.2cm}  $\beta+2$ &\hspace{0.5cm}  $\beta+2$ &\hspace{0.5cm}  $\beta+2$\\ 
\hline	
Thin shell &\hspace{0.5cm}     &\hspace{0.5cm}    &\hspace{0.5cm}     & \hspace{0.5cm}       & \hspace{0.5cm}  \\ \hline 	
 $ \nu < \nu^{\rm ssc}_{\rm m, r}$   	        & \hspace{0.5cm}  $-\frac{1}{3}$           &\hspace{0.5cm}  $\frac{(143-75p)\beta}{35(p-1)}$	                    &\hspace{0.5cm}  $-\frac{\beta}{5}$ &\hspace{0.5cm}  $\frac{5\beta(31-19p)}{21(p-1)}$ &\hspace{0.5cm}  $-\frac{5\beta}{3}$	\\	
 $ \nu^{\rm ssc}_{\rm m, r} < \nu < \nu^{\rm ssc}_{\rm cut, r}$   	                & \hspace{0.5cm}  $\frac{p-1}{2}$  &\hspace{0.5cm}  $\frac{61+6\beta}{35}$	                    &\hspace{0.5cm}  $\frac{74\beta+27}{35}$ &\hspace{0.5cm}  $\frac{5(12-\beta)}{21}$ &\hspace{0.5cm}  $\frac{5(11\beta+6)}{21}$ \\ 	
 $\nu^{\rm ssc}_{\rm cut,r} < \nu$   	                                 & \hspace{0.5cm}  $\frac{p}{2}$     &\hspace{0.5cm}  $\beta+2$	                    &\hspace{0.5cm}  $\beta+2$ &\hspace{0.5cm}  $\beta+2$ &\hspace{0.5cm}  $\beta+2$\\

%
%
\hline
\end{tabular}
\end{table}

\begin{table}
\centering
\caption{The quantity and percentage of GRBs that adhere to each of the CRs of the SSC scenario from the reverse-shock region developing in a thick- and thin-shell regime. We consider the values of spectral and temporal indeces of a PL and a BPL reported in 2FLGC.}
\label{tab:CR-Results}
\resizebox{\columnwidth}{!}{%
\begin{tabular}{cccccccccc}\cline{1-10}
\multicolumn{2}{c}{} &
  PL &
  $\beta$ &
  \begin{tabular}[c]{@{}c@{}}$\alpha(\beta)$\\ $1 < p < 2$\end{tabular} &
  \begin{tabular}[c]{@{}c@{}}$\alpha(\beta)$\\  $2 < p$\end{tabular} &
  \begin{tabular}[c]{@{}c@{}}Coincidences \\  PL \end{tabular} &
  \begin{tabular}[c]{@{}c@{}}Percent (\%) \\  PL \end{tabular} &
  \begin{tabular}[c]{@{}c@{}}Coincidences \\  BPL \end{tabular} &
  \begin{tabular}[c]{@{}c@{}}Percent (\%) \\  BPL \end{tabular} \\ \cline{1-10} 
\multicolumn{1}{c|}{\multirow{6}{*}{Thick shell}} & \multirow{3}{*}{ISM}  & $ \nu < \nu^{\rm ssc}_{\rm m, r}$ & $-\frac{1}{3}$ & $\frac{(106-59p)\beta}{24(p-1)}$ & $-\frac{\beta}{2}$ & 0 & 0.00 & 0 & 0.00 \\
\multicolumn{1}{c|}{}                       &                       & $ \nu^{\rm ssc}_{\rm m, r} < \nu < \nu^{\rm ssc}_{\rm cut, r}$ & $\frac{p-1}{2}$ & $\frac{5\beta+88}{48}$ & $\frac{99\beta+41}{48}$ & 1 & 1.16 & 4 & 4.65 \\
\multicolumn{1}{c|}{}                       &                       & $\nu^{\rm ssc}_{\rm cut,r} < \nu$ & $\frac{p}{2}$ & $\beta+2$ & $\beta+2$ & 1 & 1.16 & 4 & 4.65 \\ \cline{2-10} 
\multicolumn{1}{c|}{}                       & \multirow{3}{*}{Wind} & $ \nu < \nu^{\rm ssc}_{\rm m, r}$ & $-\frac{1}{3}$ & $\frac{(28-17p)\beta}{4(p-1)}$ & $-\frac{3\beta}{2}$ & 0 & 0.0 & 0 & 0.00 \\
\multicolumn{1}{c|}{}                       &                       & $ \nu^{\rm ssc}_{\rm m, r} < \nu < \nu^{\rm ssc}_{\rm cut, r}$ & $\frac{p-1}{2}$ & $\frac{22-\beta}{8}$ & $\frac{21\beta+11}{8}$ & 1 & 1.16 & 2 & 2.33 \\
\multicolumn{1}{c|}{}                       &                       & $\nu^{\rm ssc}_{\rm cut,r} < \nu$ & $\frac{p}{2}$ & $\beta+2$ & $\beta+2$ & 1 & 1.16 & 4 & 4.65\\ \hline
\multicolumn{1}{c|}{\multirow{6}{*}{Thin shell}}  & \multirow{3}{*}{ISM}  & $ \nu < \nu^{\rm ssc}_{\rm m, r}$ & $-\frac{1}{3}$ & $\frac{(143-75p)\beta}{35(p-1)}$ & $-\frac{\beta}{5}$ & 2 & 2.33 & 0 & 0.00 \\
\multicolumn{1}{c|}{}                       &                       & $ \nu^{\rm ssc}_{\rm m, r} < \nu < \nu^{\rm ssc}_{\rm cut, r}$ & $\frac{p-1}{2}$ & $\frac{61+6\beta}{35}$ & $\frac{74\beta+27}{35}$ & 1 & 1.16 & 5 & 5.81\\
\multicolumn{1}{c|}{}                       &                       & $\nu^{\rm ssc}_{\rm cut,r} < \nu$ & $\frac{p}{2}$ & $\beta+2$ & $\beta+2$ & 1 & 1.16 & 4 & 4.65 \\ \cline{2-10} 
\multicolumn{1}{c|}{}                       & \multirow{3}{*}{Wind} & $ \nu < \nu^{\rm ssc}_{\rm m, r}$ & $-\frac{1}{3}$ & $\frac{5\beta(31-19p)}{21(p-1)}$ & $-\frac{5\beta}{3}$ & 0 & 0.00 & 0 & 0.00 \\
\multicolumn{1}{c|}{}                       &                       & $ \nu^{\rm ssc}_{\rm m, r} < \nu < \nu^{\rm ssc}_{\rm cut, r}$ & $\frac{p-1}{2}$ & $\frac{5(12-\beta)}{21}$ & $\frac{5(11\beta+6)}{21}$ & 1 & 1.16 & 2 & 2.33 \\
\multicolumn{1}{c|}{}                       &                       & $\nu^{\rm ssc}_{\rm cut,r} < \nu$ & $\frac{p}{2}$ & $\beta+2$ & $\beta+2$ & 1 & 1.16 & 4 & 4.65 \\ \hline
\end{tabular}%
}
\end{table}

\begin{table}
\centering \renewcommand{\arraystretch}{1.6}\addtolength{\tabcolsep}{2pt}
\caption{Density parameter evolution in a stellar-wind and constant density medium ($F^{\rm ssc}_{\rm \nu, r} \propto n^{\alpha_{\rm k}} (A_{\rm W}^{\alpha_{\rm k}})$) in the the SSC scenario from the reverse shock after the shock-crossing time.}
\label{table:dens_paramet}
\begin{tabular}{cccccc}
\hline
\hline
     &               & ISM     & ISM      & wind    & wind      \\

     &               & $1<p<2$     & $2<p$       & $1<p<2$     & $2<p$       \\
\hline

 & $\beta$    & $\alpha_{\rm k}$    & $\alpha_{\rm k}$    & $\alpha_{\rm k}$    & $\alpha_{\rm k}$    \\
\hline
Thick shell &    &    &    &    &    \\
\hline
$ \nu < \nu^{\rm ssc}_{\rm m,r}$    &    $-\frac13$         &     $-\frac{15-14p}{12(p-1)}$        &      $\frac{3}{4}$       &       $-\frac{15-14p}{6(p-1)}$      &      $\frac{3}{2}$       \\
$ \nu^{\rm ssc}_{\rm m,r} < \nu < \nu^{\rm ssc}_{\rm cut,r}$    &    $\frac{p-1}{2}$           &       $\frac{11-2p}{8}$      &       $\frac{3p+5}{8}$      &      $\frac{11-2p}{4}$       &      $\frac{3p+5}{4}$       \\
$\nu^{\rm ssc}_{\rm cut,r} < \nu$    &  $\frac{p}{2}$       &         $0$         &    $0$         &       $0$      &       $0$      \\

\hline
Thin shell &  &     &     &    &    \\
\hline
$ \nu < \nu^{\rm ssc}_{\rm m,r}$    &    $-\frac13$         &     $-\frac{122-115p}{105(p-1)}$        &      $\frac{44}{45}$       &       $-\frac{164-157p}{63(p-1)}$      &      $\frac{22}{9}$       \\
$ \nu^{\rm ssc}_{\rm m,r} < \nu < \nu^{\rm ssc}_{\rm cut,r}$    &    $\frac{p-1}{2}$          &       $\frac{541-117p}{420}$      &       $\frac{425 -43p}{420}$      &      $\frac{241-53p}{84}$       &      $\frac{221 - 47p}{84}$       \\
$\nu^{\rm ssc}_{\rm cut,r} < \nu$    &  $\frac{p}{2}$             &    $0$         &    $0$         &       $0$      &       $0$      \\

\hline
\end{tabular}
\end{table}

\begin{table}
\centering \renewcommand{\arraystretch}{1.85}\addtolength{\tabcolsep}{1.5pt}
\caption{Closure relations of synchrotron ($F^{\rm syn}_{\rm \nu,r}\propto t^{-\alpha(\beta)}\nu^{-\beta}$)  from the reverse shock afterglow model evolving in a thick- and thin-shell case.}
\label{Table5}
\begin{tabular}{c c c  c c c}
 \hline \hline
&\hspace{0.5cm}     &\hspace{0.5cm}   ISM &\hspace{0.2cm}   ISM  & \hspace{0.5cm}   wind   & \hspace{0.5cm}  wind \\ 
&\hspace{0.5cm}     &\hspace{0.5cm} $1 < p <2$   &\hspace{0.5cm}   $2 < p$  & \hspace{0.2cm}    $1 < p <2$   & \hspace{0.5cm}  $2 < p$ \\
\hline
 &\hspace{0.5cm} $\beta$    &\hspace{0.5cm} $\alpha(\beta)$   &\hspace{0.5cm}   $\alpha(\beta)$  & \hspace{0.2cm}    $\alpha(\beta)$   & \hspace{0.5cm}  $\alpha(\beta)$    \\ \hline

Thick shell &\hspace{0.5cm}     &\hspace{0.2cm}    &\hspace{0.5cm}     & \hspace{0.5cm}       & \hspace{0.5cm}  \\ \hline 	
 $ \nu^{\rm syn}_{\rm m, r} < \nu < \nu^{\rm syn}_{\rm cut, r}$   	                & \hspace{0.5cm}  $\frac{p-1}{2}$  &\hspace{0.5cm}  $\frac{73\beta+47}{48}$	                    &\hspace{0.2cm}  $\frac{73\beta+47}{48}$ &\hspace{0.5cm}  $\frac{3(5\beta + 3)}{8}$ &\hspace{0.5cm}  $\frac{3(5\beta + 3)}{8}$ \\ 	
 $\nu^{\rm syn}_{\rm cut,r} < \nu$   	                                 & \hspace{0.5cm}  $\frac{p}{2}$     &\hspace{0.5cm}  $\beta+2$	                    &\hspace{0.2cm}  $\beta+2$ &\hspace{0.5cm}  $\beta+2$ &\hspace{0.5cm}  $\beta+2$\\ 
\hline	
Thin shell &\hspace{0.5cm}     &\hspace{0.5cm}    &\hspace{0.5cm}     & \hspace{0.5cm}       & \hspace{0.5cm}  \\ \hline 	
 $ \nu^{\rm syn}_{\rm m, r} < \nu < \nu^{\rm syn}_{\rm cut, r}$   	                & \hspace{0.5cm}  $\frac{p-1}{2}$  &\hspace{0.5cm}  $\frac{54\beta+34}{35}$	                    &\hspace{0.5cm}  $\frac{54\beta+34}{35}$ &\hspace{0.5cm}  $\frac{78\beta+46}{42}$ &\hspace{0.5cm}  $\frac{78\beta+46}{42}$ \\ 	
 $\nu^{\rm syn}_{\rm cut,r} < \nu$   	                                 & \hspace{0.5cm}  $\frac{p}{2}$     &\hspace{0.5cm}  $\beta+2$	                    &\hspace{0.5cm}  $\beta+2$ &\hspace{0.5cm}  $\beta+2$ &\hspace{0.5cm}  $\beta+2$\\

%
%
\hline
\end{tabular}
\end{table}

\begin{table}
\centering \renewcommand{\arraystretch}{1.5}\addtolength{\tabcolsep}{2pt}

\caption{Sample of bursts from 2FLGC described with a PL or a BPL for which at least one of the temporal indeces is greater than 2. Temporal and spectral PL indices are taken from \citet{2019ApJ...878...52A} with $\beta_{\rm L}=\Gamma_{\rm L}-1$.}
\label{tab:sample1}
\begin{tabular}{lccc}
\hline
GRB & $\alpha_{\rm L} \pm \delta_{\alpha_{\rm L}}$ & & $\beta_{\rm L} \pm \delta_{\beta_{\rm L}}$ \\
 & Decay index 1 (BPL) &  Decay index 2 (BPL)  &  \\

\hline
090510	&	$2.3\pm0.2$	&	$1.3\pm	0.2$ &	$1.15	\pm	0.08$	\\
090902B	&	$1.9\pm0.2$	&	$1.2\pm	0.2$ &	$0.92	\pm	0.06$	\\
090926A	&	$1.8\pm0.2$	&	$1.1\pm	0.2$ &	$0.86	\pm	0.07$	\\
100116A	&	$2.7\pm0.2$	&	$-$ &	$0.6	\pm	0.2$	\\
100414A	&	$1.8\pm0.2$	&	$0.3\pm	0.6$ &	$0.8	\pm	0.1$	\\
130327B	&	$1.3\pm0.9$	&	$1.7\pm	0.3$ &	$0.8	\pm	0.1$	\\
131108A	&	$1.9\pm0.5$	&	$1.2\pm	0.6$ &	$1.8	\pm	0.3$	\\
150627A	&	$3.0\pm0.2$	&	$0.41\pm0.07$ &	$0.7	\pm	0.1$	\\
160625B	&	$2.2\pm0.3$	&	$-$ &	$0.8	\pm	0.3$	\\
170115B	&	$2.0\pm2.0$	&	$0.9\pm	0.7$ &	$1.5	\pm	0.3$	\\
170214A	&	$2.0\pm2.0$	&	$1.6\pm	0.5$ &	$1.3	\pm	0.1$	\\
171010A	&	$2.2\pm0.7$	&	$1.0\pm	0.3$ &	$1.0	\pm	0.1$	\\
180720B	&	$1.5\pm0.2$	&	$3.2\pm	0.6$ &	$1.15	\pm	0.10$	\\


\hline
\end{tabular}
\end{table}

\begin{table}
\centering \renewcommand{\arraystretch}{1.5}\addtolength{\tabcolsep}{2pt}

\caption{Sample of 8 bursts with an atypical spectral PL indices with $\beta_{\rm L}=\Gamma_{\rm L}-1$.}
\label{tab:sample}
\begin{tabular}{lc}
\hline
GRB & $\beta_{\rm L} \pm \delta_{\beta_{\rm L}}$ \\
\hline
090427A	&	$-0.2\pm0.7$		\\
091127	&	$0.2	\pm	0.4$	\\
101014A	&	$0.3	\pm	0.3$	\\
101227B	&	$0.5	\pm	0.5$	\\
150514A	&	$0.1	\pm	0.4$	\\
160422A	&	$0.2	\pm	0.3$    \\
160702A	&	$0.4	\pm	0.4$    \\
160829A	&	$0.3	\pm	0.3$    \\
\hline
\end{tabular}
\label{tab:PLsample}
\end{table}

\begin{table*}
\centering \renewcommand{\arraystretch}{1.25}\addtolength{\tabcolsep}{1pt}
\caption{CRs of the forward-shock synchrotron afterglow model during the radiative regime in stellar-wind and homogeneous medium.}
\label{CR_FS_eps}
\begin{tabular}{ccccc}
\hline
 & & & Synchrotron & \\
 & & $\beta$ & $\alpha(\beta)$ & $\alpha(\beta)$ \\
 & & & $(1<p<2)$ & $(2<p)$ \\
\hline
  & & & Wind, Fast Cooling & \\
\hline
1 & $\nu^{\rm syn}_{\rm c, f}<\nu<\nu^{\rm syn}_{\rm m, f}$ & $\frac{1}{2}$ & $\frac{(\epsilon+2)\beta}{4-\epsilon}$ & $\frac{(\epsilon+2)\beta}{4-\epsilon}$ \\
2 & $\nu^{\rm syn}_{\rm m, f}<\nu$ & $\frac{p}{2}$ & $\frac{\beta+3}{4-\epsilon}$ & $\frac{\epsilon-2+\beta(6-\epsilon)}{4-\epsilon}$ \\
\hline
 & & & Wind, Slow Cooling & \\
\hline
3 & $\nu^{\rm syn}_{\rm m, f}<\nu<\nu^{\rm syn}_{\rm c, f}$ & $\frac{p-1}{2}$ & $\frac{2\beta+9-\epsilon}{2(4-\epsilon)}$ & $\frac{2+\beta(6-\epsilon)}{4-\epsilon}$ \\
4 & $\nu^{\rm syn}_{\rm c, f}<\nu$ & $\frac{p}{2}$ & $\frac{\beta+3}{4-\epsilon}$ & $\frac{\epsilon-2+\beta(6-\epsilon)}{4-\epsilon}$ \\
\hline
 & & & ISM, Fast Cooling & \\
\hline
5 & $\nu^{\rm syn}_{\rm c, f}<\nu<\nu^{\rm syn}_{\rm m, f}$ & $\frac{1}{2}$ & $\frac{4\beta(\epsilon+1)}{8-\epsilon}$ & $\frac{4\beta(\epsilon+1)}{8-\epsilon}$ \\
6 & $\nu^{\rm syn}_{\rm m, f}<\nu$ & $\frac{p}{2}$ & $\frac{3\beta+5+2\epsilon}{8-\epsilon}$ & $\frac{2(6\beta+\epsilon-2)}{8-\epsilon}$ \\
\hline
& & & ISM, Slow Cooling & \\
\hline
7 & $\nu^{\rm syn}_{\rm m, f}<\nu<\nu^{\rm syn}_{\rm c, f}$ & $\frac{p-1}{2}$ & $\frac{3(2\beta+3+2\epsilon)}{2(8-\epsilon)}$ & $\frac{3(4\beta+\epsilon)}{8-\epsilon}$ \\
8 & $\nu^{\rm syn}_{\rm c, f}<\nu$ & $\frac{p}{2}$ & $\frac{3\beta+5+2\epsilon}{8-\epsilon}$ & $\frac{2(6\beta+\epsilon-2)}{8-\epsilon}$ \\
\hline
\end{tabular}
\end{table*}

\begin{table*}
\centering \renewcommand{\arraystretch}{1.25}\addtolength{\tabcolsep}{1pt}
\caption{Summary of notation list of all parameters, the physical meaning, units and normalization convention used.}
\label{tab:sample}
\begin{tabular}{l c c c}
\hline
Symbol & Units & Definition & Convention\\
\hline
$m_p$	&	& Proton rest mass  &  \\
$m_e$	&	& Electron rest mass  &  \\
$\sigma_T$ & & Thompson cross section & \\
$T_{90}$	& (s) &  Duration of the prompt episode &\\
$p$ & & Electron PL index &\\
z	      &	& Redshift		      &  $\frac{1+Q}{x}$\\
$Y$ &  & Compton parameter &  $\frac{1+Q}{x}$\\
$g(p)$	&	& $\frac{p-2}{p-1}$   & $Q(x)=\frac{x-2}{x-1}$\\
$\tilde{g}(p)$	&  &	$\frac{2 - p}{p-1}$  &  $Q(x)=\frac{2 - x}{x-1}$\\
$\Gamma$  & & Bulk Lorentz factor & $Q_{\rm x}=\frac{Q}{10^{\rm x}}$\\
E	&	(erg) & Isotropic equivalent kinetic energy &$Q_{\rm x}=\frac{Q}{10^{\rm x}}$	\\
$E_{\rm \gamma, iso}$	& (erg) & Isotropic gamma-ray  energy & $Q_{\rm x}=\frac{Q}{10^{\rm x}}$\\
$\varepsilon_{\rm B_r}$	&	& Fraction of energy given to amplify the magnetic field & $Q_{\rm x}=\frac{Q}{10^{\rm x}}$\\
$\varepsilon_{\rm e_r}$	&	& Fraction of energy given to accelerate electrons & $Q_{\rm x}=\frac{Q}{10^{\rm x}}$\\
$\chi_{\rm e}$ &  & Amount of electrons accelerated in shock  & $Q_{\rm x}=\frac{Q}{10^{\rm x}}$ \\
$\eta$  &   & $E_{\rm \gamma, iso}/(E+E_{\rm \gamma, iso})$ (kinetic efficiency) & $Q_{\rm x}=\frac{Q}{10^{\rm x}}$\\
$\dot{M}_W$	& $M_\odot\,yr^{-1}$ & Mass lost rate  & $Q_{\rm x}=\frac{Q}{10^{\rm x}}$\\
$v_W$	&	(${\rm km\, s^{-1}}$) & Wind speed & $Q_{\rm x}=\frac{Q}{10^{\rm x}}$\\
$n$     & (${\rm cm^{-3}}$)  & Density (constant) & $Q_{\rm x}=\frac{Q}{10^{\rm x}}$\\
$A_W$	& 	& Density parameter (wind) &  $Q_{\rm x}=\frac{Q}{10^{\rm x}}$ \\
$t_x$	& ({\rm s})	& Shock crossing time  & $Q_{\rm x}=\frac{Q}{10^{\rm x}}$\\
$\Delta$	& ({\rm cm}) &	Observed width of the shell &  $Q_{\rm x}=\frac{Q}{10^{\rm x}}$  \\
$d_z$	& ({\rm cm})	& Luminosity distance  & $Q_{\rm x}=\frac{Q}{10^{\rm x}}$ \\
\hline
Derived Quantities\\
\hline
$\Gamma_{\rm c}$  & &Critical Lorentz factor\\
$\gamma_{\rm m}$	&  & Electron minimum Lorentz factor\\
$\gamma_{\rm c}$ & & Electron cooling Lorentz factor\\
$\gamma_{\rm max}$ & & Electron maximum Lorentz factor\\
$\nu^{\rm syn}_{\rm m}$	& ({\rm eV}) & Characteristic frequency of synchrotron radiation\\
$\nu^{\rm syn}_{\rm c}$	& ({\rm eV}) & Cooling frequency of synchrotron radiation\\
$\nu^{\rm syn}_{\rm cut}$	& ({\rm eV}) & Cut-off frequency of synchrotron radiation\\
$F^{\rm syn}_{\rm max}$	& ({\rm mJy}) & Maximum synchrotron flux \\
$\nu^{\rm SSC}_{\rm m}$	& ({\rm GeV}) & Characteristic frequency of SSC process\\
$\nu^{\rm SSC}_{\rm c}$	& ({\rm GeV}) & Cooling frequency of SSC process\\
$\nu^{\rm syn}_{\rm cut}$	& ({\rm GeV}) & Cut-off frequency of SSC process\\
$F^{\rm SSC}_{\rm max}$	& ({\rm mJy}) & Maximum SSC flux \\
$\nu^{\rm SSC}_{\rm KN, m}$	& ({\rm GeV}) & Characteristic frequency of SSC process in the KN regime\\
$\nu^{\rm SSC}_{\rm KN, c}$	& ({\rm GeV}) & Cooling frequency of SSC process in the KN regime\\

\hline
\end{tabular}
\end{table*}

\begin{figure}
{ \centering
\resizebox*{\textwidth}{0.6\textheight}
{\includegraphics{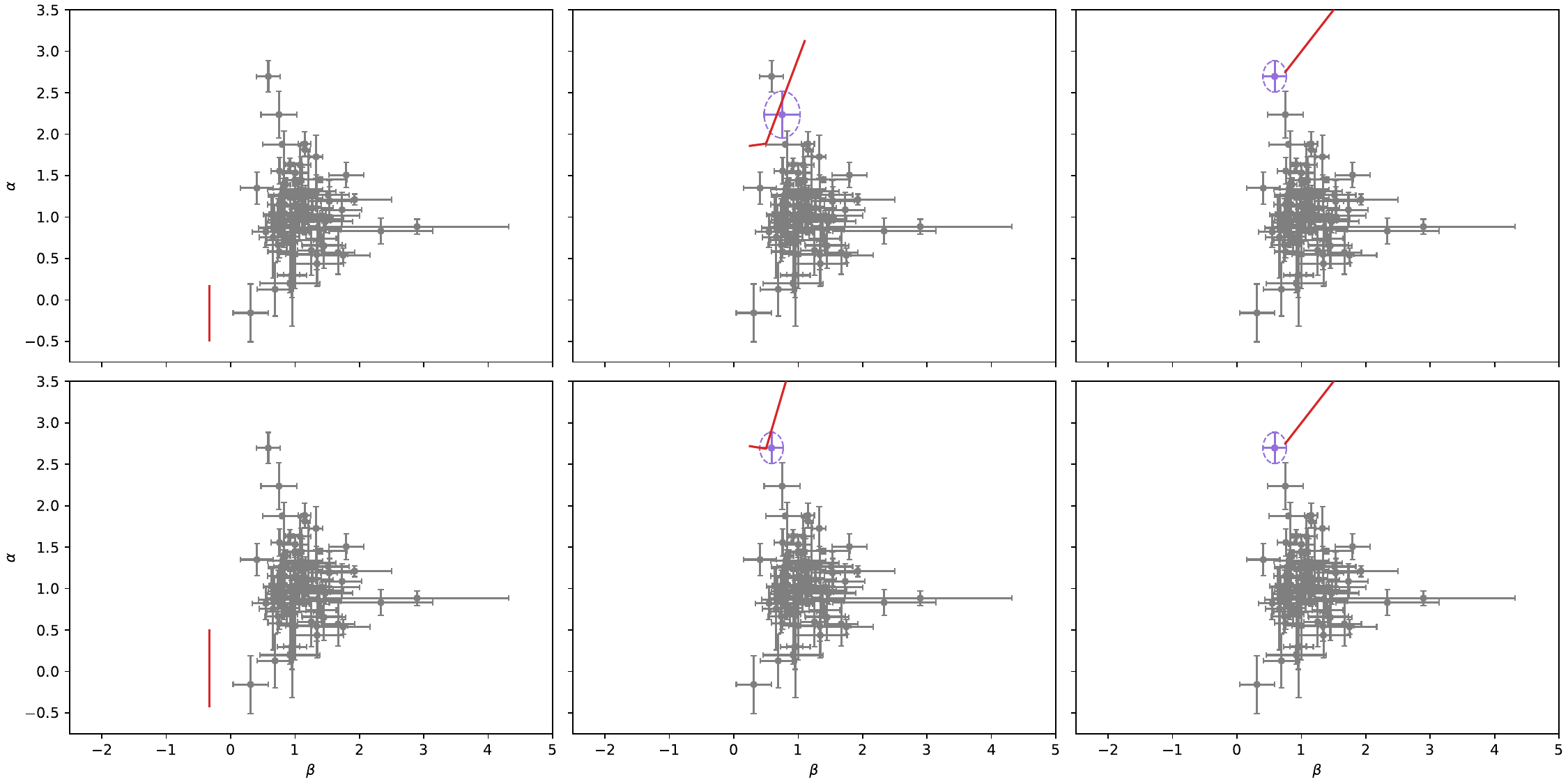}
}}
\caption{Spectral and temporal PL indexes of bursts reported in 2FLGC (gray) together with CRs of SSC from the reverse shock region evolving in the thick case (red lines). Data associated to the ellipses in purple display the cases in which the CR are fulfilled within 1 sigma error bars. Panels, from left to right, correspond to the cooling condition $ \nu < \nu^{\rm ssc}_{\rm m,r} $, $ \nu^{\rm ssc}_{\rm m,r} < \nu < \nu^{\rm ssc}_{\rm cut,r}$ and $ \nu^{\rm ssc}_{\rm cut,r} < \nu $.}
 \label{fig1:CRs}
\end{figure}

\begin{figure}
{ \centering
\resizebox*{\textwidth}{0.6\textheight}
{\includegraphics{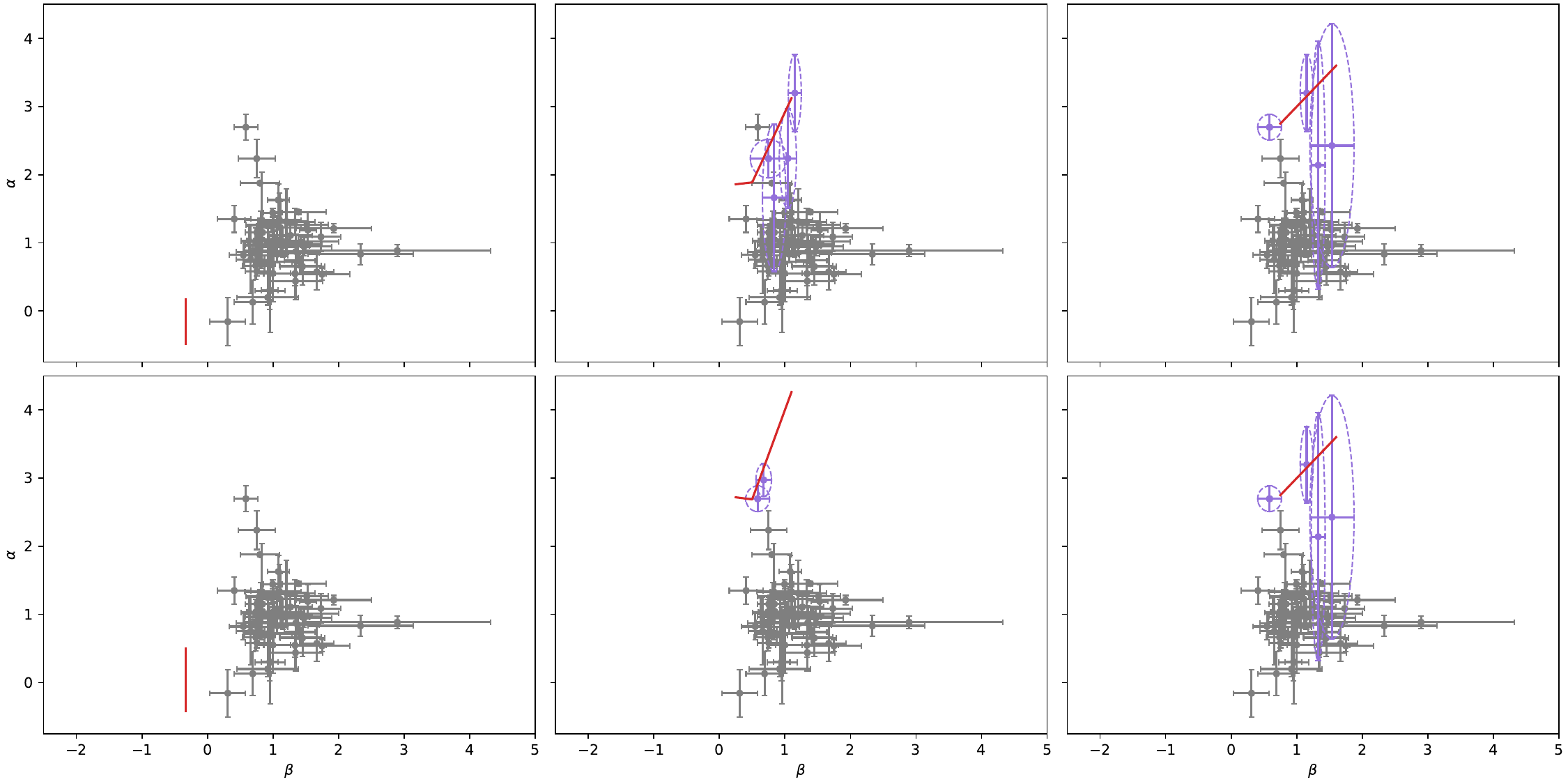}
}}
\caption{The same of Figure \ref{fig1:CRs}, but we consider the description of a BPL reported in  2FLGC for some bursts.}
 \label{fig2:CRs}
\end{figure}

\begin{figure}
{ \centering
\resizebox*{\textwidth}{0.6\textheight}
{\includegraphics{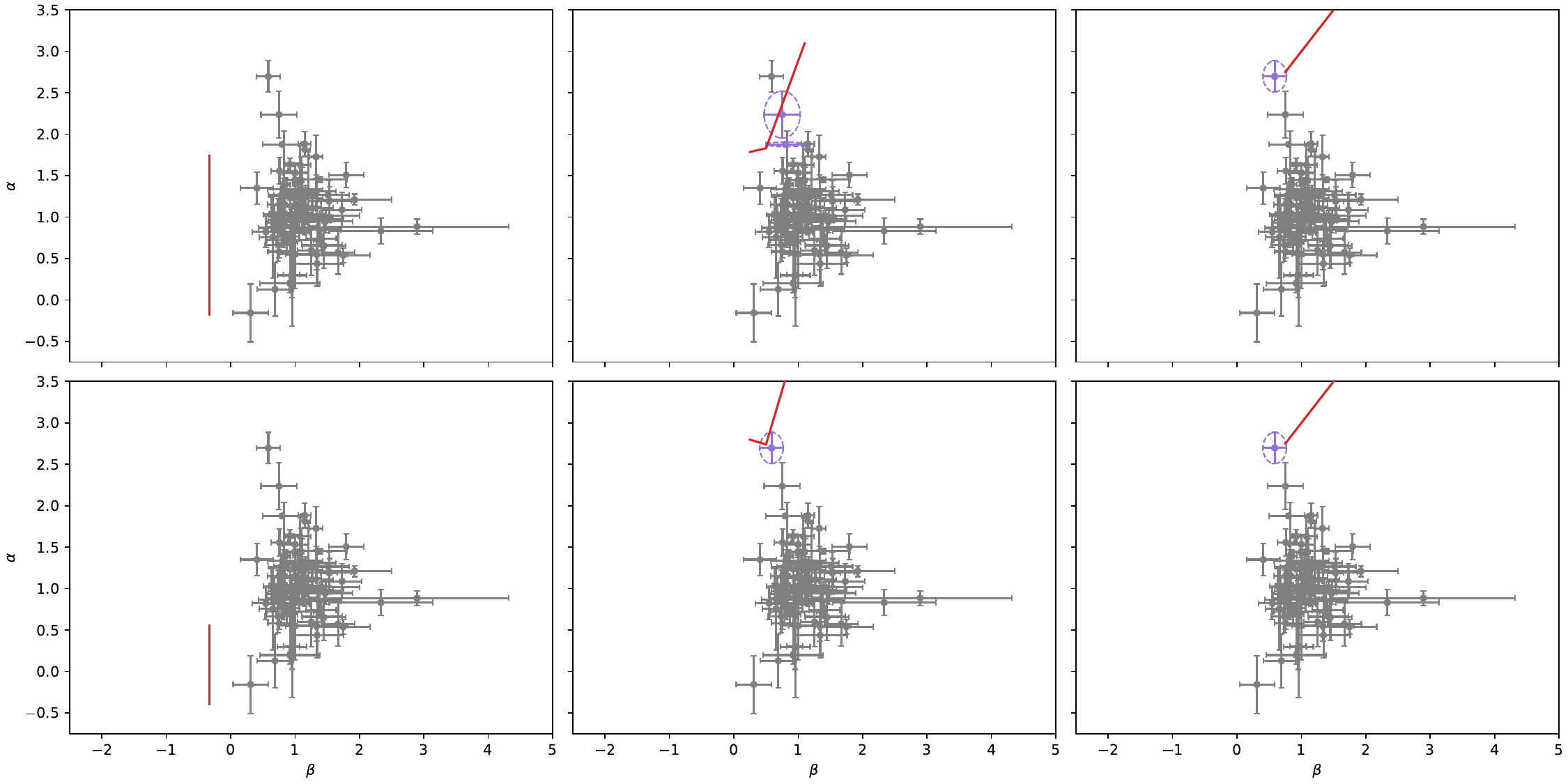}
}}
\caption{The same of Figure \ref{fig1:CRs}, but for the reverse shock evolving in the thick case.}
 \label{fig3:CRs}
\end{figure}

\begin{figure}
{ \centering
\resizebox*{\textwidth}{0.6\textheight}
{\includegraphics{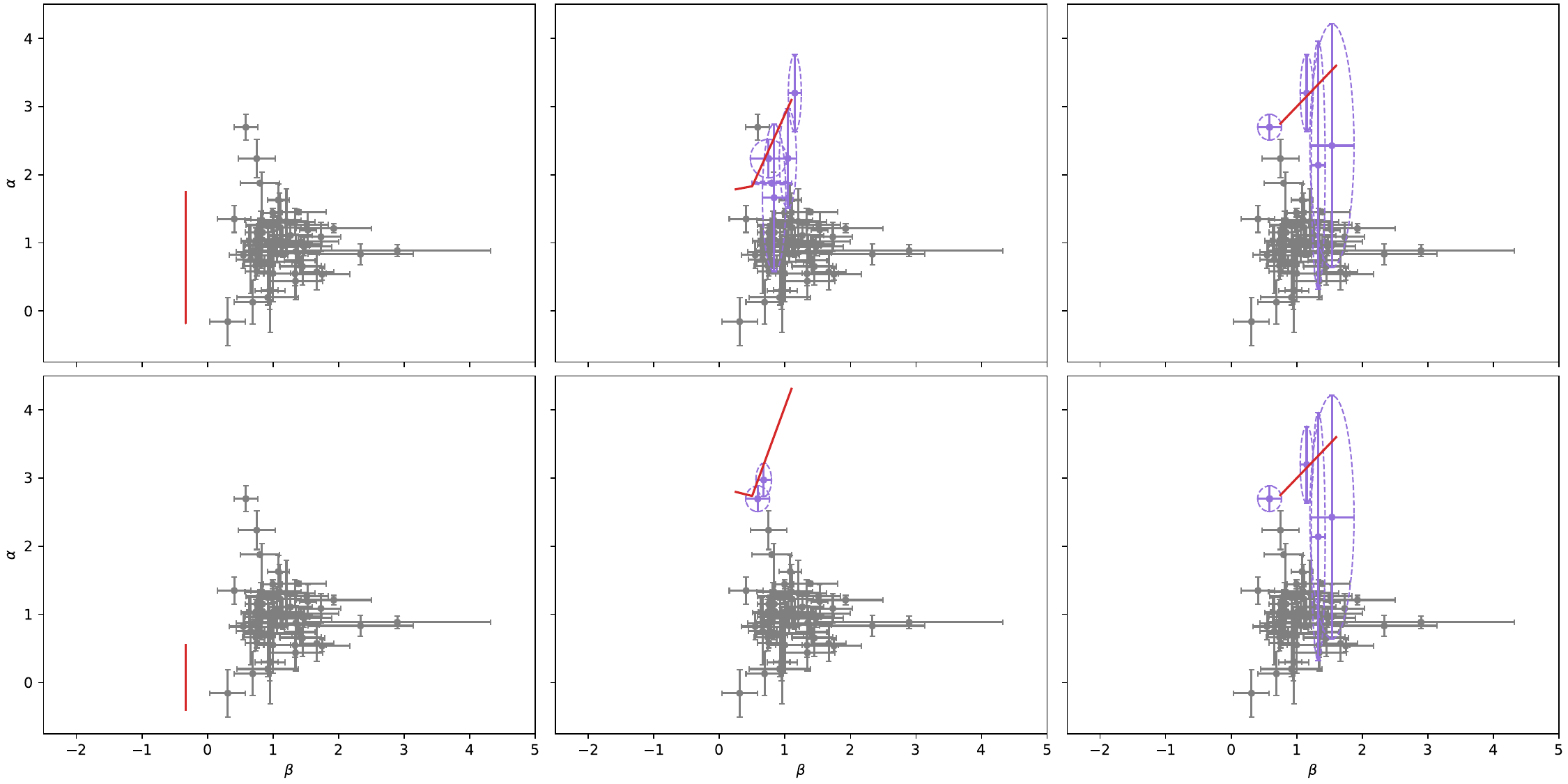}
}}
\caption{The same of Figure \ref{fig3:CRs}, but we consider the description of a BPL reported in  2FLGC for some bursts.}
 \label{fig4:CRs}
\end{figure}

\begin{figure}
{ \centering
\resizebox*{\textwidth}{0.95\textheight}
{\includegraphics{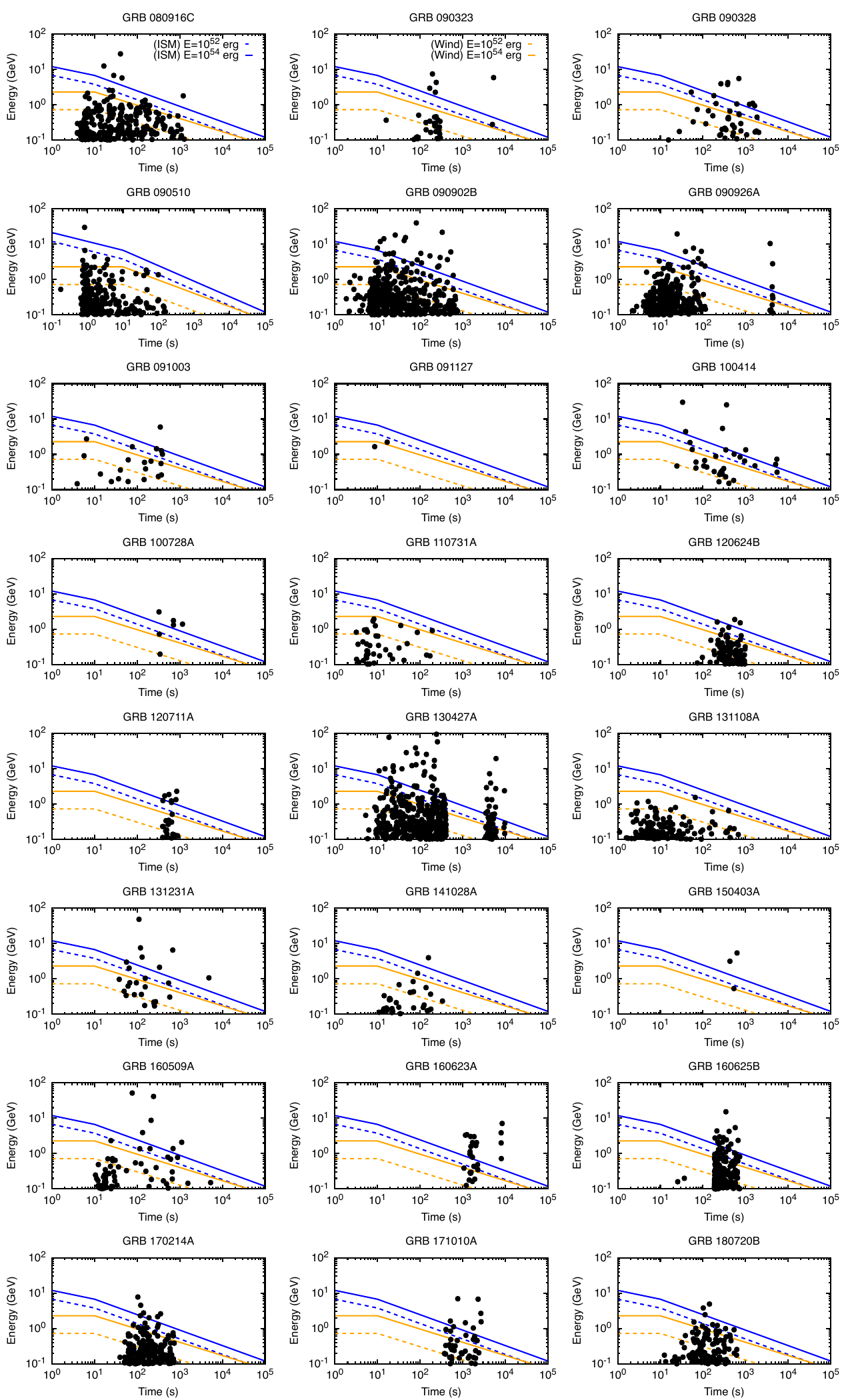}}}
\caption{Each burst in our sample is associated with high-energy photons above 100 MeV and a probability of $>90$\% to be associated. The maximal photon energies released by the synchrotron reverse-shock scenario that evolves in the thick-shell regime are shown for stellar-wind and ISM. The values used are $n=1\,{\rm cm^{-3}}$, $A_{\rm W}=1$, $\Gamma=300$, and $z=1$.}
\label{fig7:CRs}
\end{figure}

\begin{figure}
{ \centering
\resizebox*{\textwidth}{0.965\textheight}
{\includegraphics{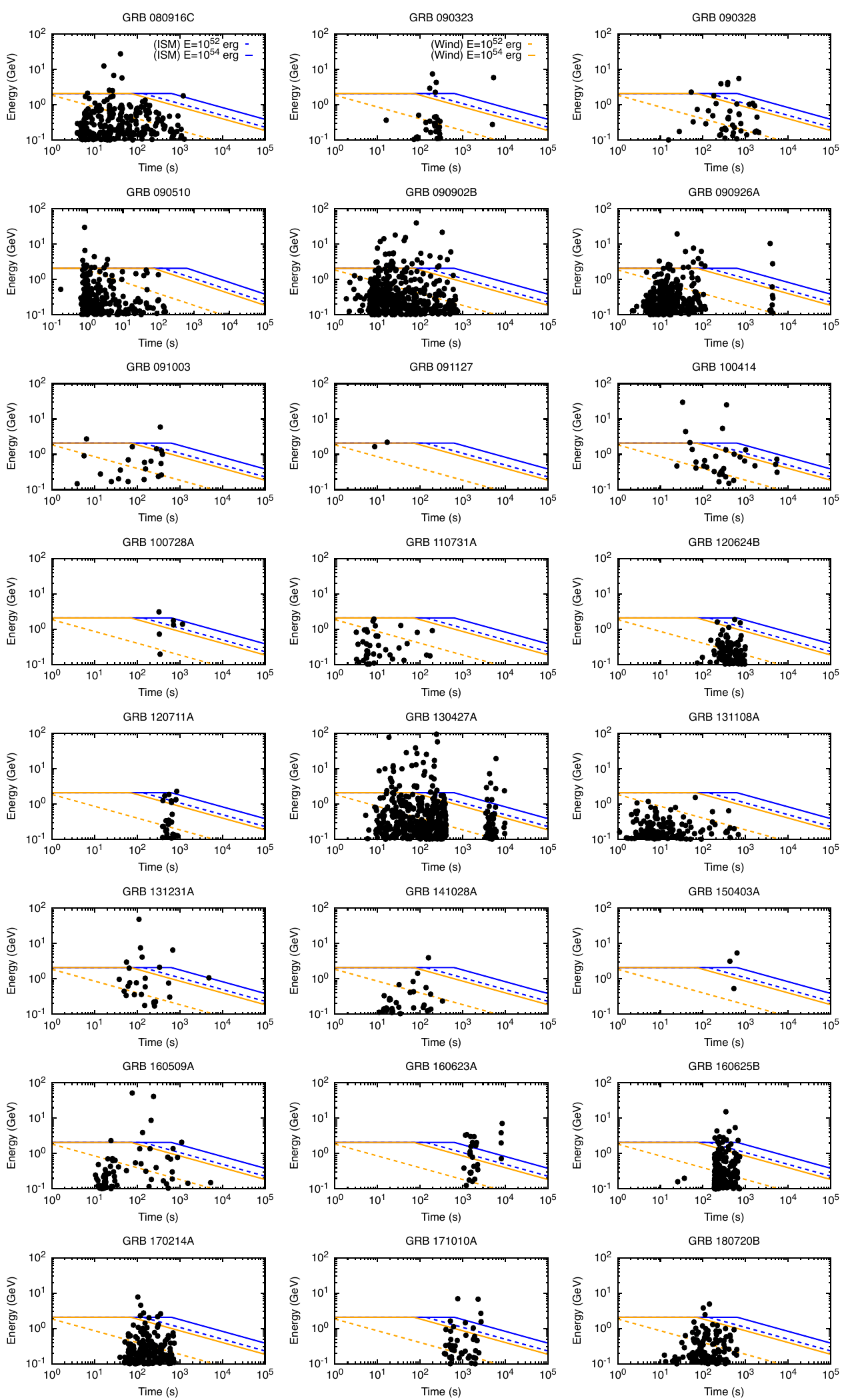}}
}
\caption{The same of Figure \ref{fig7:CRs}, but for the thin-shell regime. }
 \label{fig8:CRs}
\end{figure}

\begin{figure}
{ \centering
\resizebox*{\textwidth}{0.6\textheight}
{\includegraphics{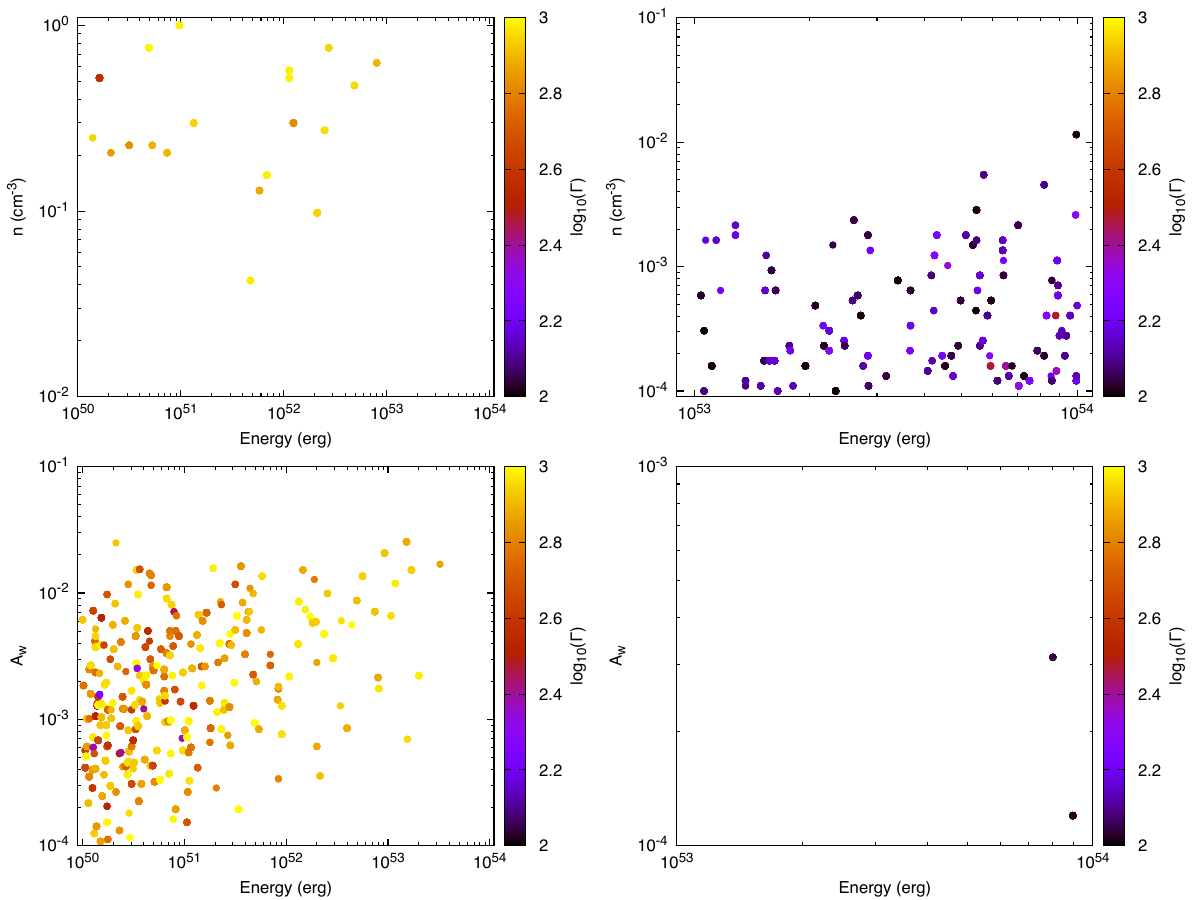}
}}
\caption{Parameter space of the density of the surrounding medium, kinetic energy and bulk Lorentz factor for which 
the CRs of SSC emission from the reverse shock satisfy the atypical spectral index ($\beta \approx 0$) and the SSC flux is above 100 MeV. The evolution of the reverse shock in the thick (left) and thin (right) regime is considered.}
 \label{fig:param_space}
\end{figure}


\begin{figure}
{ \centering
\resizebox*{\textwidth}{0.6\textheight}
{\includegraphics{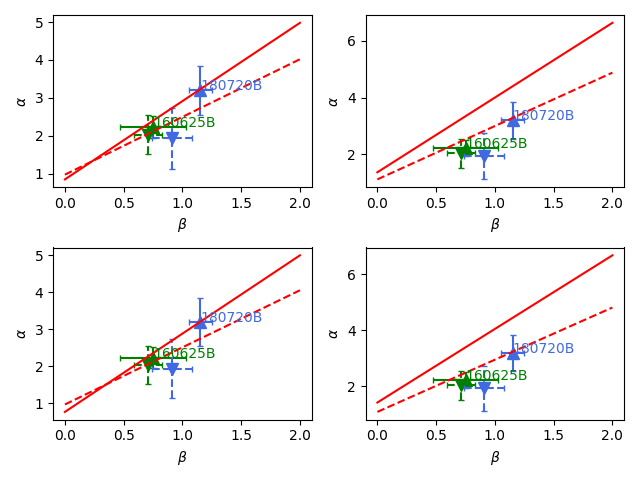}
}}
\caption{Spectral and temporal indexes of GRB 160625B and GRB 180720B in the LAT (solid) and optical (dahsed) bands together with the CRs of SSC (solid red line) and synchrotron (dashed red line) flux from the reverse-shock region in the cooling condition $ \nu^{\rm ssc}_{\rm m,r} < \nu < \nu^{\rm ssc}_{\rm cut,r}$ and $ \nu^{\rm syn}_{\rm m,r} < \nu < \nu^{\rm syn}_{\rm cut,r}$, respectively. Panels above and below correspond to the evolution of the reverse shock in thick and thin shell, respectively, for ISM (left) and stellar wind (right).}
 \label{fig5:CRs}
\end{figure}

\begin{figure}
{ \centering
\resizebox*{\textwidth}{0.45\textheight}
{\includegraphics{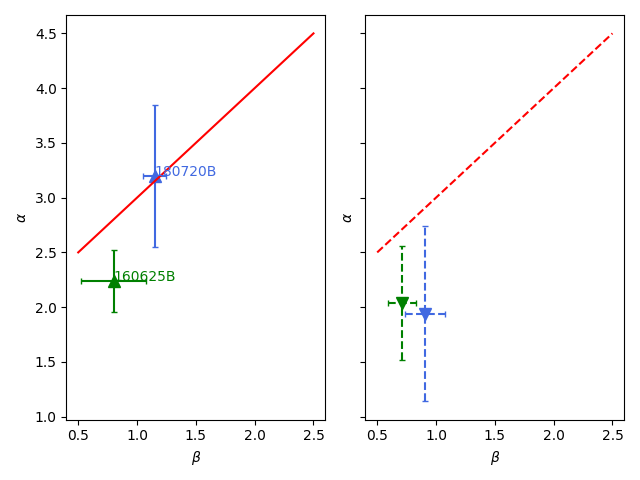}
}}
\caption{The same as Figure \ref{fig5:CRs}, but the reverse shock lies in the cooling condition $ \nu^{\rm ssc}_{\rm cut,r} < \nu $ and $ \nu^{\rm syn}_{\rm cut,r} < \nu $ for SSC (left) and synchrotron (right) flux, respectively.}
 \label{fig6:CRs}
\end{figure}

\begin{figure}
{ \centering
\resizebox*{\textwidth}{0.4\textheight}
{\includegraphics{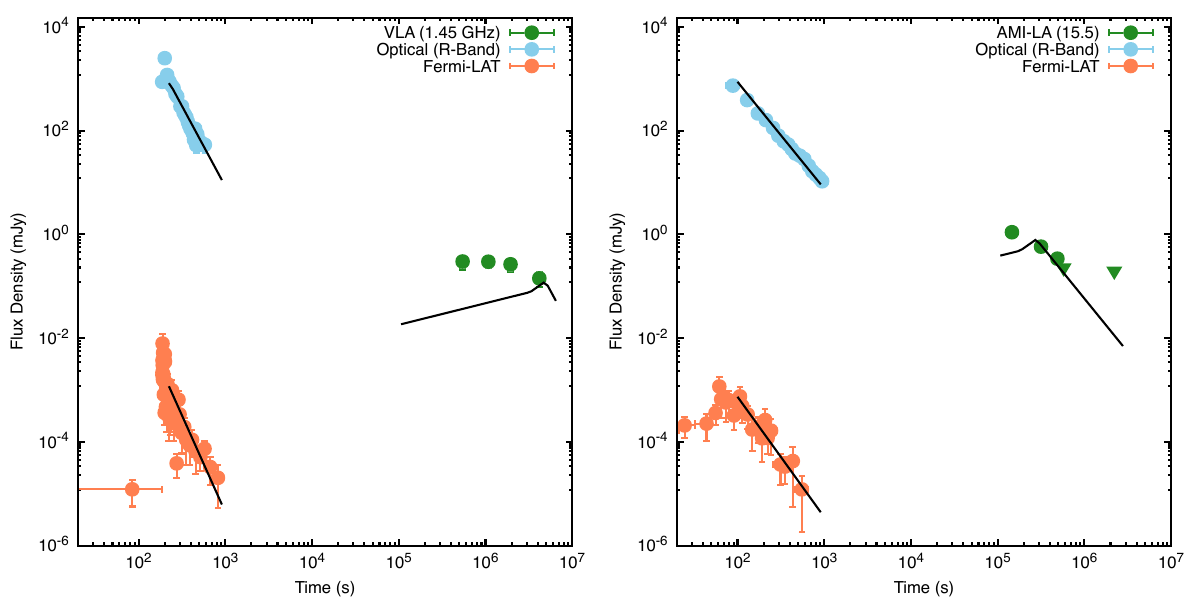}
}}
\caption{Early optical and LAT observations of GRB 160625B (left) and GRB 180720B (right) together the best-fit curve for the synchrotron and SSC processes from the reverse shock region. The best-fit values found for the GRB 160825B where; E = $3.16\times10^{53} \ \rm erg$, n = $1.00\times10^{0} \ \rm cm^{-3}$, $\rm \Gamma = 1.00\times10^{3}$, $\rm \varepsilon_{e_r} = 8.00\times10^{-1}$, $\rm \varepsilon_{B_r} = 3.23\times10^{-6}$ and p=$3.79$, whereas, for the GRB 180720B the values where E = $3.16\times10^{53} \ \rm erg$, n = $9.43\times10^{-1} \ \rm cm^{-3}$, $\rm \Gamma = 9.86\times10^{2}$, $\rm \varepsilon_{e_r} = 7.96\times10^{-1}$, $\rm \varepsilon_{B_r} = 2.80\times10^{-5}$ and p=$2.27$.  The radio observations of GRB 160625B and GRB 180720B were taken from \citeauthor{2020ApJ...904..166C} and
\citeauthor{2020MNRAS.496.3326R}, respectively.}
 \label{fig:best_fit}
\end{figure}


\bsp	
\label{lastpage}
\end{document}